\documentclass[%
 reprint,
 superscriptaddress,
 amsmath,amssymb,
 apl,
]{revtex4-2}

\usepackage[utf8x]{inputenc} % added by arXiv

\usepackage{graphicx}% Include figure files
\usepackage[colorlinks]{hyperref}% add hypertext capabilities
\usepackage{comment}

 \newcommand{\lc}{\left\{}
 \newcommand{\rc}{\right\}}
 \newcommand{\lb}{\left(}
 \newcommand{\rb}{\right)}
 \newcommand{\ls}{\left[}
 \newcommand{\rs}{\right]}
 \newcommand{\abs}[1]{\left\vert #1\right\vert}
 
 \newcommand{\di}{{\rm d}}
 \newcommand{\I}{{\rm i}}

 \usepackage{bm}

\DeclareRobustCommand{\uvec}[1]{{%
  \ifcsname uvec#1\endcsname
     \csname uvec#1\endcsname
   \else
    \bm{\hat{\mathbf{#1}}}%
   \fi
}}

\usepackage[hints]{minitoc}
\renewcommand \thepart{}
\renewcommand \partname{}
\mtcsetfeature{parttoc}{open}{\vspace{5mm}}
\mtcsetfeature{parttoc}{close}{\vspace{1mm}}

\begin{document}

\doparttoc % Tell minitoc to generate a toc for the parts
\faketableofcontents % Run a fake tableofcontents command for the partocs
%\part{} % Start the document part
%\parttoc % Insert the document TOC

\preprint{APS/123-QED}

\title{A simple test for causality in complex systems}

\author{Kristian Agasøster Haaga}

 \affiliation{Department of Earth Science, University of Bergen, PO Box 7803, NO-5020 Bergen, Norway}
 \affiliation{K.G. Jebsen Centre for Deep Sea Research, PO Box 7803, NO-5020 Bergen, Norway}
 \affiliation{Bjerknes Centre for Climate Research, PO Box 7803, NO-5020 Bergen, Norway}
\email{kristian.haaga@uib.no}

\author{David Diego}
 \affiliation{Department of Earth Science, University of Bergen, PO Box 7803, NO-5020 Bergen, Norway}
 \affiliation{K.G. Jebsen Centre for Deep Sea Research, PO Box 7803, NO-5020 Bergen, Norway}

\author{Jo Brendryen}
 \affiliation{Department of Earth Science, University of Bergen, PO Box 7803, NO-5020 Bergen, Norway}
 \affiliation{K.G. Jebsen Centre for Deep Sea Research, PO Box 7803, NO-5020 Bergen, Norway}
 \affiliation{Bjerknes Centre for Climate Research, PO Box 7803, NO-5020 Bergen, Norway}

\author{Bjarte Hannisdal}
 \affiliation{Department of Earth Science, University of Bergen, PO Box 7803, NO-5020 Bergen, Norway}
 \affiliation{K.G. Jebsen Centre for Deep Sea Research, PO Box 7803, NO-5020 Bergen, Norway}
 \affiliation{Bjerknes Centre for Climate Research, PO Box 7803, NO-5020 Bergen, Norway}
\date{\today}

\begin{abstract}
\textbf{
We provide a new solution to the long-standing problem of inferring causality from observations without modeling the unknown mechanisms. We show that the evolution of any dynamical system is related to a predictive asymmetry that 
quantifies causal connections from limited observations.
A built-in significance criterion obviates surrogate testing and drastically improves computational efficiency. 
We validate our test on numerous synthetic systems exhibiting behavior commonly occurring in nature, from linear and nonlinear stochastic processes to systems exhibiting non-linear deterministic chaos, and on real-world data with known ground truths. Applied to the controversial problem of glacial-interglacial sea level and CO$_2$ evolving in lock-step, our test uncovers empirical evidence for CO$_2$ as a driver of sea level over the last 800 thousand years. 
Our findings are relevant to any discipline where time series are used to study natural systems.
}
\end{abstract}

%Use showkeys class option if keyword
\maketitle

%\tableofcontents

\section{\label{sec:introduction}Introduction}
Natural systems, such as ecosystems, Earth systems, or the human brain, pose formidable challenges to causal inference.  Complex underlying dynamics may render traditional statistical means of correlation powerless to resolve causal relationships in real-world data. If process modeling is impractical or if model parameters cannot be constrained, then the prospect of non-parametric detection of causality directly from observations becomes tantalizing. 
Hence, several methods aimed at quantifying causal interactions from observed time series have been proposed 
\cite{Granger1969,
Chen2004extendedgranger,
Marinazzo2008kernelgranger,
Schreiber2000, 
Palus2007cmi, Rulkov1995generalizedsync_closeness,
Schiff1996mutualprediction,
Arnhold1999robustinterdependences,
Quiroga2002synchronization,
Chicharro2009directionrankstats,
Sugihara2012,
Wiesenfeldt2001mixedstateanalysis,
Feldmann2004predictabilityimprovement,
Krakovska2016mixredprediction,
Liang2013information,
McCracken2016}. 

Widely used methods include Granger causality \cite{Granger1969}, its information-theoretic cousin, transfer entropy \cite{Schreiber2000}, and geometrically inspired methods such as convergent cross mapping \cite{Sugihara2012, Ye2015}. Despite their promise, attempts to quantify the strength and directionality of causal interactions from observed time series, without recourse to modeling, remain controversial. For example, the applicability of these causality detection methods is system-dependent, can suffer from biases and low statistical power \cite{Krakovska2018, Smirnov2013spurious}, and usually requires additional statistical testing against surrogate data \cite{Theiler1992, Lancaster2018}, which inevitably introduces subjectivity when it comes to surrogate data design. Moreover, numerical estimates of information-theoretic quantities such as transfer entropy may not converge to zero for uncoupled systems, may overestimate or fail to quantify information flow, or underestimate dynamical influence \cite{Smirnov2013spurious, James2016information}. In appendix \ref{sec_supp:comparisons_te_pa}, we demonstrate some of these issues using transfer entropy (TE) as an exemplar.

Here, we present the predictive asymmetry --- a simple and robust causality test that by construction overcomes many of these issues. Our test is based on a difference between two information-theoretic functionals computed directly from observed time series. We prove that this simple difference is directly related to the flow of the underlying dynamics. 
By quantifying the \textit{difference} between forwards-in-time and backwards-in-time prediction, our predictive asymmetry test unequivocally determine the correct causation in cases where TE alone fails (Figs. 
\hyperref[fig_supp:comparison_te_pa_chen2d_nonlinearsystem_nocoupling]{\ref{fig_supp:comparison_te_pa_chen2d_nonlinearsystem_nocoupling}}, \hyperref[fig_supp:comparison_te_pa_chen2dnonlinearcoupling]{\ref{fig_supp:comparison_te_pa_chen2dnonlinearcoupling}}, \hyperref[fig_supp:comparison_te_pa_rosslerlorenz_nocoupling]{\ref{fig_supp:comparison_te_pa_rosslerlorenz_nocoupling}}, \hyperref[fig_supp:comparison_te_pa_rosslerlorenz_withcoupling]{\ref{fig_supp:comparison_te_pa_rosslerlorenz_withcoupling}}). 
Our test provides a statistic that is zero when there is no coupling, positive in the causal direction (driver $\to$ response) when directional coupling exists, and negative in the non-causal direction (response $\to$ driver) when directional coupling exists. 
Simultaneously, by using an intrinsic, dynamically informed significance test, our method alleviates computational demands associated with surrogate testing, raising the prospect of fast quantification of causal networks from large datasets.

In the following, we formally derive the predictive asymmetry statistic, and present analytical and numerical results demonstrating its robustness as a quantifier of directional causality. 
We explore test performance both in the small-data frontier, as well as its asymptotic behavior for time series with more observations. 
Then we verify the method on multiple real-world datasets with known ground truths, showcasing our recommended workflow for data with uncertainties and a limited number of observations. Finally, we apply the method to paleoclimate time series, 
identifying atmospheric CO$_2$ as a key driver of global sea level on glacial-interglacial time scales.

\section{\label{sec:pred_test}A predictive test for causality}

\subsection{A causality statistic linked to the flow of dynamical systems}

Consider a generic coupled dynamical system generated by the vector field 

{
\begin{subequations}
\begin{align}
    \dot{\chi} &= f(\xi, \chi), \qquad \chi \in \mathbb{R}^{n_1} \label{eq:dynamical_system_vector_field_1}\\
    \dot{\xi} &= g(\xi, \chi), \qquad \xi \in \mathbb{R}^{n_2} \label{eq:dynamical_system_vector_field_2}
\end{align}
\label{eq:dynamical_system_vector_field}
\end{subequations}
}

\noindent
and denote by $\phi(t; \chi, \xi)$ its evolution operator. Without loss of generality, assume that $\partial_{\chi_1} g_1 \neq 0$, so that there is coupling between the variables, and define the time series $x(t) = \chi_1(t)$ and $y(t) = \xi_1(t)$. We introduce the following difference between information-theoretic functionals as a causality statistic:

\begin{equation}
    \mathbb{A}_{x \to y}(\eta) := \int_{0}^{\eta} \left( TE_{x\to y}(\nu) - TE_{x\to y}(-\nu)\right) \di \nu,
    \label{eq:asymmetry}
\end{equation}

\noindent for some prediction lag $\eta > 0$, where  $TE_{x\to y}(\eta)$ is the TE corresponding to the prediction lag $\eta$ (Appendix \ref{sec:supp_experimental_setup}). This quantity, which is a difference in predictability forwards and backwards in time using TE, measures the evolution of the system through the equality $\mathbb{A}_{x \to y}(\eta) = \int_{0}^{\eta} \di \nu \int_{\mathbb{E}(\nu)} P \ln{(K)}$. Similar identities are also found using mutual information (Appendix \ref{sec:supp_formal_derivation}). We show that 

{
\begin{eqnarray}
TE_{x \to y}(|\eta|) - TE_{x \to y}(|-\eta|) = \int_\mathbb{E} P \ln{K},
\end{eqnarray}
}

\noindent where $\mathbb{E}$ is a generalized delay reconstruction of the dynamics from $x(t)$ and $y(t)$,
$P$ is the invariant distribution over $\mathbb{E}$ and $K$ is a quantity closely related to the flow of the system $\phi(t; \chi, \xi)$ as

{\begin{equation}
K \propto |\partial_{\phi_{\chi_1}(\eta; \chi, \xi)} \phi_{\chi_1}(-2\eta; \chi, \xi)|. \label{eq:K}
\end{equation}
}

\noindent We show that if there is no direct coupling $x \to y$, then  $\int_{\mathbb{E}(\nu)} P \ln{(K)} = 0$ when using an appropriate delay reconstruction $\mathbb{E}(\nu)$ (Appendix \ref{sec:supp_formal_derivation}). The positivity or negativity of $\int_{\mathbb{E}(\nu)} P \ln{(K)}$ in the general case is not obvious, but for a widely used family of stochastic systems, we can show that if a coupling $x \to y$ exists, then the sign of $\mathbb{A}_{x \to y} > 0$ (while $\mathbb{A}_{y \to x} < 0$), and that $\mathbb{A}_{x \to y} = 0$ when no coupling exists.

\subsection{Sign and magnitude of predictive asymmetry reflects underlying coupling}

For systems of random variables, marginal entropies can be computed directly from the covariance matrix of the system \cite{Hahs2013}. This allows us to obtain exact expressions for the predictive asymmetries (eq. \ref{eq:asymmetry}) for stochastic processes with known parameters. Here, we demonstrate predictive asymmetries on the following  unidirectionally coupled, stationary autoregressive system with $|a| < 1$, $c_{xy} \geq 0$,  and innovations $w_t \thicksim N(0, \sigma_x)$ and $v_t \thicksim N(0, \sigma_y)$ (Appendix \ref{sec_supp:analytical_asymmetry_ar1}):

{
\begin{subequations}
\begin{align}
x_{t} &= a x_{t-1} + w_{t} \label{eq:ar1_example_y} \\
y_{t} &= c_{xy} x_{t-1} + v_{t}. \label{eq:ar1_example_x}
\end{align}
\label{eq:ar1_example}
\end{subequations}
}

\begin{figure}[h]
\includegraphics[width=1.0\linewidth]{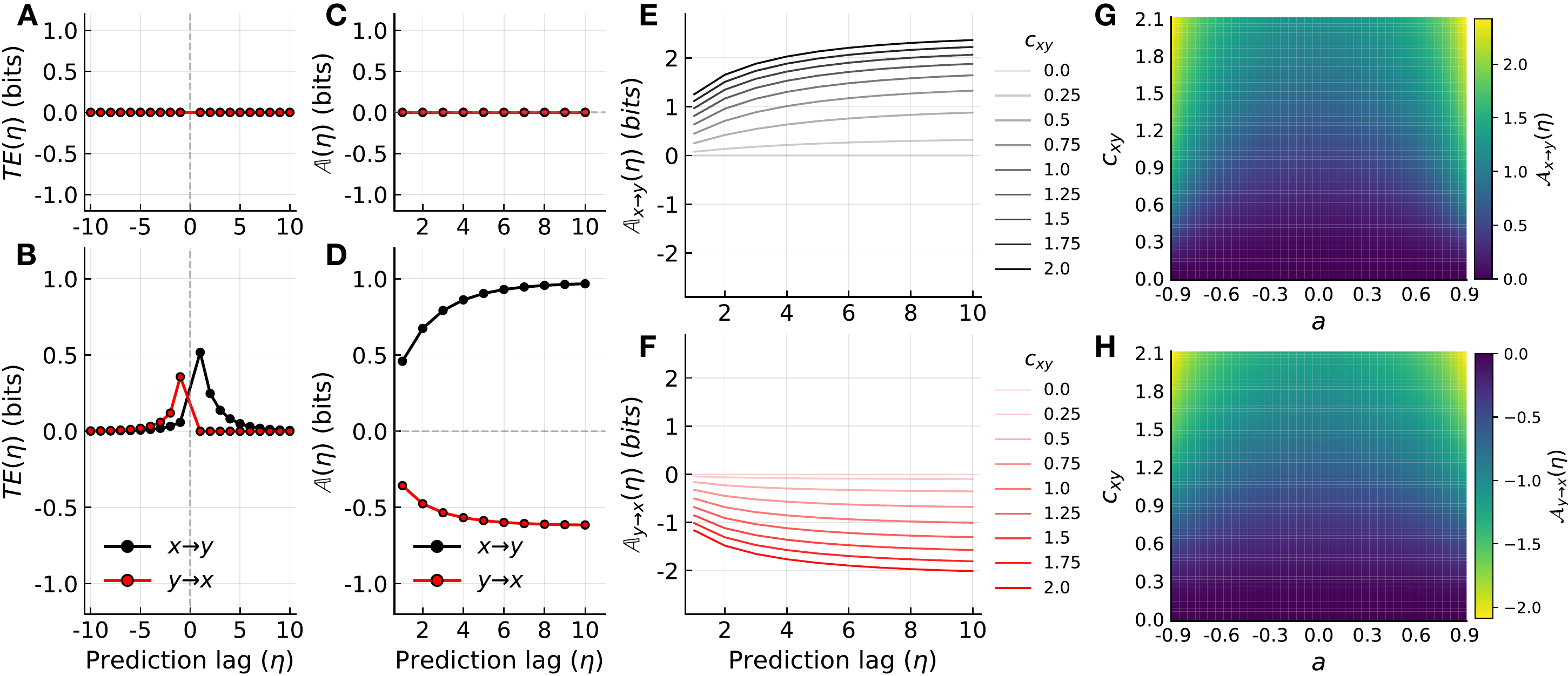}
\caption{Exact transfer entropy (A, B) and exact predictive asymmetry (eq. \ref{eq:asymmetry}; C, D) for a bivariate order-one autoregressive system (eq. \ref{eq:ar1_example}). When the random variables are decoupled ($c_{xy} = 0$), both transfer entropy and predictive asymmetry are zero (A, C). Non-zero coupling (here $c_{xy} = 0.8$) introduces non-zero transfer entropy (B) both in the causal ($x \to y$) and non-causal direction ($y \to x$), resulting in distinctive predictive asymmetries that reach a plateau with increasing predicting lag $\eta$ (D). The magnitude of the predictive asymmetry varies with coupling strength (E, F) (here $a = 0.8$) and the value of the parameter $a$ (the prediction lag is fixed $\eta = 10$ for the heat maps). Values are computed as described in appendix \ref{sec_supp:analytical_asymmetry_ar1}. 
}
\label{fig:analytical_asymmetry}
\end{figure}

When the dynamical variables are decoupled, predictive asymmetries are zero in both directions (Fig. \ref{fig:analytical_asymmetry}). 
When coupling exists, $\mathbb{A}_{x \to y}(\eta)$ is positive and increases monotonically with $\eta$. Conversely, $\mathbb{A}_{y \to x}(\eta)$ is negative and decreases with increasing $\eta$. 
Contributions to $\mathbb{A}$ are most pronounced at low $\eta$, and diminish for higher $\eta$ (Fig. \ref{fig:analytical_asymmetry}D). 
The predictive asymmetry therefore plateaus at some system-specific threshold value of $\eta$, which is the time horizon beyond which information about the forcing is no longer detectable in the response.
Lagged information beyond this threshold does not contribute substantially to the predictive asymmetry, because beyond some system-specific and data resolution specific time lag, the extra information is irrelevant to the interaction. 
In other words, the influence of $x(t)$ on $y(t+\eta)$ fades due to vanishing covariance between states that are further apart in time.

For dynamical systems in general, this convergence can be understood as follows (Appendix \ref{sec:supp_heuristic}). 
If there is a dynamical link in the direction $x\to y$, then the influence that current values of $x$ have on future values of $y$ (forwards-in-time prediction) 
is expected to be stronger than the influence that current values of $x$ have on past values of $y$ (backwards-in-time prediction), 
i.e. $TE_{x\to y}(\nu) > TE_{x\to y}(-\nu)$ for every $\nu > 0$. Hence, we expect  $\mathbb{A}_{x\to y}(\eta) > 0$ if $x$ influences $y$.
In the case of bidirectional influence $x \leftrightarrow y$, we expect that both $\mathbb{A}_{x\to y}(\eta), \mathbb{A}_{y\to x}(\eta) > 0$. 
Moreover, the relative magnitudes of $\mathbb{A}_{x \to y}(\eta)$ and $\mathbb{A}_{y \to x}(\eta)$ preserve the rank order of the underlying coupling strength (Fig. \ref{fig:analytical_asymmetry}E-H).

\subsection{\label{sec:statistical_robustness_applications_to_synthetic_systems}An intrinsic significance criterion: the normalized predictive asymmetry test}

In practice, a strict criterion of $\mathbb{A} > 0$ is not ideal for testing the hypothesis of directional causality. 
Governing equations of observed processes are rarely available in practice, so exact values for the statistic are unobtainable.
Predictive asymmetries must therefore be approximated from phase space reconstructions from observed time series data (Appendix \ref{sec:supp_experimental_setup}).  In the limit of few observations and low coupling strength, statistical fluctuations will cause numerical estimates of $\mathbb{A}$ to deviate from the true value. 

As demonstrated in appendix \ref{sec_supp:comparisons_te_pa}, the value of TE estimates can vary greatly depending on the system. Here we leverage this system-specific TE as a dynamically informed, intrinsic significance criterion. 
We thus introduce a normalized causality statistic by dividing the predictive asymmetry on the TE of the observed time series integrated across the spectrum of prediction lags:
{
\begin{eqnarray}
\mathcal{A}_{x \to y}^{f}(\eta) :=  \dfrac{\mathbb{A}_{x \to y}(\eta)}{\frac{f}{\eta} \int_{-\eta}^{\eta} TE_{x \to y}(\nu) \di \nu}. \label{eq:normalized_asymmetry_criterion}
\end{eqnarray}
}

This normalization to the system-intrinsic TE allows the relative magnitude of predictive asymmetry to be compared across systems. Furthermore, $\mathcal{A}_{x \to y}^{f}$ can be treated as a binary classifier of directional causality, so that $\mathcal{A}_{x \to y} > f$ indicates a positive detection of an influence from $x$ to $y$. The stringency of the test is then determined by the constant $f$. 
We use the criterion $\mathcal{A}^{f}_{x \to y} > 1$ with $f=1$ (i.e. normalizing to the mean TE; appendix \ref{sec_supp:significance_test}) to determine the statistical robustness of the test, defining $\mathcal{A}^{f}_{x \to y} > 1$ as a detection of directional coupling from $x$ to $y$ (i.e. a "positive"), and $\mathcal{A}^{f}_{x \to y} \leq 1$ as a detection of a non-interaction (i.e. a "negative").

Non-parametric approaches to detecting causality from time series typically rely on the method of surrogates \citep{Theiler1992, Lancaster2018} to avoid spurious results. A surrogate time series is a randomized or modeled version of the original time series, designed to establish a baseline for significance testing. The value of a causality statistic is deemed significant if it exceeds some threshold value obtained in a large ensemble of surrogates. An ideal surrogate for causality testing preserves all statistical properties of the original signal, except the property of being causally connected. The design of appropriate surrogate data remains a thorny problem.

Our predictive asymmetry approach solves this problem by design. The $\mathbb{A}$ statistic compares the magnitude of forward-in-time TE to its complementary backward-in-time TE, \textit{computed on the same time series}. The backward-in-time TE can we viewed as a system-intrinsic and estimator-specific "reversed-time surrogate", and is part of the test by construction. This time reversal preserves all properties of the signal, but explicitly breaks causality by reversing the arrow of time.

Estimator specific bias, which arises due to the fact that the asymptotic distribution of the sample statistic for TE is not known analytically \cite{Barnett2009}, and due to disparate frequencies in the time series (which are known to plague TE; \cite{Palus2007cmi, hannisdal2011non}), are thus equally encoded in both the backward-in-time predictions and the forward-in-time predictions for stationary systems. Any asymmetry that remains is due to forwards-in-time information flow in the presence of directional coupling. 
With time series recorded from variables that are not connected, time-reversal has no effect, so forwards-in-time and backwards-in-time predictions are balanced (Fig. \ref{fig:analytical_asymmetry}; 
formally proved in appendices \ref{sec:supp_formal_derivation} and \ref{sec_supp:analytical_asymmetry_ar1}). 
Predictive asymmetry thus arises intrinsically from causal connectivity in the underlying system.

Because of its built-in significance test, predictive asymmetry does not require explicit surrogate testing, which drastically reduces its computational demands. A potentially powerful application of the method would thus be to initially screen for the presence of causal relationships in large time series ensembles. Our method could also be applied in continuous monitoring of real systems, to detect time-variable dynamical interactions.

\section{Identifying directional causation from time series}

Here, we apply the method to multiple synthetic coupled stochastic and dynamical systems with known governing equations, and show that the predictive asymmetry yields a stand-alone criterion for the detection of directional causality from time series. 

\subsection{Two example systems}
\begin{figure}
\includegraphics[width=1.0\linewidth]{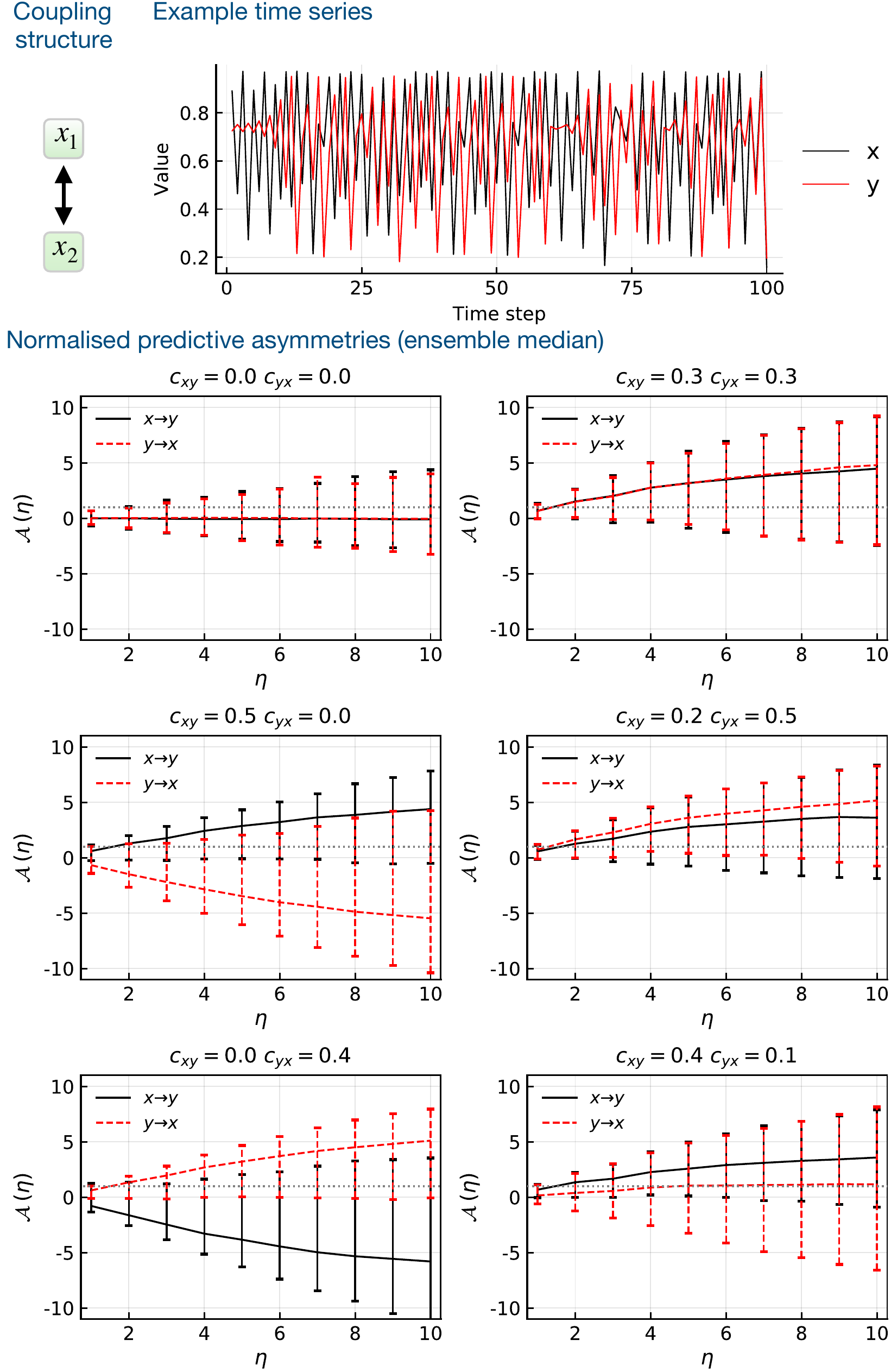}
\caption{
Normalized predictive asymmetry $\mathcal{A}^{f=1}$ (eq. \ref{eq:normalized_asymmetry_criterion}) for a bidirectional logistic map model (eq. \ref{eq:logistic_bidir}). Values and error bars are the median and 80th percentile ranges of $\mathcal{A}$ over unique 1000 realizations of the model with parameters randomized as described in Appendix \ref{sec:supp_test_systems_logistic_bidir}, and time series consisting of 500 observations. According to eq. \ref{eq:normalized_asymmetry_criterion}, values above 1 (dotted gray lines) are significantly positive. Generalized embeddings were constructed with $k = l = m = 1$ (see Appendix \ref{sec:supp_Generalized_embedding}).
}
\label{fig:res_asymmetries_logistic_bidir}
\end{figure}

\begin{figure}
\includegraphics[width=1.0\linewidth]{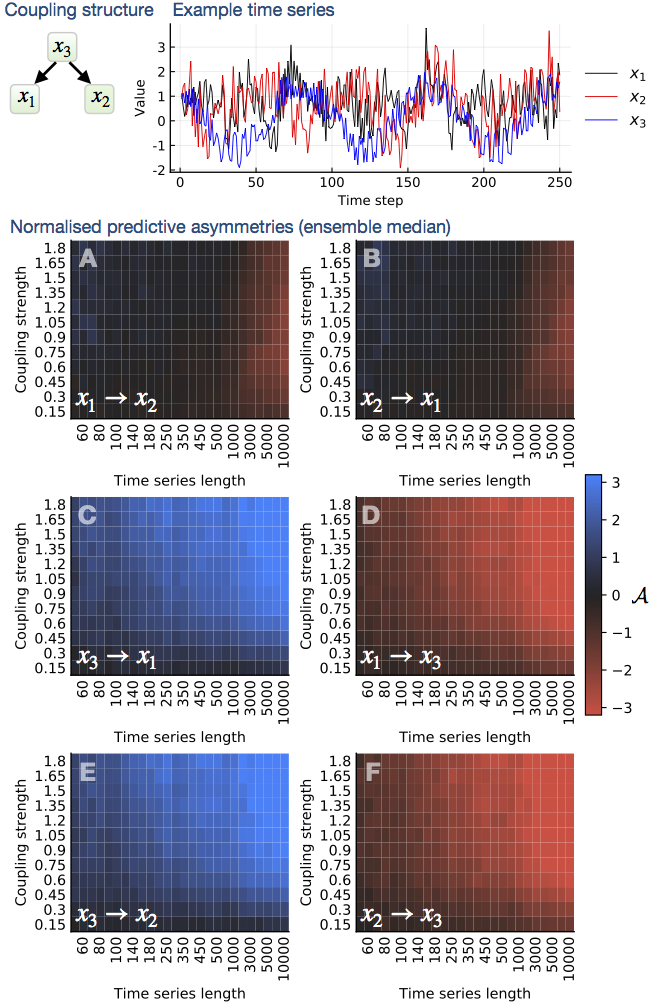}
\caption{
Normalized predictive asymmetry $\mathcal{A}^{f=1}$ (eq. \ref{eq:normalized_asymmetry_criterion}) for a nonlinear common-cause model of two non-interacting variables $x_1$ and $x_2$, both independently forced by an external driver $x_3$ (eq. \ref{eq:sys_noninter_ext}) at different forcing magnitudes. All variables have a deterministic component, a cyclic component and a stochastic component. Outputs from this model resemble paleoclimate time series, which often consist of high-frequency variability over lower-frequency periodic signals. With a model time step of 1 $kyr$, periods of the cyclic components of the signals are chosen randomly between 20 $kyr$ and 100 $kyr$, which is within the range of typical orbital-type frequencies that occur in real paleoclimate time series. Periodic signal components are phase-shifted randomly relative to those of the other variables. Values in heat map cells, for each  combination of coupling strength and time series length, are the median normalized predictive asymmetries computed from 300 unique realizations of the model with parameters randomized as described in section \ref{sec:supp_test_systems_noninteracting} and $\eta = 15$.  According to eq. \ref{eq:normalized_asymmetry_criterion}, $\mathcal{A}^{f=1}$ values above 1 indicate the presence of directional coupling. Generalized embeddings were constructed with $k = l = m = 1$ (see Appendix \ref{sec:supp_Generalized_embedding}).
}
\label{fig:res_asymmetries_commoncause}
\end{figure}

Consider a chaotic interaction model for two species, $X$ and $Y$ \cite{Diego2018}, where coupling can be absent, unidirectional, or bidirectional (Fig. \ref{fig:res_asymmetries_logistic_bidir}), and a 
common-cause model where two non-interacting variables with nonlinear deterministic dynamics, $x_1$ and $x_2$, respond to the same external forcing $x_3$ (Fig. \ref{fig:res_asymmetries_commoncause}). 
These examples serve to illustrate two important hurdles in causality testing that our statistic should reliably overcome: (1) distinguishing between uncoupled, unidirectional, and bidirectional relationships, and (2) distinguishing correlation from causation in systems with uncoupled variables responding to a common external driver, which may introduce strong correlation between the uncoupled variables. The common-cause model specifically also targets the issue of identifying causation in time series with relatively strong periodicities, which are pervasive in paleoclimate time series like the ones we analyze below. 

\textit{Characteristic asymmetries for uncoupled variables}. 
Absence of coupling consistently yields predictive asymmetry distributions centered around zero. 
Hence, when applied to the common-cause scenario, the normalized test reveals no evidence of directional coupling between the non-interacting variables, despite the common external forcing (Fig. \ref{fig:res_asymmetries_commoncause}A-B). 
The same holds for two-species chaotic model when there is no underlying coupling (Fig. \ref{fig:res_asymmetries_logistic_bidir}).
If two time series $x$ and $y$ are recorded from independent systems, then predicting future values of $y$ from present values of $x$, and predicting past values of $y$ from present values of $x$, are equally (un)informative. 

As we expand the prediction window, statistical noise is introduced by the inclusion of more non-informative history, which results in increasing variability of $\mathcal{A}^f$ for increasing $\eta$.
Access to more observations counteracts this effect, reducing the dispersion of $\mathcal{A}$, and yielding more narrow-tailed, zero-centered distributions (Appendix \ref{sec_supp_significance_test_demonstration_of_criterion}). As expected, having more information about two unrelated variables increases our ability to reject a coupling between them.
In summary, when no underlying coupling exists, predictions backwards and forwards in time are on average of similar magnitude, and $\mathcal{A}$ is thus centered around zero across a range of prediction lags.

\textit{Characteristic asymmetries for unidirectionally coupled variables}. In contrast, unidirectional coupling manifests as positive predictive asymmetry in the causal direction (driver $\to$ response), and negative predictive asymmetry in the non-causal direction (response $\to$ driver) (Appendix \ref{sec:characteristic_asymmetries_unidircoupling}). 

If we reverse the direction of coupling, then the signs of $\mathcal{A}^f$ will follow suit (Fig. \ref{fig:res_asymmetries_logistic_bidir}).

Why does this happen? If a unidirectional coupling from $x$ to $y$ exists, forwards-in-time prediction (values of $x$ predict future values of $y$) is stronger than backwards-in-time prediction (values of $x$ predicts past values of $y$). The opposite happens in the non-causal direction ($y \to x$): backwards-in-time prediction (future values of $y$ predicts past values of $x$) becomes more successful than forwards-in-time prediction (past values of $y$ predicts future values of $x$).  
In uncoupled systems, on the other hand, there is no shared information that improves prediction neither forwards in time nor backwards in time. 
Time-asymmetric predictability is thus  characteristic of systems with directional coupling. 

\textit{Characteristic asymmetries for bidirectionally coupled variables}.

Bidirectional coupling between variables yields predictive asymmetries that are on average positive in both directions (Fig. \ref{fig:res_asymmetries_logistic_bidir}; Appendix \ref{sec:characteristic_asymmetries_bidircoupling}),
and with relative magnitudes reflecting the underlying coupling strengths.
If the difference between the underlying relative coupling strengths for a bidirectional system is substantial, 
then values of $\mathcal{A}^f$
in the direction of the weaker forcing may approach zero. Thus, bidirectional coupling is most likely detected if coupling strengths in both directions are roughly equal, whereas if coupling strengths are significantly different, then the system may appear unidirectional in the eyes of the test (Appendix \ref{sec_supp:significance_test_bidircoupling}). Unlike unidirectionally coupled stochastic AR systems, for which the predictive asymmetry works remarkably well, the results of the predictive asymmetry test for bidirectionally coupled AR systems are sensitive to model parameters (as is TE alone). We stress, however, that our approach rests on the existence of a fundamental connection between the predictive asymmetry based on attractor reconstruction and the flow of the underlying dynamical system. Whether an equivalent fundamental connection exists for stochastic systems is a topic for future research.

\textit{Characteristic asymmetries for causal chains}. Relative positive/negative magnitudes of $\mathcal{A}^f$ for the model systems depend on the underlying coupling strength. Pairwise application to variables of multidimensional systems with chained unidirectional coupling shows that magnitude of $\mathcal{A}^f$ also depends on the number of intermediate variables: 
The magnitude of $\mathcal{A}^f$ is greater for adjacent nodes in the interaction network and decreases with an increasing number of intermediate, indirect links (Appendix \ref{sec:characteristic_asymmetries_unidir_chains}).

\subsection{\label{sec:statistical_robustness}Statistical robustness}

How robust is the predictive asymmetry as a causality detection criterion? For time series generated from synthetic systems, we compare the results of the normalized predictive asymmetry test as a binary classifier (eq. \ref{eq:normalized_asymmetry_criterion}) with the known ground truths. We repeat this process for a large number of parameterisations of different systems with varying coupling strengths, each parameterisation yielding a distinct dynamical system and a unique set of corresponding time series with varying statistical properties. Thus, we obtain counts of false positive, false negative, true positive and true negative detections. Their corresponding rates are summarized in confusion matrices, from which we compute Matthews' correlation coefficient (MCC) \cite{Matthews1975corr, Chicco2017tenmachinetips} and other test performance indicators.
The MCC takes on values on $[-1, 1]$, where MCC = 1 indicates perfect agreement between actual values and predictions, and MCC = 0 indicates no correlation between predictions and actual values. 

\begin{figure*}
\centering
\includegraphics[width=0.8\linewidth]{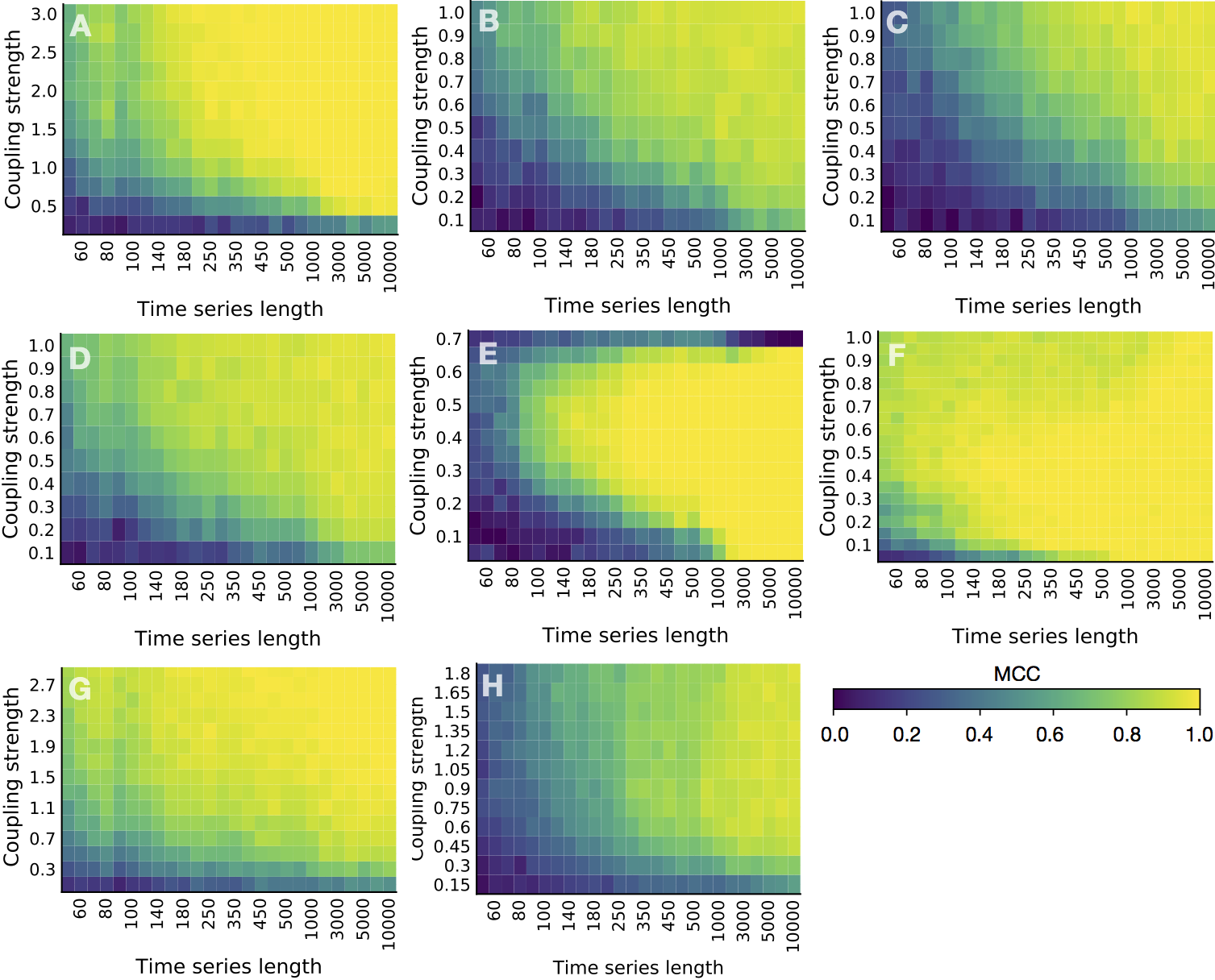}
\caption{
Statistical robustness of the predictive asymmetry causality test for different coupled dynamical and stochastic synthetic systems, as measured by the Matthews correlation coefficient (MCC). For every combination of time series lengths and coupling strength (varies over physically meaningful system-specific ranges), we first generate 300 unique randomized system realizations, sample 300 orbits and record relevant pairs of time series. Then we compute the predictive asymmetry between time series pairs, using $f=1$ as the normalization factor, resulting in 300 values of $\mathcal{A}$ for each heat map cell. Because the normalized predictive asymmetry is binary classifier  ($\mathcal{A} > 1 \rightarrow \text{detection of causality}$ and no detectable causality otherwise) and we know the ground truths (the underlying causal networks), we can compute confusion matrices for each heat map cell. Each confusion matrix is then summarized by the MCC, which provides a balanced measure of overall statistical robustness. If there is perfect correlation between test predictions and ground truths, then $MCC = 1$. On the other hand, $MCC = 0$ means that there is no correlation between test predictions and ground truths. In appendix \ref{sec_supp:significance_test}, we have analyzed a more comprehensive suite of statistical robustness measures for the same systems. Details on analysis parameters can be found in the corresponding supplementary figure captions. A: Periodic autoregressive variables with strongly nonlinear coupling (eq. \ref{eq:system_chain_autoregressiveperiodic_stronglynonlinearcoupling}); B: Nonlinear system with linear coupling (eq. \ref{eq:system_chain_nonlinear_chen_linearcoupling}); C: Nonlinear system with nonlinear coupling (eq. \ref{eq:system_chain_nonlinear_chen_nonlinearcoupling}); D: Nonlinear system with periodic component and linear coupling (eq. \ref{eq:system_chain_chen_nonlinearperiodic_linearcoupling}); E: Logistic map system with dynamical noise, variable interaction lags, variable internal lags, and dynamical noise (eq. \ref{eq:system_chain_logistic_chain_variablelags}); F: Henon map (eq. \ref{eq:system_chain_henonchain}); G: Unidirectionally coupled autoregressive systems of maximum order 5 with 30\% observational noise (section \ref{sec:supp_test_systems_var}). H: Common-cause model (eq. \ref{eq:sys_noninter_ext}). 
}
\label{fig:MCCs}
\end{figure*}

In our sensitivity tests, we find that for sufficient coupling strength and time series length, MCC converges to high values (> 0.8) for all test systems, including stochastic, periodic and nonlinear dynamics with different types of coupling, and chaotic dynamics where coupling strengths are below synchronization thresholds (Fig. \ref{fig:MCCs}). For the common-cause model, which is strongly periodic, MCC converges to values above $0.8$ for time series with 500 observations or more. We emphasize that these results are obtained by applying the normalized predictive asymmetry test alone, without any surrogate testing. 

\section{\label{sec:applications_real_data}Application to real data}

An ensemble approach is needed to statistically characterize the predictive asymmetry in a system, as demonstrated above for synthetic systems. 
In real-world applications, time series typically represent a single realization of the system observed over a limited time window, and the governing equations are not known.
Nonetheless, we can estimate "ensemble statistics" for empirical data in two ways: 
(1) We generate a distribution of $\mathcal{A}^f$ values computed on random segments of the time series, varying the length and position of each segment. 
(2) If we have information on the uncertainty of the observed data (e.g. standard errors), then we generate a distribution of $\mathcal{A}^f$ values by random resampling within the uncertainty bounds of the data. 
In certain situations we need to resample also within the uncertainties in the time index (e.g. age estimates in paleoclimate proxy records; see example below). 
Note that these two resampling approaches can be combined \cite{Haaga2019uncertaindata}. A sliding-window approach is also possible, but we limit the present study to the analysis of ensemble predictive asymmetry averaged over the total window of observation.

In appendix \ref{sec_supp:application_to_real_data} we characterize causal interactions from multiple real-world data sets for which the true causality is known. In each case where we know the ground truth, the ensemble predictive asymmetry correctly determines the underlying directional coupling. Here, we address the
causal interactions between the key climate system parameters of atmospheric CO$_2$, global sea level and summer insolation at 65$^{\circ}$N during the last 800 thousand years.

\begin{figure}[!h]
\centering
\includegraphics[width=1.0\linewidth]{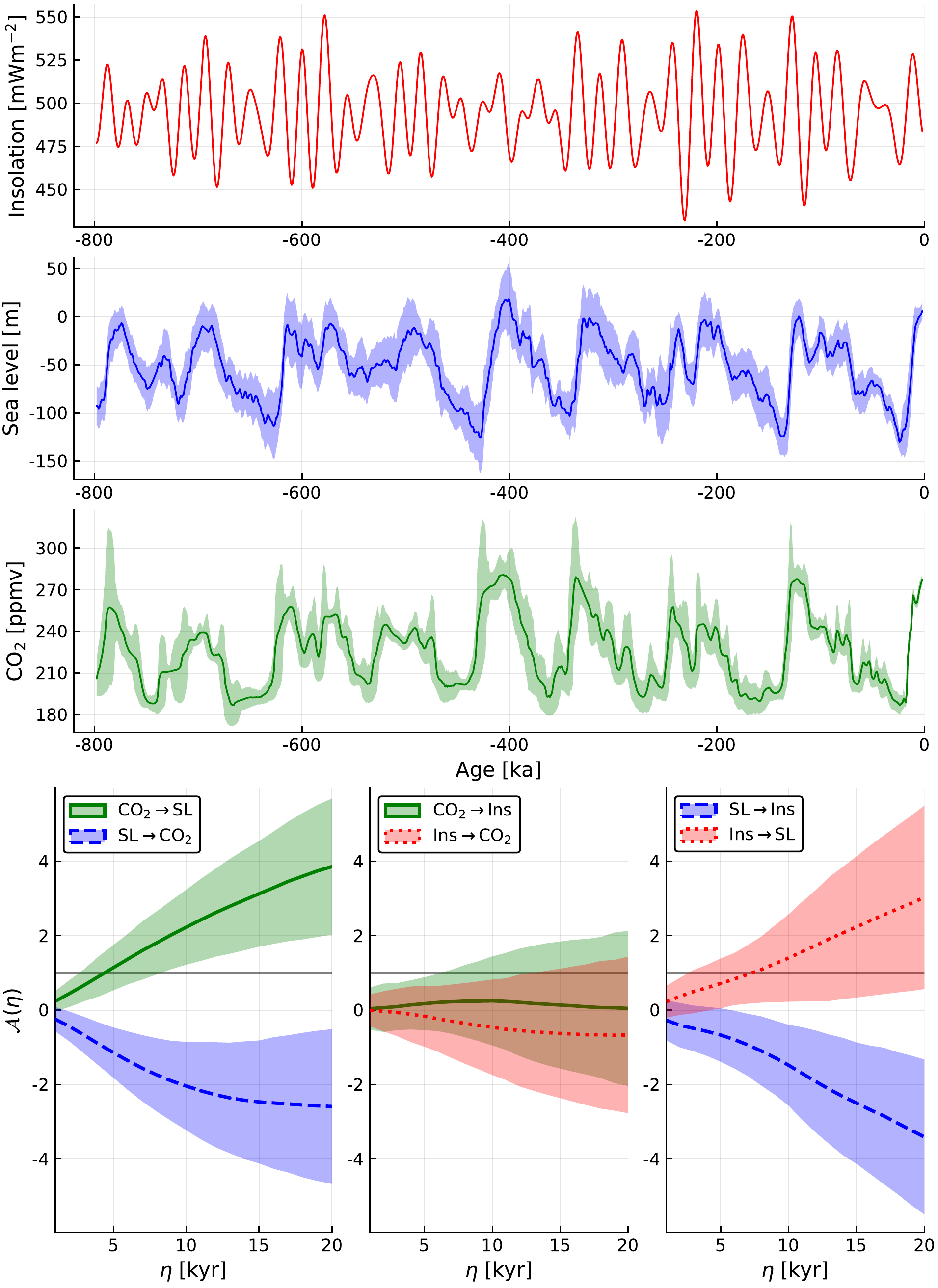}
\caption{Predictive asymmetry analysis of key climate variables in the late Pleistocene. (A) The Laskar 2004 \cite{Laskar2004long} solution for June 21 insolation at 65$^{\circ}$N.  (B) Global sea level stack \cite{Spratt2016}. Values are the first principal component with 95\% confidence ribbon representing uncertainty in both sea level estimates and ages, as reported by Spratt and Lisiecki \cite{Spratt2016}. (C) Composite ice core record of atmospheric CO$_2$ \cite{bereiter2015revision}, with AICC2012 age model uncertainty \cite{bazin2013optimized,veres2013antarctic}. Values are medians and  95\% confidence ribbon representing uncertainty in both CO$_2$ values and ages, computed by Monte Carlo resampling in 1-kyr bins using the UncertainData.jl Julia package \cite{Haaga2019uncertaindata}. (D-F) Mean normalized predictive asymmetry $\mathcal{A}(\eta)$ with $f=1$, computed over 1,000 randomly positioned segments, each of length ranging from 600 to 800 kyr. Ribbons represent 95\% confidence intervals from resampling within uncertainties across the ensemble of segments. 
}
\label{fig:real_data_quaternary}
\end{figure}

Since the discovery of the ice ages in the early 19th century \cite{esmark1824bidrag}, many hypotheses have been put forward to explain the recurrent waxing and waning of Pleistocene ice sheets. During glacial intervals, these ice sheets sequestered huge volumes of fresh water, thus controlling global mean sea level, which has varied by up to $130$ m during the past 800 kyr (Fig. \ref{fig:real_data_quaternary}). Chief among the causal explanations is variability in Earth's orbit \cite{esmark1824bidrag, croll1875climate, Milankovitch1941, Hays1976}, which affects the seasonal distribution of solar energy reaching the Earth. Difficulties remain, however, in explaining the $\sim$100 kyr saw-tooth pattern characteristic of the major late Pleistocene ice ages. Despite a similar periodicity, the energy forcing associated with changes in orbital eccentricity is negligible, hence several hypotheses have been proposed to explain the deep glacial maxima and their abrupt terminations \cite{pisias1981evolution, Raymo1997, tzedakis2017simple, denton2010last, Wolff2009}. 

When ancient air bubbles trapped in Antarctic ice cores revealed that fluctuations in atmospheric CO$_2$ were tightly linked to ice volume changes throughout the glacial-interglacial cycles (Fig. \ref{fig:real_data_quaternary}C), changes in radiative forcing due to greenhouse gases were implicated in the dynamics of ice ages \cite{petit1999climate, shackleton2000100}. Proxy reconstructions and transient modelling point to CO$_2$ as a 
forcing of global temperature rise during the last glacial termination \cite{shakun2012global}. However, the drive-response relationship between CO$_2$ and global ice volume remains controversial. On the one hand, coupled ice sheet and general circulation models are able to recreate the saw-tooth pattern by internal feedbacks without CO$_2$ forcing \cite{Abe-Ouchi2013}. 
On the other hand, it has been proposed that glacial terminations were a response to CO$_2$ release from warming southern oceans and associated changes in atmospheric and oceanic circulation \citep{Wolff2009,denton2010last}. The impetus for 
this mechanism is thought to be the long-lasting impact of meltwater from the massive circum-North Atlantic ice sheets that formed during glacial maxima, which created an ice sheet -- CO$_2$ feedback loop mediated by ocean circulation. In this view, the orbital variability acts as a "pacemaker" for the ice sheet -- CO$_2$ system rather than being the primary driver of the $\sim$100 kyr ice age cycles \cite{Raymo1997}.    

Here we test the causal pathways among key climate system variables in the late Pleistocene: insolation, atmospheric CO$_2$ concentration, and global sea level (ice volume). As an external variable we use the canonical June 21 insolation at 65$^{\circ}$N \cite{Laskar2004long} (Fig. \ref{fig:real_data_quaternary}A), which is typically used as a proxy for the astronomical forcing linked to the growth and decay of large ice sheets in the Northern Hemisphere through the Pleistocene epoch \cite{Milankovitch1941, Hays1976, Raymo1997, Huybers2005, Haaga2018}. 
We use a composite ice core record of atmospheric CO$_2$ \cite{bereiter2015revision} with reported means and standard errors for the CO$_2$ measurements, and the AICC2012 ice core chronology with associated age uncertainties \cite{bazin2013optimized,veres2013antarctic} (Fig. \ref{fig:real_data_quaternary}B). 
A global sea level stack \cite{Spratt2016}, reported as first principal component scores in 1-kyr bins, with a 95 \% confidence envelope accommodating uncertainty in both sea level estimates and ages, serves as a proxy for ice volume (Fig. \ref{fig:real_data_quaternary}C). By combining the random segment and uncertainty resampling, we obtain ensembles of predictive asymmetries for the three pairwise comparisons (Fig. \ref{fig:real_data_quaternary}D-F).

The dynamical evidence in the data shows that climate-intrinsic radiative forcing has a significant influence on the long-term evolution of global sea level (Fig. \ref{fig:real_data_quaternary}D). 
Changes in the planet's energy budget and seasonal energy distribution caused by oscillations of Earth's orbit also seem to be a strong direct driver of sea level as represented by the global stack (Fig. \ref{fig:real_data_quaternary}F). 
Relatively speaking, the influence of CO$_2$-driven radiative forcing on sea level is greater than that of northern summer insolation. 
Furthermore, insolation is not a significant driver of atmospheric CO$_2$ (Fig. \ref{fig:real_data_quaternary}E), indicating that the CO$_2$ forcing of sea level is independent from the insolation forcing of sea level and not a mutual response to orbital variability. 

Our analysis takes into account the reported uncertainty in the CO$_2$ and sea level estimates as well as uncertainty in the associated ages. 
One important caveat, however, is that the age model of the global sea-level stack inherently assumes a lagged response to orbital forcing \cite{Spratt2016}. 
To assess the impact of this assumption on the dynamical information in the paleoclimate records, we repeated our analysis on a 500-kyr record of sea level from Grant \textit{et al.} \cite{grant2014sea}, which is chronologically independent of orbital parameters (Appendix \ref{sec_supp:pleistocene_grant}). 
In the orbitally independent sea level record, the evidence for insolation forcing of global ice volume is drastically reduced (Fig. \ref{fig_supp:pleistocene_grant}), highlighting the importance of age model assumptions in determining dynamical information in geological records. 
Nevertheless, the results clearly confirm the strong influence of CO$_2$ on global ice volume (Fig. \ref{fig_supp:pleistocene_grant}). We thus conclude that state-of-the-art paleoclimate records, despite uncertainty in estimates and ages, strongly suggest that CO$_2$ was a major dynamical driver of glacial-interglacial ice volume variability. 
\section{\label{sec:concluding_remarks}Concluding Remarks}
Inferring the strength and directionality of causal interactions from observed time series, without recourse to mechanistic modeling, is a subject of controversy. 
Transfer entropy, or generalized Granger causality, has seen widespread application across many disciplines \cite{Bossomaier2016}. Related approaches to predictive causality based on dynamical systems reconstruction have also garnered considerable attention, including the geometric prediction method of convergent cross mapping \cite{Sugihara2012}. 
On a practical level, however, both transfer entropy and cross mapping have serious limitations. Both require \textit{ad hoc} interpretation of prediction skill to distinguish non-causal from causal coupling and both rely on the method of surrogate testing as a bulwark against false positives.
On a more fundamental level, and to the best of our knowledge, neither transfer entropy nor cross mapping have an explicit, precise relation to the flow of a dynamical system. 
The novelty of our contribution lies in showing that a simple difference between transfer entropy for forwards and backwards prediction lags quantifies the flow of the underlying dynamical system. The resulting predictive asymmetry provides a theoretically founded causality statistic, 
with robust numerical performance for deterministic and stochastic systems. 
Hence, our test can unambiguously resolve causal relationships in nonlinear systems where transfer entropy alone fails (Appendix \ref{sec_supp:comparisons_te_pa}). Moreover, our test easily characterizes the type of systems originally used to demonstrate cross mapping (Fig. \ref{fig:res_asymmetries_logistic_bidir}), where the latter requires additional hypothesis testing.
With its built-in significance criterion, our method eliminates costly surrogate testing, which raises the prospect of causal network reconstruction in large data sets and monitoring applications.
By linking predictive asymmetry to dynamical causality, our work represents a major advance in the causal analysis of observations, with implications for any field where time series are used to study the dynamics of natural systems.

%\nocite{*}

\clearpage
\widetext

\begin{center}
\textbf{{\LARGE Supplementary materials}} \\
\vspace{\baselineskip}
\textbf{K. A. Haaga \textit{et al.}\\ A simple test for causal asymmetry in complex systems}
\end{center}

\vspace{-2\baselineskip}

%%%%%%%%%% Merge with supplemental materials %%%%%%%%%%
%%%%%%%%%% Prefix a "S" to all equations, figures, tables and reset the counter %%%%%%%%%%
\setcounter{equation}{0}
\setcounter{figure}{0}
\setcounter{table}{0}
\setcounter{page}{1}
\makeatletter
\renewcommand*{\theequation}{S.\thesection\arabic{equation}}
\renewcommand*{\thefigure}{S.\thesection\arabic{figure}}
\renewcommand*{\bibnumfmt}[1]{[S#1]}
\renewcommand*{\citenumfont}[1]{S#1}
% the following are needed to tweak the alignment of the ToC
% The numbers correspond to (defaults in comments):
% 1)  logical depth (cp. to tocdepth or minitocdepth)
% 2)  the indentation
% 3)  the width reserved for the section/subsection/... number
\renewcommand*\l@section{\@dottedtocline{1}{1.0em}{2.0em}} %{1.5em}{2.3em}}
\renewcommand*\l@subsection{\@dottedtocline{2}{3.8em}{1.8em}} %{3.8em}{3.2em}}
\renewcommand*\l@subsubsection{\@dottedtocline{3}{6.0em}{4.1em}} %{7.0em}{4.1em}}
\renewcommand*\l@paragraph{\@dottedtocline{4}{10em}{5em}} %{10em}{5em}}
\renewcommand*\l@subparagraph{\@dottedtocline{5}{12em}{6em}} %{12em}{6em}}
\makeatother

\appendix
\hypersetup{linkcolor=black}

\addcontentsline{toc}{section}{Appendix} % Add the appendix text to the document TOC
\part{} % Start the appendix part
\parttoc 
\addtocontents{toc}{\protect\setcounter{tocdepth}{-1}}

\clearpage 

\section{\label{sec_supp:comparisons_te_pa}Comparison of transfer entropy and predictive asymmetries}

Here, we demonstrate some common issues associated with time series causality methods (exemplified using TE) and how our predictive asymmetry method overcomes these issues.

\begin{figure*}[h]
\centering
\includegraphics[width=1.0\linewidth]{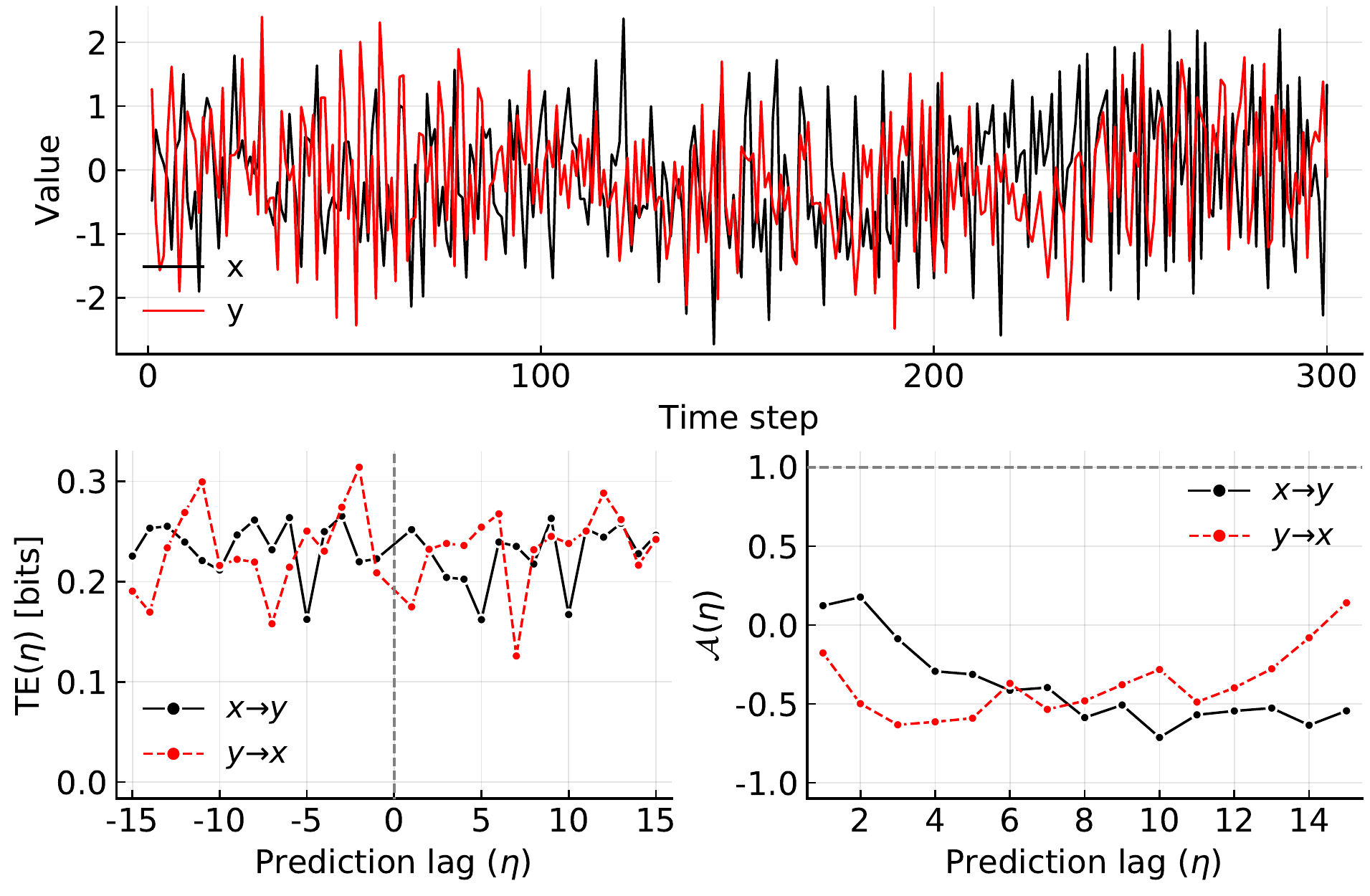}
\caption{Transfer entropy (lower left panel; eq. \ref{eq:te_simple}) and normalized predictive asymmetry (lower right panel; eq. \ref{eq:normalized_asymmetry_criterion}) as a function of prediction lag $\eta$ for another nonlinear 2D system with with no coupling between $x$ to $y$  (eq. \ref{eq:nonlinear_withoutdynamicalnoise_withoutperiodicity} with $c_{xy} = 0$). The statistics were computed for time series consisting of 1000 points (upper panel; only the first 300 points are plotted). Parameters are set as follows: $a_1 = 3.4$, $a_2 = 0.8$, $b_1 = 3.4$, and $b_2 = 0.8$, Internal lags were set to $\tau_{x_1} = 1$, $\tau_{x_2} = 7$, $\tau_{y_1} = 5$, and $\tau_{y_2} = 5$, while the interaction delay is set to $\tau_{c_{xy}} = 5$. 
Observational noise was added to the time series after sampling them, and a was sampled from two independent normal distributions $\mathcal{N}(0, \sigma_x)$ and $\mathcal{N}(0, \sigma_y)$, where $\sigma_x$ and $\sigma_x$ were chosen as 0.5 times the empirical standard deviation of the sampled time series. 
Generalized embeddings were constructed with $k = l = m = 1$, and $\mathcal{A}$ was computed with normalization factor $f=1.0$. The dashed line in (C) indicates the significance threshold;  according to eq. \ref{eq:normalized_asymmetry_criterion}, only values above this line are significant and counts as a positive detection of directional coupling.
}
\label{fig_supp:comparison_te_pa_chen2d_nonlinearsystem_nocoupling}
\end{figure*}

\begin{figure*}[h]
\centering
\includegraphics[width=1.0\linewidth]{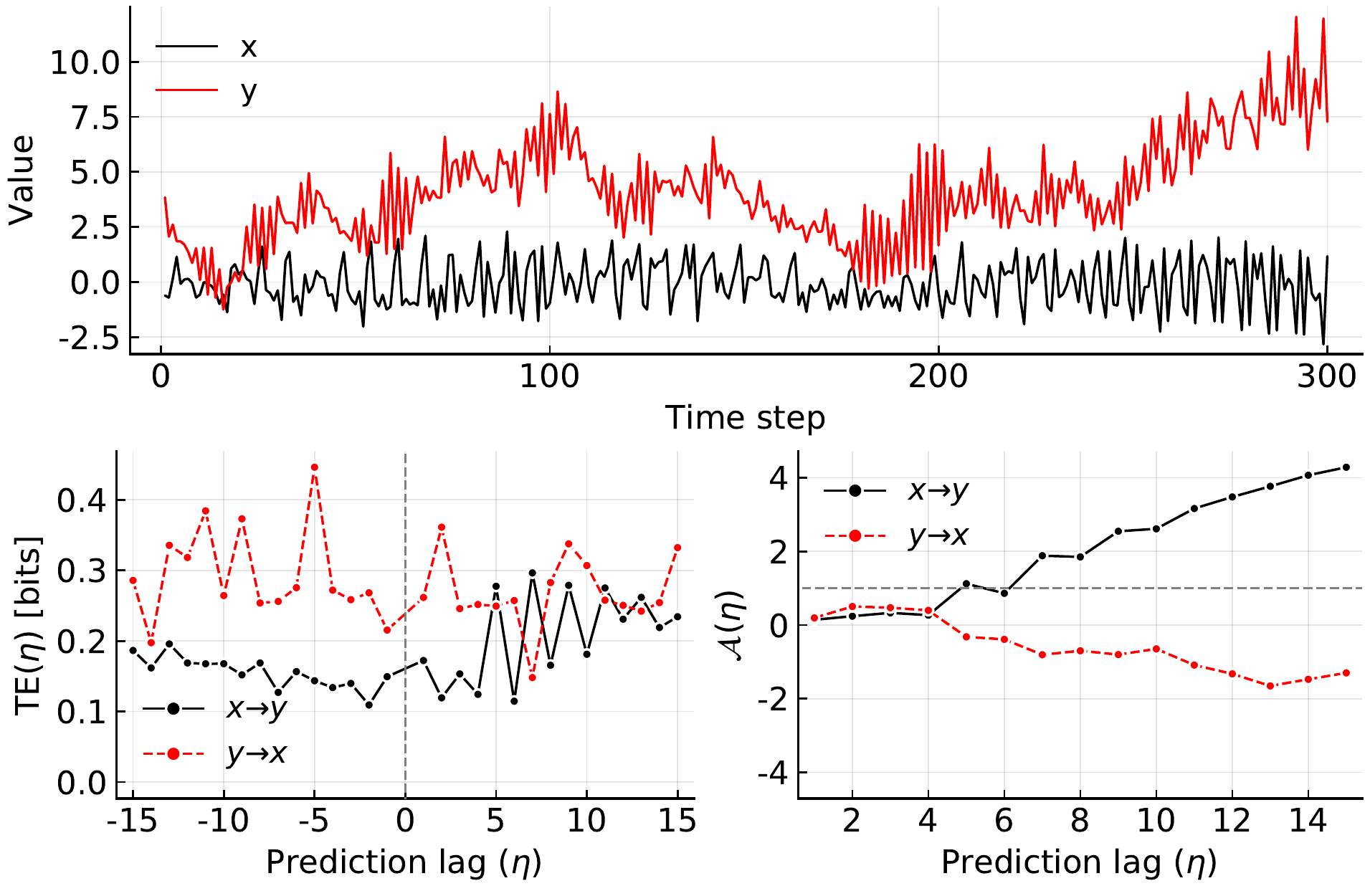}
\caption{Transfer entropy (lower left panel; eq. \ref{eq:te_simple}) and normalized predictive asymmetry (lower right panel; eq. \ref{eq:normalized_asymmetry_criterion}) as a function of prediction lag $\eta$ for another nonlinear 2D system with with coupling from $x$ to $y$  (eq. \ref{eq:nonlinear_withoutdynamicalnoise_withoutperiodicity} with $c_{xy} = 0.8$). The statistics were computed for time series consisting of 1000 points (upper panel; only the first 300 points are plotted). Parameters are set as follows: $a_1 = 3.4$, $a_2 = 0.8$, $b_1 = 3.4$, and $b_2 = 0.8$, Internal lags were set to $\tau_{x_1} = 1$, $\tau_{x_2} = 7$, $\tau_{y_1} = 3$, and $\tau_{y_2} = 2$, while the interaction delay is set to $\tau_{c_{xy}} = 5$. 
Observational noise was added to the time series after sampling them, and a was sampled from two independent normal distributions $\mathcal{N}(0, \sigma_x)$ and $\mathcal{N}(0, \sigma_y)$, where $\sigma_x$ and $\sigma_x$ were chosen as 0.5 times the empirical standard deviation of the sampled time series. 
Generalized embeddings were constructed with $k = l = m = 1$, and $\mathcal{A}$ was computed with normalization factor $f=1.0$. The dashed line in (C) indicates the significance threshold;  according to eq. \ref{eq:normalized_asymmetry_criterion}, only values above this line are significant and counts as a positive detection of directional coupling.
}
\label{fig_supp:comparison_te_pa_chen2dnonlinearcoupling}
\end{figure*}

\begin{figure*}[h]
\centering
\includegraphics[width=1.0\linewidth]{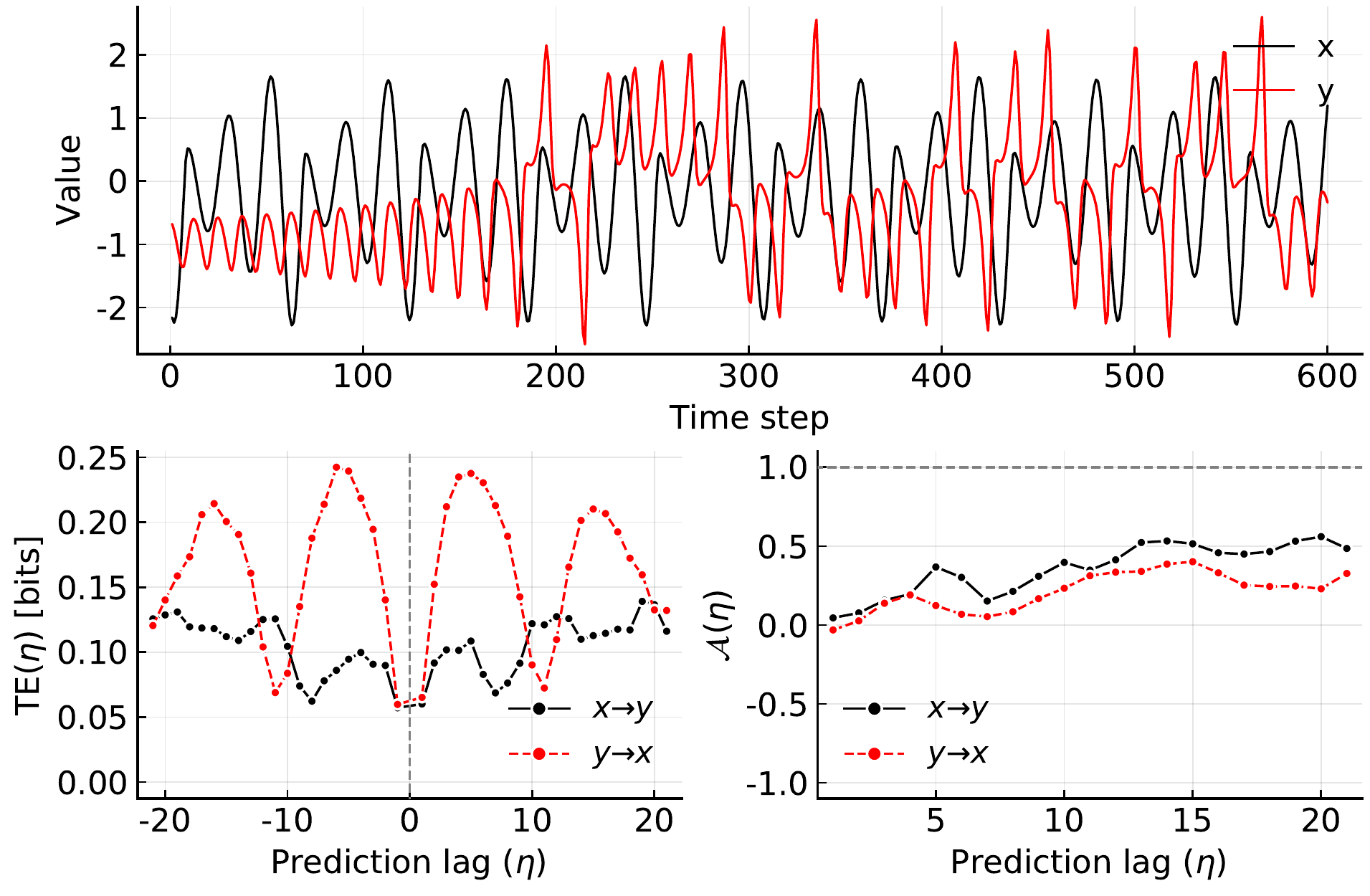}
\caption{Transfer entropy (lower left panel; eq. \ref{eq:te_simple}) and normalized predictive asymmetry (lower right panel; eq. \ref{eq:normalized_asymmetry_criterion}) as a function of prediction lag $\eta$ for a unidirectionally coupled Rössler-Lorenz system, where the Rössler subsystem drives the Lorenz subsystem (eq. \ref{eq:system_rosslerlorenz}), but here with $c_{xy} = 0.0$, so that the subsystems are decoupled. 
The statistics were computed over 30 randomly selected sub-segments of a 3000 points long time series, where each segment has a length of 70\% of the original time series (upper right panel; only the first 500 points are plotted). We show the median ensemble predictive asymmetry. The time series was generated with randomized parameters in the range that yield good attractors.
Observational noise was added to the time series after sampling them, and a was sampled from two independent normal distributions $\mathcal{N}(0, \sigma_x)$ and $\mathcal{N}(0, \sigma_y)$, where $\sigma_x$ and $\sigma_x$ were chosen as 0.1 times the empirical standard deviation of the sampled time series. 
Generalized embeddings were constructed with $k = l = m = 1$, and $\mathcal{A}$ was computed with normalization factor $f=1.0$.  The dashed line in (C) indicates the significance threshold;  according to eq. \ref{eq:normalized_asymmetry_criterion}, only values above this line are significant and counts as a positive detection of directional coupling.
}
\label{fig_supp:comparison_te_pa_rosslerlorenz_nocoupling}
\end{figure*}

\begin{figure*}[h]
\centering
\includegraphics[width=1.0\linewidth]{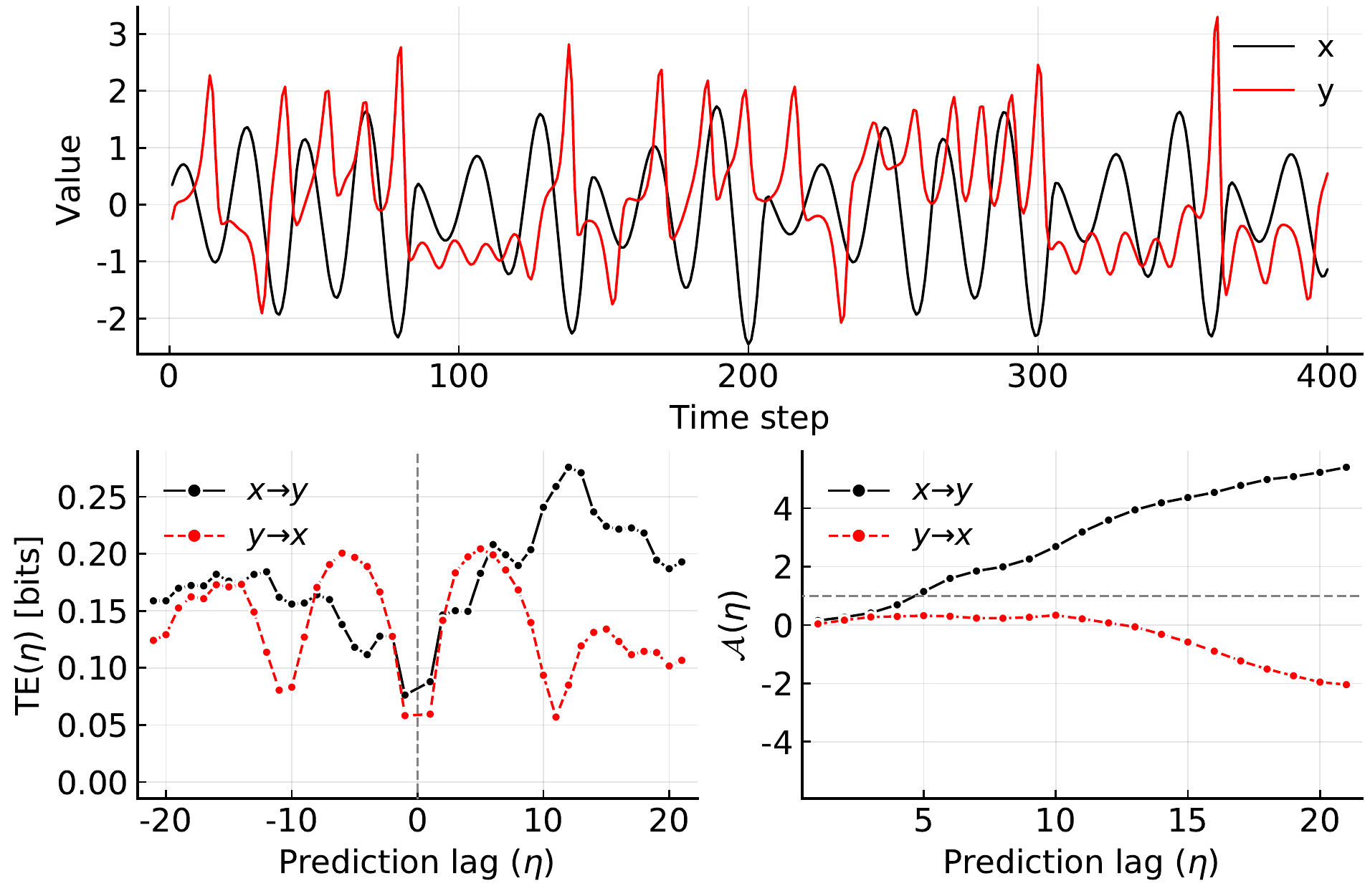}
\caption{Transfer entropy (lower left panel; eq. \ref{eq:te_simple}) and normalized predictive asymmetry (lower right panel; eq. \ref{eq:normalized_asymmetry_criterion}) as a function of prediction lag $\eta$ for a unidirectionally coupled Rössler-Lorenz system, where the Rössler subsystem drives the Lorenz subsystem (eq. \ref{eq:system_rosslerlorenz}). The statistics were computed over 30 randomly selected sub-segments of a 3000 points long time series, where each segment has a length of 70\% of the original time series (upper panel; only the first 500 points are plotted). The time series was generated with randomized parameters in the range that yield good attractors and with $c_{xy} = 1.2$. We show the median ensemble predictive asymmetry. 
Observational noise was added to the time series after sampling them, and a was sampled from two independent normal distributions $\mathcal{N}(0, \sigma_x)$ and $\mathcal{N}(0, \sigma_y)$, where $\sigma_x$ and $\sigma_x$ were chosen as 0.1 times the empirical standard deviation of the sampled time series. 
Generalized embeddings were constructed with $k = l = m = 1$, and $\mathcal{A}$ was computed with normalization factor $f=1.0$.  The dashed line in (C) indicates the significance threshold;  according to eq. \ref{eq:normalized_asymmetry_criterion}, only values above this line are significant and counts as a positive detection of directional coupling.
}
\label{fig_supp:comparison_te_pa_rosslerlorenz_withcoupling}
\end{figure*}

Time series generated from unrelated variables 
may give nonzero TE, which can be of equal magnitude in both 
directions (Fig. \ref{fig_supp:comparison_te_pa_chen2d_nonlinearsystem_nocoupling}). 
Even for systems with unidirectional causation between variables, 
TE may be of similar average magnitude both for the causal and for the non-causal direction (Fig. \hyperref[fig_supp:comparison_te_pa_chen2dnonlinearcoupling]{\ref{fig_supp:comparison_te_pa_chen2dnonlinearcoupling}}). 
Without additional information, both these cases could be erroneously interpreted 
as a bidirectional, equal-strength influence between the variables. 
In a more favorable scenario, unidirectional coupling yields forwards-in-time TE that is higher in the causal direction than in the non-causal direction. 
One could misinterpret this result as bidirectional coupling with 
dominant control from one variable to the other, but at least the preferred direction of information flow is correctly detected.
More disturbingly, TE values can be greater in the non-causal than in the causal direction in a unidirectionally coupled system at certain prediction lags (Fig. \ref{fig_supp:comparison_te_pa_chen2dnonlinearcoupling}),
which at face value would seem to imply bidirectional interaction with
dominant control in the \textit{non-causal} direction. Another challenge emerges in strongly periodic systems, both coupled and decoupled, which also yield oscillating TE values that are hard to interpret and could lead to the erroneous inference of two-way coupling (Figs. \hyperref[fig_supp:comparison_te_pa_rosslerlorenz_nocoupling]{\ref{fig_supp:comparison_te_pa_rosslerlorenz_nocoupling}}, \hyperref[fig_supp:comparison_te_pa_rosslerlorenz_withcoupling]{\ref{fig_supp:comparison_te_pa_rosslerlorenz_withcoupling}}). Similar false causalities also obtain with other causality statistics, and are difficult to remedy, even with external hypothesis testing using surrogate data \cite[e.g. ][]{Krakovska2018}.

%
% Pitfalls in time series causality estimation
% 

\clearpage

\section{\label{sec:supp_formal_derivation}Formal proofs}

\subsection{Relation between predictive asymmetry and the underlying dynamics of the system}

 In this section we provide a formal derivation of the relation between the causality asymmetry and the underlying dynamics. The result applies for both discrete and continuous systems. A generic continuous system is generated by a vector field of the form
 
 {\small
 \begin{align}
     & \dot x = f(x,y),\qquad x\in\mathbb R^n\label{A:dotx}\\
     &\dot y = g(x,y),\qquad y\in\mathbb R^m \label{A:doty}
 \end{align}
 }

 while a generic discrete system is generated by the map
 
 {\small
 \begin{align}
     & x(k+1) = F_x(x(k),y(k)),\qquad x\in\mathbb R^n\label{A:x(n)}\\
     &y(k+1) = F_y(x(k),y(k)),\qquad y\in\mathbb R^m \label{A:y(n)}
 \end{align}
 }
 
 where $F:=(F_x,F_y)$ is  smoothly invertible. Let $\phi(t;x,y)$, generically denote the evolution operator of the system. In the discrete case, $t\in\mathbb Z$ and $\phi(t+1;x,y) = F\circ \phi(t;x,y)$ and $\phi(0;x,y) = (x,y)$, that is: $\phi(t;x,y)$ is the $t$-fold composition of $F$. In the continuous case, $\phi(t;x,y)$ is the solution to the differential equation $\partial_t \phi = F(\phi)$ with the initial condition $\phi(0;x,y)=(x,y)$, for all $(x,y)\in\mathbb R^{n+m}$, that is: $\phi(t;x,y)$ is the flow of the vector field. Let $\phi_{x_i}$ 
 and $\phi_{y_j}$ denote the function components of $\phi$ corresponding to the $x_i$ and $y_j$ axes, respectively, and  w.l.o.g., assume that $\partial_{x_1} g_1 \neq 0$. For fixed $\eta >0$, define the variables
 
 {\small
  \begin{align}
      &\alpha_+ := \phi_{y_1}(\eta;x,y)\,,\label{A:alpha+}\\
      &\alpha_- := \phi_{y_1}(-\eta;x,y)\,,\label{A:alpha-}\\
      &a:= (y_1,\phi_{y_1}(-\tau_1;x,y),\cdots ,\phi_{y_1}(-n_1\tau_1;x,y))\,,\label{A:a}\\
      & b:= (x_1,\phi_{x_1}(-\tau_2;x,y),\cdots ,\phi_{x_1}(-n_2\tau_2;x,y))\label{A:b}\,,
  \end{align}
 }
 
 and assume that the maps 
 {\small
 \begin{align}
     & F(x,y)= (\alpha_+,a,b)\,,\\
     & G(x,y)= (\alpha_-,a,b)\,,\\
 \end{align}
 }
 
 are diffeomorphisms over $\mathbb R^{n+m}$ and call $\mathbb E_\eta := F(\mathbb R^{n+m})$ and $\mathbb E_{-\eta} := G(\mathbb R^{n+m})$. 
 Notice that $\alpha_-=\phi_{y_1}(-\eta;x,y) = \phi_{y_1}(-2\eta;\phi_x(\eta;x,y),\phi_y(\eta;x,y)) =: h(\alpha_+,a,b)$ and thus the map
 
 {\small
 \begin{equation}
     f(\alpha_+,a,b):=(h(\alpha_+,a,b),a,b)\label{A:change_var}
 \end{equation}
 }
 
 generates the desired change of coordinates.  Equivalently, $f(\mathbb E_\eta) = \mathbb E_{-\eta}$. Its inverse map is of the same form, that is:
 
 {\small
 \begin{equation}
     f^{-1}(\alpha_-,a,b):=(j(\alpha_-,a,b),a,b)
 \end{equation}
 }

  Let $\rho(x,y)$ be an invariant distribution of the system and let $P(\alpha_+,a,b)$ and $Q(\alpha_-,a,b)$ be the expression of $\rho$ on the coordinates corresponding to $\mathbb E_\eta$ and $\mathbb E_{-\eta}$, respectively. Therefore, for any measurable set $A\subset \mathbb E_\eta$ it must hold that
 
 {\small
 \begin{equation}
    \int_A \di\mu\,P = \int_{f(A)} \di\nu\,Q = \int_A \di\mu\,K\,P\circ f 
 \end{equation}
 }
 
 where $\di\mu := \di \alpha_+\di a\di b$, $\di\nu := \di \alpha_-\di a\di b$ and we have used that $\di\nu = K\,\di \mu$, with $K(\alpha_+,a,b) := \abs{\frac{\partial h}{\partial \alpha_+}}$. In our case this implies
 
 {\small
 \begin{align}
    P(\alpha_+,a,b) &= K(\alpha_+,a,b)\, Q(h(\alpha_+,a,b),a,b)\,,\\
    Q(\alpha_-,a,b) &= \frac{P(j(\alpha_-,a,b),a,b)}{K(j(\alpha_-,a,b),a,b)}
 \end{align}
 }

 The transfer entropies $TE_{x_1\to y_1}(\eta)$ and
 $TE_{x_1\to y_1}(-\eta)$ are given by
 
 {\small 
 \begin{align}
     TE_{x_1\to y_1}(\eta) &:= \int_{\mathbb E_\eta}\di\mu\, P(\alpha_+,a,b)\,\log_{2}\frac{P(\alpha_+|a,b)}{P(\alpha_+|a)}\\
     TE_{x_1\to y_1}(-\eta) &:= \int_{\mathbb E_{-\eta}}\di\nu\, Q(\alpha_-,a,b)\,\log_{2}\frac{Q(\alpha_-|a,b)}{Q(\alpha_-|a)}\,.\label{-eta}
 \end{align}
 }
 
 It is convenient to re express the above TE in terms of mutual information. In particular, defining the quantities
 
 {\small
 \begin{align}
    &I_{R,\mathbb M}(A;(B,C)) := \int_{\mathbb M} \di\mu_{\mathbb M}\, R(A,B,C)\,\log_{2}\frac{R(A,B,C)}{R(A) R(B,C)} \\
    &I_{R,\mathbb M}(A;B) := \int_{\mathbb M} \di\mu_{\mathbb M}\, R(A,B,C)\,\log_{2}\frac{R(A,B)}{R(A) R(B)}
 \end{align}}
 
 and using the identities 
 
 {\small
 \begin{align}
\frac{P(\alpha_+|a,b)}{P(\alpha_+|a)} &= \frac{P(\alpha_+,a,b)}{P(\alpha_+)P(a,b)}\,\frac{P(\alpha_+)P(a)}{P(\alpha_+,a)}\\
\frac{Q(\alpha_-|a,b)}{Q(\alpha_-|a)} &= \frac{Q(\alpha_-,a,b)}{Q(b)P_-(\alpha_-,a)}\,\frac{Q(a)Q(b)}{Q(a,b)}
 \end{align}
 }
 
 one easily checks that

 {\small
 \begin{align}
   TE_{x_1\to y_1}(\eta) &= I_{P,\mathbb E_\eta}(\alpha_+;(a,b)) - I_{P,\mathbb E_\eta}(\alpha_+;a)\\
   TE_{x_1\to y_1}(-\eta) &= I_{Q,\mathbb E_{-\eta}}(b;(\alpha_-,a)) - I_{Q,\mathbb E_{-\eta}}(a;b)
 \end{align}
 }
 
 and hence their difference is expressed as
 
 {\small
 \begin{align}
TE_{x_1\to y_1}(\eta) - TE_{x_1\to y_1}(-\eta) &=  I_{P,\mathbb E_\eta}(\alpha_+;(a,b)) + I_{Q,\mathbb E_{-\eta}}(a;b)\nonumber\\
 & - \ls I_{Q,\mathbb E_{-\eta}}(b;(\alpha_-,a)) + I_{P,\mathbb E_\eta}(\alpha_+;a)\rs
 \end{align}
 }
 
  We claim that $I_{Q,\mathbb E_{-\eta}}(a;b) = I_{P,\mathbb E_{\eta}}(a;b)$. To see this, we first show that the marginals $P(a,b)$ and $Q(a,b)$ coincide:
  
  {\small
  \begin{align}
      Q(a,b) &:= \int_{{\mathbb E_{-\eta}}_{(a,b)}} \di\alpha_-\,Q(\alpha_-,a,b)\nonumber\\
      &= \int_{f^{-1}\lb {\mathbb E_{-\eta}}_{(a,b)}\rb} \di\alpha_+\,K\,Q(h(\alpha_+,a,b),a,b)\nonumber\\
      &=\int_{{\mathbb E_\eta}_{(a,b)}} \di\alpha_+\,P(\alpha_+,a,b)=:P(a,b)
  \end{align}
  }
  
  where ${\mathbb E_{ \eta}}_{(a,b)}:=\lc \alpha\in\mathbb R\,|\, (\alpha,a,b)\in \mathbb E_{\eta}\rc$ and analogously for ${\mathbb E_{ -\eta}}_{(a,b)}$. From this it follows that
  
  {\small
  \begin{align}
      I_{Q,\mathbb E_{-\eta}}(a,b) &= \int_{\mathbb E_{-\eta}} \di\nu\,Q\,\log_{2}\frac{Q(a,b)}{Q(a) Q(b)} \\ 
      & = \int_{\mathbb E_{-\eta}} \di\nu\,Q\,\log_{2}\frac{P(a,b)}{P(a) P(b)}  \nonumber\\
      & = \int_{\mathbb E_{\eta}} \di\mu\,K\,Q\circ h\,\log_{2}\frac{P(a,b)}{P(a) P(b)} \\
      & = \int_{\mathbb E_{\eta}} \di\mu\,P\,\log_{2}\frac{P(a,b)}{P(a) P(b)} = I_{P,\mathbb E_{\eta}}(a,b)  
  \end{align}
  }
  
  In addition, for a fixed $(\alpha_-,a,b)\in\mathbb E_{-\eta}$, the straight line ${\mathbb E_{-\eta}}_{(\alpha_-,a;b)}:=\lc (\alpha_-,a,b+\beta)\in\mathbb E_{-\eta}\,|\,\beta\in\mathbb R\rc$ corresponds to the curve $\xi:\mathbb R\to \mathbb E_\eta$ given by $\xi(\beta) :=(j(\alpha_-,a,b+\beta),a,b+\beta)$. Notice that neither the set ${\mathbb E_{-\eta}}_{(\alpha_-,a;b)}$ nor the curve $\xi$ do really depend on $b$. The presence of $b$ in their definitions is a pure formality so that the pair $(\alpha_-,a)$ can be assigned a unique pair $(\alpha_+,a)$ and vice versa. More specifically, $\alpha_+ = j(\alpha_-,a,b)$ and $\alpha_-=h(\alpha_+,a,b)$. 
  It then holds that 
  
  {\scriptsize
  \begin{align}
  Q(\alpha_-,a) &:= \int_{{\mathbb E_{-\eta}}_{(\alpha_-,a;b)}} \di \beta\,Q(\alpha_-,a,\beta) \\
  &= \int_{{\mathbb E_{-\eta}}_{(\alpha_-,a;b)}} \di \beta\,\frac{1}{K(j(\alpha_-,a,\beta),a,\beta)}P(j(\alpha_-,a,\beta),a,\beta)\label{A:Qmarginal}
    \end{align}
    }
  
  Defining the quantity 
  {\scriptsize
  \begin{align}
      \frac{1}{\hat K(\alpha_+,a)} &:=  
       \frac{1}{P(\alpha_+,a)} \nonumber \\
       & \left( \cdot \int_{{\mathbb E_{-\eta}}_{(\alpha_-,a;b)}} \di \beta\,   \frac{1}{K(j(\alpha_-,a,\beta),a,\beta)}  P(j(\alpha_-,a,\beta),a,\beta) \right)
  \end{align}}
  
  where $P(\alpha_+,a):=\int_{{\mathbb E_\eta}_{(\alpha_+,a)}} \di b\,P(\alpha_+,a,b)$ denotes the marginal of $P$ on the plane $(\alpha_+,a)$, 
  we have that 

  {\small \begin{equation}
  Q(\alpha_-,a) = \frac{1}{\hat K} P(j(\alpha_-,a,b),a) =\frac{1}{\hat K(\alpha_+,a)} P(\alpha_+,a) 
  \end{equation}}
  
  From this it follows that 
  
  {\small
  \begin{align}
      I_{Q,\mathbb E_{-\eta}}(b;(\alpha_-,a)) &= \int_{\mathbb E_{-\eta}} \di\nu\, Q\,\log_{2}\frac{Q(\alpha_-,a,b)}{Q(\alpha_-,a) Q(b)} \nonumber\\
      & = \int_{\mathbb E_\eta} \di\mu\, P\,\log_{2}\frac{\frac{1}{K(\alpha_+,a,b) } P(\alpha_+,a,b)}{\frac{1}{\hat K(\alpha_+,a)} P(\alpha_+,a) P(b)}\nonumber\\
      &= I_{P,\mathbb E_{\eta}}(b;(\alpha_+,a)) - \int_{\mathbb E_\eta} \di\mu\,P\,\log_{2}\frac{K}{\hat K}\label{A:IQ=IP+log}
  \end{align}
  }
  
  Thus the difference between the TE is given by
  
  {\small
  \begin{align}
    &TE_{x_1\to y_1}(\eta) - TE_{x_1\to y_1}(-\eta) =\nonumber \\
    &I_{P,\mathbb E_\eta}(\alpha_+;(a,b)) + I_{P,\mathbb E_\eta}(a;b) - \ls I_{P,\mathbb E_\eta}(b;(\alpha_+,a)) + I_{P,\mathbb E_\eta}(\alpha_+;a)\rs\nonumber\\
    &+\int_{\mathbb E_\eta} \di\mu\, P\,\log_{2} \frac{K}{\hat K}\,.
  \end{align}
  }
  
  Now we may use the identity
  {\small
  \begin{align}I_{R,\mathbb M}(A;(B,C)) + I_{R,\mathbb M}(B;C) &= \nonumber \\
  \int_{\mathbb M} \di\mu_{\mathbb M} \,R(A,B,C)\,\log_{2}\frac{R(A,B,C)}{R(A) R(B) R(C)}
  \end{align}
  }
  
  to deduce that $I_{P,\mathbb E_\eta}(\alpha_+;(a,b)) + I_{P,\mathbb E_\eta}(a;b) = I_{P,\mathbb E_\eta}(b;(\alpha_+,a)) + I_{P,\mathbb E_\eta}(\alpha_+;a)$ and from here
  
  {\small
  \begin{equation}
      TE_{x_1\to y_1}(\eta) - TE_{x_1\to y_1}(-\eta) = \int_{\mathbb E_\eta} \di\mu\, P\,\log_{2} \frac{K}{\hat K}\,.
  \end{equation}
  }
  
  Finally, let $\mathbb P$ denote the original phase space of the system with $\di\lambda$ denoting the Lebesgue volume element in $\mathbb P$. Then the above expression can be written as
  
  {\small
  \begin{align}
      \Delta TE_{x_1\to y_1}(\eta) &:= TE_{x_1\to y_1}(\eta) - TE_{x_1\to y_1}(-\eta) \\
      &= \int_{\mathbb P} \di\lambda\, \rho\,\log_{2} \frac{K}{\hat K}\,.\label{A:asymmetry}
  \end{align}
  }
  
  \noindent Notice that the function $K$ carries the dependence on the embedding as 
  $K =\abs{\frac{\partial \phi_{y_1}(-\eta;\cdot)}{\partial \phi_{y_1}(\eta;\cdot)}}$. 
  
\subsection{$\Delta TE_{x_1\to y_1}(\eta)$ for unidirectionally coupled systems}
  
  Suppose the dynamical system is generated by the vector field
  
  {\small 
  \begin{align}
      \dot x&= f(x,y)\\
      \dot y&= g(y)\,, 
  \end{align}
  }
  
  respectively, by the map 
  
  {\small 
  \begin{align}
      x(k+1)&= f(x(k),y(k))\\
      y(k+1)&= g(y(k))\,, 
  \end{align}
  }
  
  The change of variables given in equations [\ref{A:alpha+}]-[\ref{A:b}] now reduces to
  
  {\small
  \begin{align}
      &\alpha_+ = \phi_{y_1}(\eta;y)\,,\\
      &\alpha_- = \phi_{y_1}(-\eta;y)\,,\\
      &a= (y_1,\phi_{y_1}(-\tau_1;y),\cdots ,\phi_{y_1}(-n_1\tau_1;y))\,,\\
      & b= (x_1,\phi_{x_1}(-\tau_2;x,y),\cdots ,\phi_{x_1}(-n_2\tau_2;x,y))\,,
  \end{align}
 }
 
 Assume that the map $H(y):=(\phi_{y_1}(\eta;y),y_1,\phi_{y_1}(-\tau_1;y),\cdots ,\phi_{y_1}(-r\,\tau_1;y))$, for some $r\leq n_1$, is (smoothly) invertible, then we can write $y=H^{-1}(\alpha_+,\tilde a)$, where $\tilde a := (a_1,\cdots,a_r)$. The invertibility of $H$ is equivalent to saying that the delay reconstruction $\lc y_1(t+\eta),y_1(t),y_1(t-\tau_1),\cdots,y_1(t-r\,\tau_1)\rc$ reproduces the dynamics generated by $\dot y= g(y)$ (respectively, by $y(k+1)=g(y(k))$). In this case it follows that
 $\alpha_- = \phi_{y_1}\lb -\eta; H^{-1}(\alpha_+,\tilde a)\rb$ which we can generically denoted as $\alpha_- = h(\alpha_+,a)$ and hence the map generating the change of variables (equation [\ref{A:change_var}]) reduces in this case to
 
 {\small
 \begin{equation}
     f(\alpha_+,a,b) = \lb h(\alpha_+,a),a,b\rb
 \end{equation}
 }
 
 The key difference now is that the variable $b$ is decoupled from the pair $(\alpha_+,a)$ with respect to the map connecting the embeddings $\mathbb E_\eta$ and $\mathbb E_{-\eta}$. In other words, the lack of coupling $x\to y$ is transmitted into the embeddings $\mathbb E_\eta$ and $\mathbb E_{-\eta}$ as a decoupling between $b$ and $(\alpha_+,a)$. Such a decoupling implies that $K=\abs{\partial_{\alpha_+} h}$ will be a function of $(\alpha_+,a)$ alone. In addition, the set ${\mathbb E_{-\eta}}_{(\alpha_-,a;b)}:=\lc (\alpha_-,a,b+\beta)\in\mathbb E_{-\eta}\,|\,\beta\in\mathbb R\rc$ is mapped by $f^{-1}$ into the set
  $\lc (j(\alpha_-,a),a,b+\beta)\in\mathbb E_{\eta}\,|\,\beta\in\mathbb R\rc=:{\mathbb E_{\eta}}_{(\alpha_+,a;b)}$, where we have used that $f^{-1}(\alpha_-,a,b) = (j(\alpha_-,a),a,b)$. Therefore equation [\ref{A:Qmarginal}] reduces to
 
 {\small
 \begin{equation}
     Q(\alpha_-,a) = \frac{1}{K(\alpha_+,a)}\,P(\alpha_+,a)\label{A:trivialQmarginal}
 \end{equation}
 }
  
   and from this it follows that 
  
  {\small
  \begin{equation}
      I_{Q,\mathbb E_{-\eta}}(b;(\alpha_-,a)) = I_{P,\mathbb E_{\eta}}(b;(\alpha_+,a))
  \end{equation}
  }
  
  and hence 
  
  {\small 
  \begin{equation}
      \Delta TE_{x_1\to y_1}(\eta) = 0\,.
  \end{equation}\label{eq:supp_formal_unidir_zerononcausal}
  }
  
  This result tells us that for unidirectionally coupled systems in the direction, say $x \to y$, the difference $\Delta TE_{y\to x}(\eta)$ yields 0. 
  
  \subsection{Predictive asymmetry based on mutual information} 
  
  In this section we show that a similar relation with the underlying flow is found for a predictive asymmetry based on mutual information. More precisely, we consider again a generic dynamical system as in equations [\ref{A:dotx}] and [\ref{A:doty}], also assuming that there is a direct coupling $x_1\to y_1$, and we use the same definitions as in equations [\ref{A:alpha+}]-[\ref{A:b}]. We may consider, for instance, the asymmetry  
 
  {\small
   \begin{equation}
       \Delta I_{x_1\to y_1}(\eta) = I_{P,\mathbb E_\eta}((\alpha_+,a);b) - I_{Q,\mathbb E_{-\eta}}((\alpha_-,a);b)
   \end{equation}
  }
  
  In this case, the equality in equation [\ref{A:IQ=IP+log}] also implies that 
  
  {\small
   \begin{equation}
       \Delta I_{x_1\to y_1}(\eta) = \int_{\mathbb E_\eta} \di\mu\,P\,\log_{2}\frac{K}{\hat K}
   \end{equation}
  }
  
  A less repetitive example could be the asymmetry
  
  {\small
   \begin{equation}
       \tilde{\Delta I}_{x_1\to y_1}(\eta) = I_{P,\mathbb E_\eta}(\alpha_+;(a,b)) - I_{Q,\mathbb E_{-\eta}}(\alpha_-;(a,b))
   \end{equation}
  }
  
  In this case, for a fixed $(\alpha_-,a,b)\in\mathbb E_{-\eta}$, one finds  
  
  {\small
  \begin{align}
      Q(\alpha_-) &= \int_{{\mathbb E_{-\eta}}_{\alpha_-}} \di a'\di b'\, Q(\alpha_-,a',b')\nonumber\\
      &= \int_{{\mathbb E_{-\eta}}_{\alpha_-}} \di a'\di b'\,\frac{P(j(\alpha_-,a',b'),a',b')}{K(j(\alpha_-,a',b'),a',b')} \\
      &= \frac{1}{\bar K(\alpha_+)}P(\alpha_+)
  \end{align}
  }
  
  where 
 
 {
 \begin{align}
 \alpha_+ &= j(\alpha_{-},a,b) \\
 P(\alpha_{+}) &= \int_{{\mathbb E_\eta}_{\alpha_{+}}} \di a'\di b' \, P(j(\alpha_{-},a,b),a',b')
 \end{align}
}
  
  and 
  
  {\small
  \begin{equation}
     \frac{1}{\bar K(\alpha_+)} = \frac{1}{P(\alpha_+)}\int_{{\mathbb E_{-\eta}}_{\alpha_-}} \di a'\di b'\,\frac{P(j(\alpha_-,a',b'),a',b')}{K(j(\alpha_-,a',b'),a',b')}
  \end{equation}
  }
  
  Therefore,
  
  {\small
  \begin{align}
     I_{Q,\mathbb E_{-\eta}}(\alpha_-;(a,b)) &= \int_{\mathbb E_{-\eta}} \di\nu\,Q\,\log_{2}\frac{Q(\alpha_-,a,b)}{Q(\alpha_-) Q(a,b)}\nonumber\\
     &= \int_{\mathbb E_{\eta}} \di\mu\,P\,\log_{2}\frac{\frac{1}{K(\alpha_+,a,b)}P(\alpha_+,a,b)}{\frac{1}{\bar K(\alpha_+)}P(\alpha_+) P(a,b)}\nonumber\\
     &= I_{P,\mathbb E_{\eta}}(\alpha_+;(a,b)) - \int_{\mathbb E_{\eta}} \di\mu\,P\,\log_{2}\frac{K}{\bar K} 
  \end{align}
  }
  
  and from this it follows
  
  {\small
  \begin{equation}
      \tilde{\Delta I}_{x_1\to y_1}(\eta) = \int_{\mathbb E_{\eta}} \di\mu\,P\,\log_{2}\frac{K}{\bar K} 
  \end{equation}
  }

\section{\label{sec_supp:analytical_asymmetry_ar1}Exact expressions for the predictive asymmetry for autoregressive systems}

We will expand on the approach from \cite{Hahs2013} to compute the predictive asymmetry for arbitrary prediction lags for a coupled bivariate autoregressive system. For such systems, marginal entropies may be computed analytically.

\subsection{Covariance matrix for unidirectionally coupled AR1 system}\label{app:concepts}

Consider a simple unidirectionally coupled bivariate AR system with coefficients chosen such that the system is stationary (this is the same system as the main text's eq. \ref{eq:ar1_example}). Let $\sigma_x$ and $\sigma_y$ be the standard deviations of two independent normal distributions, where the noise draws $w_{t} \thicksim N(0, \sigma_x)$ and $v_{t} \thicksim N(0, \sigma_y)$ are independent at each time step.

{
\begin{align}
x_{t} &= a x_{t-1} + w_{t} : w_{t} \thicksim N(0, \sigma_x) \\
y_{t} &= c x_{t-1} + v_{t} : v_{t} \thicksim N(0, \sigma_y).
\end{align}
}

\subsubsection{Variances for $x_{t}$ and $y_{t}$}

\noindent Due to stationarity, which we have by definition, $E[x_{t}] = E[x_{t+k}] = 0$ and $E[y_{t}] = E[y_{t+k}] = 0$. We also find 

$$
\begin{aligned}
Var(x_t) &= Var(a x_{t-1} + w_{t}) = a^2 Var(x_{t-1}) + Var(w_t) = a^2 Var(x_{t}) + Var(w_t) \\
Var(x_t) &= Var(w_t)/(1-a^2)
\end{aligned}
$$

and 

$$
\begin{aligned}
Var(y_t) &= Var(c x_{t-1} + v_{t}) = c^2 Var(x_{t-1}) + Var(v_t) \\
Var(y_t) &= c^2 Var(x_{t}) + Var(v_t) 
\end{aligned}
$$

\subsubsection{Auto-covariances for $x_{t+k}$ and $y_{t+l}$}

 The covariances between observations of $x_t$ separated by $k$ time steps are therefore given by the following expectations

$$
\begin{aligned}
Cov(x_{t+k}, x_t) &= E\left[(x_{t+k} - E[x_{t+k}])(x_t - E[x_t]) \right] = E[x_{t+k}x_t] \\
Cov(y_{t+k}, y_t) &= E\left[(y_{t+k} - E[y_{t+k}])(y_t - E[y_t]) \right] = E[y_{t+k}y_t].
\end{aligned}
$$ 

with 

$$
\begin{aligned}
  x_{t+k} & = a^k \,x_t + \sum_{l=0}^{k-1} a^{l} \omega_{t+k+1-l}\\
  y_{t+k} & = c x_{t+k-1} +v_{t+k} = c\,a^{k-1} \,x_t + c \sum_{l=0}^{k-2} a^{l} \omega_{t+k-l} + v_{t+k}
\end{aligned}
$$

thus we find

\begin{align}
Cov(x_{t+k},x_t)&=E[x_{t+k} x_t]= E\ls \lb a^k x_t+\sum_{i=1}^k a^{k-i} w_{t+i}\rb x_t\rs \nonumber\\
&= a^k E[(x_t)^2]=a^k Var(x_t). \nonumber 
\end{align}

To evaluate the autocovariance of $y$, we use that 
\begin{align}
    y_{t+l} &= c a^{l-1} x_t + c\sum_{i=1}^{l-1-i} a^{l-1-i}w_{t+i} + v_{t+l}= \nonumber\\
    & = c a^{l-1} \lb a x_{t-1} + w_t \rb + c\sum_{i=1}^{l-1-i} a^{l-1-i}w_{t+i} + v_{t+l}= \nonumber\\
    & c a^{l} x_{t-1} + c\sum_{i=0}^{l-1-i} a^{l-1-i}w_{t+i} + v_{t+l} \nonumber
\end{align}
and also that 
$$y_t = c x_{t-1}+v_t$$
Then,
\begin{align}
    Cov(y_{t+l},y_t)&= E\ls\lb c a^{l} x_{t-1} + c\sum_{i=0}^{l-1-i} a^{l-1-i}w_{t+i} + v_{t+l}\rb \lb c x_{t-1} + v_t\rb\rs=\nonumber\\
    &= c^2 a^l Var(x_{t-1}) = \frac{c^2 a^l}{1-a^2} Var(w_t)\nonumber 
\end{align}
If $l=0$, then $Cov(y_t,y_t)=E\ls(y_t)^2\rs = Var(y_t) = \frac{c^2}{1-a^2} Var(w_t)+Var(v_t)$
 so, altogether we find that 
 \begin{align}
     Cov(y_{t+l},y_t) =  \frac{c^2 a^l}{1-a^2} Var(w_t) + \hat\theta(l) Var(v_t)\nonumber,
 \end{align}
 
where $\hat{\theta}(l) = 1$ if $l = 0$ and $\hat{\theta}(l) = 0$ otherwise.

\subsubsection{Cross-covariances for $x_t$ and $y_t$}

Using that $y_{t+l}= c x_{t+l-1}+v_{t+l}$ we find that
\begin{align}
    Cov(x_{t+k}, y_{t+l}) = Cov\lb x_{t+k}, c x_{t+l-1}+v_{t+l}\rb = E\ls x_{t+k} (c x_{t+l-1}+v_{t+l})\rs = c E\ls x_{t+k} x_{t+l-1}\rs\nonumber 
\end{align}
This expression is symmetric in the time labels, so it does not matter which one of $k$ and $l-1$ is the smallest. With no loss of generality we may thus suppose that $k\geq l-1$. Defining $r:= \abs{l-1-k}$, we have that $k = l - 1 + r$ and therefore $x_{t+k} = x_{t+l-1+r} = a^r x_{t+l-1} + \sum_{i=1}^r a^{r-i} w_{t+l-1-i}$. Accordingly,
\begin{align}
  Cov(x_{t+k},y_{t+l}) &= c E\ls x_{t+k} x_{t+l-1}\rs = c E\ls \lb a^r x_{t+l-1} + \sum_{i=1}^r a^{r-i} w_{t+l-1-i}\rb x_{t+l-1}\rs\nonumber\\
  & = c a^r Var (x_t) = \frac{c a^{\abs{l-1-k}}}{1-a^2} Var(w_t)\nonumber 
\end{align}

\subsection{Filling the covariance matrix}

Now that we have established the dependence of the covariance between time steps spaced arbitrary far from each other on time on the coefficients $a$ and $c$, we can proceed with predictive asymmetry computations. Let $\eta_{max}$ be the maximum prediction lag, and let 

$$
\vec{x} = [x_t\ x_{t+1}\ x_{t+2}\ \cdots x_{t+2\eta_{max}}\ \ y_{t}\ y_{t+1}\ y_{t+2} \cdots y_{t+2\eta_{max}}]^T
$$

\noindent and let $C_{\vec{x}}$ denote the covariance matrix for $\vec{x}$. Equivalently, if choosing $\eta_{max}$ odd, we may shift the time indices and consider 

$$\vec{x} = [x_{t-\eta_{max}} \ \cdots \, \ x_{t-1}\ x_{t}\ x_{t+1} \ \cdots \, x_{t+\eta_{max}} \ \, y_{t-\eta_{max}} \ \cdots \, \ y_{t-1} \ \, y_{t}\ y_{t+1} \ \cdots \, y_{t+\eta_{max}} ]^T
$$ 

\noindent Now, $C_{\vec{x}}$ provides sufficient information to compute transfer entropy for maximum prediction lag $2\eta_{max}$, alternatively, the predictive asymmetry for maximum  prediction lag $\eta_{max}$, assuming the lags included in the history for the target variable does not exceed $\eta_{max}$.

\noindent Computing the predictive asymmetry for a maximum prediction lag $\eta_{max}$, the covariance matrix will have thus dimensions $N$-by-$N$, where $N = 2(2\eta_{max}+1)$, accounting for $\eta_{max}$ lags for $x$ and $\eta_{max}$ lags for $y$, plus the zero lag cases. If dealing with a random system such as an autoregressive order-1 system (AR1), then this is all we need for transfer entropy computations. We just need to subset the relevant portions of the covariance matrix, compute relevant entropies, and from that compute the predictive asymmetry.

\subsubsection{Computing the predictive asymmetry}

Let $S$ and $T$ denote two generic source and target process. For convenience of notation, let $S_{pp}$ and $T_{pp}$ denote the present and past of the source and target variables (time series). During computation, $S_{pp}$ and $T_{pp}$ are kept fixed. Next, let $T_{\eta}$ denote the time series $T$, but lagged $\eta$ time steps into the future ($\eta > 0$) or into the past ($\eta < 0$). Say we want to compute the predictive asymmetry from $S$ to $T$. We then have 

$$
\begin{aligned}
\mathbb{A}_{S \to T}(\eta) &= \sum_{\nu = 1}^{\eta} I(S_{pp}, T_{\nu} | T_{pp}) - \sum_{\nu = -1}^{-\eta} I(S_{pp}, T_{\nu} | T_{pp})
\end{aligned}
$$

\noindent In terms of entropies, the conditional mutual information (CMI) terms are 

$$
\begin{aligned}
I(S_{pp}, T_\nu | T_{pp}) &= h(S_{pp} | T_{pp}) + h(T_\nu | T_{pp}) + h(S_{pp}; T_\nu |T_{pp}) \\
&= \left[h(S_{pp}, T_{pp}) - h(T_{pp}) \right] + \left[ h(T_\nu, T_{pp}) - h(T_{pp}) \right] - \left[ h(S_{pp}, T_\nu, T_{pp}) - h(T_{pp}) \right] \\
&= h(S_{pp}, T_{pp}) + h(T_\nu, T_{pp}) - h(T_{pp}) - h(S_{pp}, T_{\eta}, T_{pp}),
\end{aligned}
$$

and these entropies can be computed exactly from the covariance matrix, following the approach of \cite{Hahs2013}.

\subsection{Covariance matrix for bidirectionally coupled AR1 systems}\label{app:bivariateAR}

\subsubsection{General setting}

Consider the general linear AR system 

\begin{align}
    &x_{t+1} = a x_{t} + b y_t + u_{t+1}\,,\\
    &y_{t+1} = c x_{t} + d y_t + v_{t+1}\,,
\end{align}
where $u_t\sim N(0,\sigma_u)$ and $v_t\sim N(0,\sigma_v)$. Expressed more compactly, the above system reads
\begin{equation}
    X_{t+1} = A\cdot X_{t} + W_{t+1}\,,\label{eq:bivariate}
\end{equation}
with $X_t:=\lb\begin{array}{c}
        x_t\\ y_t 
\end{array}\rb$ and $W_t:=\lb\begin{array}{c}
        u_t\\ v_t 
\end{array}\rb$. The matrix $A:=\lb\begin{array}{cc}
    a & b \\
    c & d
\end{array}\rb$, contains the coefficients of the model. We will assume that all the eigenvalues of $A$ have norm strictly less than 1. In addition $E[W_t]:=\lb\begin{array}{c}
        E[u_t]\\ E[v_t] 
\end{array}\rb = \lb\begin{array}{c}
        0\\ 0 
\end{array}\rb$ and for the system to be stationary the column vector of expectation values of the variables must verify
$\mathcal E:=E[X_t]=E[X_{t+1}]$ and therefore, from equation (\ref{eq:bivariate}), $\mathcal E = A\cdot\mathcal E$. Since, in particular, $1$ can not be an eigenvalue of $A$, it must follow that $\mathcal E=0$. 

\subsubsection{Change of variables}

By redefining linearly the variables as %
\begin{equation}
    \begin{array}{c}
         \hat x_t := \alpha x_t +\beta y_t\,,\\
       \hat y_t := \gamma x_t +\delta y_t\,,
    \end{array}
     \qquad \forall\,t
\end{equation}
with $U:=\lb\begin{array}{cc}
    \alpha & \beta \\
    \gamma &\delta 
\end{array}\rb$ being an invertible matrix (independent of time), the system in equation (\ref{eq:bivariate}) is transformed as
\begin{align}
    &\hat X_{t+1} = \hat A \cdot \hat X_t + \hat W_{t+1}\,,\\
    & \hat A := U\cdot A\cdot U^{-1}\,,\\
    & \hat X_t:=\lb\begin{array}{c}
         \hat x_t  \\
         \hat y_t 
    \end{array}\rb\,,\\
    & \hat W_t:= U\cdot W_t =\lb\begin{array}{c}
         \alpha u_t + \beta v_t  \\
         \gamma u_t + \delta v_t
    \end{array}\rb\,.
\end{align}
The variances and covariances between the new noise terms verify that 
\begin{align}
    Var(\hat u_t) &= E\ls\lb \alpha u_t + \beta v_t\rb \lb \alpha u_t + \beta v_t\rb\rs = \alpha^2 \sigma_u^{\,2} + \beta^2 \sigma_v^{\,2}\,,\label{eq:varhatu}\\
    Var(\hat v_t) &= E\ls\lb \gamma u_t + \delta v_t\rb \lb \gamma u_t + \delta v_t\rb\rs = \gamma^2 \sigma_u^{\,2} + \delta^2 \sigma_v^{\,2}\,,\label{eq:varhatv}\\
    Cov(\hat u_t,\hat v_t) &= Cov\lb \alpha u_t + \beta v_t, \gamma u_t + \delta v_t\rb = \alpha \gamma \sigma_u^{\,2} + \delta\beta  \sigma_v^{\,2}\,,\label{eq:covhatuhatv}
\end{align}
where we have used that $Var(u_t) = \sigma_u^{\,2}$ and $Var(v_t)=\sigma_v^{\,2}$. In addition, it is clear that
\begin{align}
    & Cov(\hat u_{t+k},\hat u_t) = Cov(\hat v_{t+k},\hat v_t) = Cov(\hat u_{t+k},\hat v_t)= Cov(\hat v_{t+k},\hat u_t) = 0\,, \qquad \forall\, k\neq 0\,,
\end{align}
and all the covariances between the variables and the noise terms do vanish.

Suppose that in the variables $\hat X$, the matrix $\hat A$ adopts a particularly simple form so that the covariances 
\begin{eqnarray}
  &Cov(\hat x_{t+k},\hat x_t)\,, & Cov(\hat y_{t+k},\hat y_t) \,,\\
  &Cov(\hat x_{t+k},\hat y_t)\,, & Cov(\hat y_{t+k},\hat x_t) \,,
\end{eqnarray}
are easily computed. In addition, the entries of $U^{-1}$ are
\begin{equation}
    U^{-1} = \frac{1}{\alpha \delta-\beta\gamma} \lb\begin{array}{cc}
     \delta& -\beta  \\
     -\gamma& \alpha 
\end{array}\rb\,,
\end{equation}
and hence 
\begin{align}
    x_t &= \frac{1}{\alpha \delta-\beta\gamma} \lb \delta \hat x_t - \beta \hat y_t\rb\,,\\
    y_t &= \frac{1}{\alpha \delta-\beta\gamma} \lb \alpha \hat y_t - \gamma \hat x_t\rb\,.
\end{align}
Therefore we find
\begin{align}
    Cov(x_{t+k}, x_t) &= \frac{1}{\lb\alpha \delta-\beta\gamma\rb^2}  Cov\lb\delta \hat x_{t+k} - \beta \hat y_{t+k},\delta \hat x_{t} - \beta \hat y_{t}\rb\nonumber \\
    & = \frac{1}{\lb\alpha \delta-\beta\gamma\rb^2} \ls \delta^2 Cov(\hat x_{t+k},\hat x_t)+\beta^2 Cov(\hat y_{t+k},\hat y_t)-\delta \beta \lb Cov(\hat x_{t+k},\hat y_t) + Cov(\hat y_{t+k},\hat x_t)\rb\rs\,,\label{eq:Covxx}\\
    \nonumber\\
    Cov(y_{t+k}, y_t) &= \frac{1}{\lb\alpha \delta-\beta\gamma\rb^2}  Cov\lb\alpha \hat y_{t+k} - \gamma \hat x_{t+k},\alpha \hat y_{t} - \gamma \hat x_{t}\rb\nonumber \\
    & = \frac{1}{\lb\alpha \delta-\beta\gamma\rb^2} \ls \alpha^2 Cov(\hat y_{t+k},\hat y_t)+\gamma^2 Cov(\hat x_{t+k},\hat x_t)-\alpha \gamma \lb Cov(\hat x_{t+k},\hat y_t) + Cov(\hat y_{t+k},\hat x_t)\rb\rs\,,\label{eq:Covyy}\\
    \nonumber\\
    Cov(y_{t+k}, x_t) &= \frac{1}{\lb\alpha \delta-\beta\gamma\rb^2}  Cov\lb\alpha \hat y_{t+k} - \gamma \hat x_{t+k},\delta \hat x_{t} - \beta \hat y_{t}\rb\nonumber \\
    & = \frac{1}{\lb\alpha \delta-\beta\gamma\rb^2} \ls \alpha\delta Cov(\hat y_{t+k},\hat x_t)+\gamma\beta Cov(\hat x_{t+k},\hat y_t)-\alpha \beta Cov(\hat y_{t+k},\hat y_t) -\gamma\delta Cov(\hat x_{t+k},\hat x_t)\rs\,,\label{eq:Covyx}\\
    \nonumber\\
    Cov(x_{t+k}, y_t) &= \frac{1}{\lb\alpha \delta-\beta\gamma\rb^2}  Cov\lb\delta \hat x_{t+k} - \beta \hat y_{t+k},\alpha \hat y_{t} - \gamma \hat x_{t}\rb\nonumber \\
    & = \frac{1}{\lb\alpha \delta-\beta\gamma\rb^2} \ls \alpha\delta Cov(\hat x_{t+k},\hat y_t)+\gamma\beta Cov(\hat y_{t+k},\hat x_t)-\alpha \beta Cov(\hat y_{t+k},\hat y_t) -\gamma\delta Cov(\hat x_{t+k},\hat x_t)\rs\,,\label{eq:Covxy}
\end{align}
In the following, we will apply these formulas to two inequivalent examples of bivariate AR models.

\subsubsection{Example 1. The matrix $A$ has two distinct real eigenvalues.}

In that case there is an invertible matrix $U=\lb\begin{array}{cc}
    \alpha &\beta  \\
    \gamma &\delta 
\end{array}\rb$ such that $\hat A=U\cdot A\cdot U^{-1} = \lb \begin{array}{cc}
    \lambda_1 & 0 \\
    0 & \lambda_2
\end{array}\rb$ and the system reduces to
\begin{align}
  &\hat x_{t+1} = \lambda_1 \hat x_t + \hat u_{t+1}\,,\label{eq:hatx}\\
  &\hat y_{t+1} = \lambda_2 \hat y_t + \hat v_{t+1}\,,\label{eq:haty}
\end{align}
From equations (\ref{eq:hatx}) and (\ref{eq:haty}) we deduce that $Var(\hat x_{t})=Var(\hat x_{t+1})=Var\lb \lambda_1 \hat x_t + \hat u_t\rb = \lambda_1^{\,2} Var(\hat x_t) + Var(\hat u_t)$ and similarly for $Var(\hat y_t)$. In addition, 
$Cov(\hat x_t,\hat y_t)= Cov(\hat x_{t+1},\hat y_{t+1}) = Cov\lb \lambda_1 \hat x_t+\hat u_{t+1},\lambda_2 \hat y_t+\hat v_{t+1}\rb = \lambda_1\lambda_2 Cov(\hat x_t,\hat y_t)+ Cov(\hat u_{t+1},\hat v_{t+1})$.
From these equalities we find
\begin{align}
    & Var(\hat x_t) = \frac{\alpha^2 \sigma_u^{\,2}+\beta^2\sigma_v^{\,2}}{1-\lambda_1^{\,2}}\,,\\
    & Var(\hat y_t) = \frac{\gamma^2 \sigma_u^{\,2}+\delta^2\sigma_v^{\,2}}{1-\lambda_2^{\,2}}\,,\\
    & Cov(\hat x_t,\hat y_t) = \frac{\alpha \gamma \sigma_u^{\,2} + \delta\beta  \sigma_v^{\,2}}{1-\lambda_1\lambda_2}\,,
\end{align}
where we have used the results found in equations (\ref{eq:varhatu})-(\ref{eq:covhatuhatv}). It also clearly holds that
\begin{align}
  &\hat x_{t+k} = \lambda_1^{\,k} \hat x_t +  \sum_{l=0}^{k-1} \lambda_1^{\,k-1-l} \hat u_{t+1-l}\,,\\
  &\hat y_{t+k} = \lambda_2^{\,k} \hat y_t +  \sum_{l=0}^{k-1} \lambda_2^{\,k-1-l} \hat v_{t+1-l}\,,
\end{align}
and therefore
\begin{align}
    Cov(\hat x_{t+k},\hat x_t) & = \lambda_1^{\,k} Var(\hat x_t) = \frac{\lambda_1^{\,k}}{1-\lambda_1^{\,2}} \lb \alpha^2\sigma_u^{\,2} + \beta^2\sigma_v ^{\,2}\rb\,,\label{eq:cxx}\\
    \nonumber\\
    Cov(\hat y_{t+k},\hat y_t) & = \lambda_2^{\,k} Var(\hat y_t) = \frac{\lambda_2^{\,k}}{1-\lambda_2^{\,2}} \lb \gamma^2\sigma_u^{\,2} + \delta^2\sigma_v^{\,2}\rb\,,\label{eq:cyy}\\
    \nonumber\\
    Cov(\hat y_{t+k},\hat x_t) & = \lambda_2^{\,k} Cov(\hat y_t,\hat x_t) = \frac{\lambda_2^{\,k}}{1-\lambda_1\lambda_2} \lb \alpha \gamma \sigma_u^{\,2} + \delta\beta  \sigma_v^{\,2}\rb\,,\label{eq:cyx}\\
    \nonumber\\
    Cov(\hat x_{t+k},\hat y_t) & = \lambda_1^{\,k} Cov(\hat x_t,\hat y_t) = \frac{\lambda_1^{\,k}}{1-\lambda_1\lambda_2} \lb \alpha \gamma \sigma_u^{\,2} + \delta\beta  \sigma_v^{\,2}\rb\,.\label{eq:cxy}
\end{align}
As an example of such a case, we will consider the system
\begin{align}
    & x_{t+1} = a x_t + sb y_t + u_{t+1}\,,\\
    & y_{t+1} = a y_t + sc x_t + v_{t+1}\,,
\end{align}
with $b, c >0$ and $s^2 = 1$. In this case, the matrix $A=\lb\begin{array}{cc}
     a&sb  \\
     sc&a 
\end{array}\rb$ has eigenvalues $\lambda_{\pm}:=a\pm \sqrt{bc}$, and hence we demand that both $\abs{a+\sqrt{bc}}$ and $\abs{a-\sqrt{bc}}$, be strictly less than 1. It is easy to check that the non singular matrix 
\begin{equation}
    U  =\lb\begin{array}{cc}
        \frac{1}{2\sqrt{b}} & \frac{s}{2\sqrt{c}} \\
         \frac{-1}{2\sqrt{b}} & \frac{s}{2\sqrt{c}} 
    \end{array}\rb=:\lb\begin{array}{cc}
    \alpha &\beta  \\
    \gamma &\delta 
\end{array}\rb\,,
\end{equation}
verifies that $U\cdot A\cdot U^{-1} = \lb\begin{array}{cc}
    \lambda_+ &0  \\
    0 &\lambda_- 
\end{array}\rb$. Therefore, by substituting the corresponding $\alpha,\beta,\gamma$ and $\delta$ parameters in equations (\ref{eq:cxx})-(\ref{eq:cxy}), we find 
\begin{align}
    Cov(\hat x_{t+k},\hat x_t) &=  \frac{\lb a+\sqrt{bc}\rb^k}{1-\lb a+\sqrt{bc}\rb^2} \lb \frac{1}{4 b}\sigma_u^{\,2} + \frac{1}{4c}\sigma_v^{\,2}\rb\,,\\
    \nonumber\\
    Cov(\hat y_{t+k},\hat y_t) & = \frac{\lb a-\sqrt{bc}\rb^k}{1-\lb a-\sqrt{bc}\rb^2} \lb \frac{1}{4 b}\sigma_u^{\,2} + \frac{1}{4c}\sigma_v^{\,2}\rb\,,\\
    \nonumber\\
    Cov(\hat y_{t+k},\hat x_t) & = \frac{\lb a-\sqrt{bc}\rb^k}{1+ bc - a^2} \lb -\frac{1}{4 b} \sigma_u^{\,2} + \frac{1}{4 c} \sigma_v^{\,2}\rb\,,\\
    \nonumber\\
    Cov(\hat x_{t+k},\hat y_t) & = \frac{\lb a+\sqrt{bc}\rb^k}{1+bc - a^2}\lb -\frac{1}{4 b} \sigma_u^{\,2} + \frac{1}{4 c}  \sigma_v^{\,2}\rb\,,
\end{align}
and the results on the covariances between the original variables are obtained from equations (\ref{eq:Covxx})-(\ref{eq:Covxy}) with $\alpha = \frac{1}{2\sqrt{}b}$, $\beta = \frac{s}{2\sqrt{c}}$, $\gamma=\frac{-1}{2\sqrt{b}}$ and $\delta = \frac{s}{2\sqrt{c}}$.

\subsubsection{Example 2. The matrix $A$ has only one eigenvalue $\lambda$ (and therefore real).}

The trivial case in which $A$ is proportional to the identity is already contained in the previous case example. The non trivial instance of $A$ is therefore not diagonalizable. In that case there is an invertible matrix $U=\lb\begin{array}{cc}
    \alpha &\beta  \\
     \gamma & \delta
\end{array}\rb$ such that 
\begin{equation}
    U\cdot A\cdot U^{-1} = \lb\begin{array}{cc}
        \lambda &0  \\
        1 & \lambda
    \end{array}\rb\,,
\end{equation}
which corresponds to the Jordan normal form. Hence with the redefinitions 
\begin{align}
    &\hat x_t = \alpha x_t +\beta y_t \,,\\
    &\hat y_t = \gamma x_t +\delta y_t \,,\\
    &\hat u_t = \alpha u_t +\beta v_t \,,\\
    &\hat v_t = \gamma u_t +\delta v_t \,,
\end{align}
the system in equation (\ref{eq:bivariate}) reduces to
\begin{align}
    &\hat x_{t+1} = \lambda \hat x_t + \hat u_{t+1}\,,\label{eq:hatxlambda}\\
  &\hat y_{t+1} = \lambda \hat y_t + \hat x_t + \hat v_{t+1}\,,\label{eq:hatylambda}
\end{align}
and still it holds that 
\begin{align}
    Var(\hat u_t) &= \alpha^2 \sigma_u^{\,2} + \beta^2 \sigma_v^{\,2}\,,\\
    Var(\hat v_t) & = \gamma^2 \sigma_u^{\,2} + \delta^2 \sigma_v^{\,2}\,,\\
    Cov(\hat u_t,\hat v_t) &= \alpha \gamma \sigma_u^{\,2} + \delta\beta  \sigma_v^{\,2}\,.
\end{align}
From equations (\ref{eq:hatxlambda})-(\ref{eq:hatylambda}) we find, 
\begin{align}
 Var(\hat x_t) &= Var(\hat x_{t+1}) = Cov\lb\lambda\hat x_t+\hat u_{t+1}, \lambda\hat x_t+\hat u_{t+1}\rb \nonumber\\
  & = \lambda^2 Var(\hat x_t) + Var(\hat u_t)\,, 
  \end{align}
  and hence
  \begin{align}
  Var(\hat x_t)  &= \frac{Var(\hat u_t)}{1-\lambda^2} =   \frac{\alpha^2 \sigma_u^{\,2} + \beta^2 \sigma_v^{\,2}}{1-\lambda^2}=:\psi\,.   
\end{align}
Also 
\begin{align}
    Cov(\hat x_t,\hat y_t) &= Cov(\hat x_{t+1},\hat y_{t+1}) = Cov(\lambda \hat x_t+\hat u_{t+1},\lambda \hat y_t +\hat x_t + \hat v_{t+1}) \nonumber\\
    & = \lambda^2 Cov(\hat x_t,\hat y_t) + \lambda Var(\hat x_t) + Cov(\hat u_t,\hat v_t)\,, \nonumber
\end{align}
and hence 
\begin{align}
    Cov(\hat x_t,\hat y_t) &= \frac{\lambda}{1-\lambda^2} Var(\hat x_t) + \frac{Cov(\hat u_t,\hat v_t)}{1-\lambda^2} \nonumber\\
    &=\frac{\lambda}{1-\lambda^2} \psi + \frac{\alpha \gamma \sigma_u^{\,2} + \delta\beta  \sigma_v^{\,2}}{1-\lambda^2}=: \phi\,.
\end{align}

Finally,
\begin{align}
    Var(\hat y_t) &= Var(\hat y_{t+1}) =  Cov(\lambda \hat y_t+\hat x_t +\hat v_{t+1},\lambda \hat y_t+\hat x_t +\hat v_{t+1})\nonumber\\
    &= \lambda^2 Var(\hat y_t) + 2\lambda Cov(\hat x_t,\hat y_t) + Var(\hat x_t) + Var(\hat v_t)\,,
\end{align}
and therefore
\begin{align}
    Var(\hat y_t) &=
    \frac{2\lambda}{1-\lambda^2}Cov(\hat x_t,\hat y_t)  + \frac{1}{1-\lambda^2} Var(\hat x_t) + \frac{Var(\hat v_t)}{1-\lambda^2}\nonumber\\
    &=\frac{2\lambda}{1-\lambda^2} \phi + \frac{1}{1-\lambda^2}\psi + \frac{\gamma^2 \sigma_u^{\,2} + \delta^2 \sigma_v^{\,2}}{1-\lambda^2} =: \theta\,. 
\end{align}
To compute the $k$-th time step covariances, it is more convenient to express the system in matrix form as
\begin{equation}
    \hat X_{t+1} = \hat A\cdot \hat X_t+\hat W_{t+1}\,.
\end{equation}
From here we easily deduce that 
\begin{equation}
    \hat X_{t+k} = \hat A^k\cdot \hat X_t+\sum_{l=0}^{k-1} \hat A^{k-1+l}\cdot \hat W_{t+1+l}\,.
\end{equation}
In addition, one easily checks that $\hat A^k = \lb \begin{array}{cc}
    \lambda^k & 0 \\
    k \lambda^{k-1} &\lambda^k 
\end{array}\rb$, hence we have
\begin{align}
    & \hat x_{t+k} = \lambda^k \hat x_t + ({\rm noise\,terms})\,,\\
    &\hat y_{t+k} = \lambda^k \hat y_t + k\lambda^{k-1}\hat x_t + ({\rm noise\,terms})\,. 
\end{align}
From here we deduce that
\begin{align}
 Cov(\hat x_{t+k},\hat x_t)  &= \lambda^k Var(\hat x_t) = \lambda^k \psi\,,\\
 Cov(\hat y_{t+k},\hat y_t)  &= Cov(\lambda^k \hat y_t+k\lambda^{k-1} \hat x_t,\hat y_t) = \lambda^k Var(\hat y_t)+k\lambda^{k-1} Cov(\hat x_t,\hat y_t)\nonumber\\
 & = \lambda^k \theta+k\lambda^{k-1} \phi\,,\\
 Cov(\hat y_{t+k},\hat x_t) & = Cov(\lambda^k \hat y_t+k\lambda^{k-1} \hat x_t,\hat x_t) = \lambda^k Cov(\hat y_t,\hat x_t)+k\lambda^{k-1} Var(\hat x_t)\nonumber\\
 & = \lambda^k \phi+k\lambda^{k-1} \psi\,,\\
 Cov(\hat x_{t+k},\hat y_t)  &= \lambda^k Cov(\hat x_t,\hat y_t) = \lambda^k \phi\,.
\end{align}
In summary,
\begin{align}
    &Cov(\hat x_{t+k},\hat x_t)   = \lambda^k \psi\,,\label{eq:covlambdaxkx}\\
   & Cov(\hat y_{t+k},\hat y_t)  = \lambda^k \theta+k\lambda^{k-1} \phi\,,\label{eq:covlambdayky}\\
    &Cov(\hat y_{t+k},\hat x_t)  = \lambda^k \phi+k\lambda^{k-1} \psi\,,\label{eq:covlambdaykx}\\
    & Cov(\hat x_{t+k},\hat y_t)  =\lambda^k \phi\,,\label{eq:covlambdaxky}\\
    \nonumber\\
    &\psi = \frac{\alpha^2 \sigma_u^{\,2} + \beta^2 \sigma_v^{\,2}}{1-\lambda^2}\,,\nonumber\\
    &\phi  = \frac{\lambda}{1-\lambda^2} \psi + \frac{\alpha \gamma \sigma_u^{\,2} + \delta\beta  \sigma_v^{\,2}}{1-\lambda^2}\,,\nonumber\\
    &\theta = \frac{2\lambda}{1-\lambda^2} \phi + \frac{1}{1-\lambda^2}\psi + \frac{\gamma^2 \sigma_u^{\,2} + \delta^2 \sigma_v^{\,2}}{1-\lambda^2}\,.\nonumber
\end{align}
A generic case of this class of systems is given by
\begin{align}
    x_{t+1} = (\lambda+a) x_t - b y_t + u_{t+1}\,,\\
    y_{t+1} = (\lambda-a) y_t + \frac{a^2}{b} x_t + v_{t+1}\,,
\end{align}
 with $\abs{\lambda}<1$, $b\neq 0$ and $a\in\mathbb R$. One checks that the matrix $A=\lb\begin{array}{cc}
     \lambda+a & -b  \\
      \frac{a^2}{b}& \lambda - a 
 \end{array}\rb$ is brought into 
 $\hat A = \lb\begin{array}{cc}
     \lambda &0  \\
      1&\lambda 
 \end{array}\rb$ with the linear transformation
 \begin{equation}
     U = \lb\begin{array}{cc}
         \frac{a}{b} &-1  \\
         \frac{b}{a^2+b^2} & \frac{a}{a^2+b^2} 
     \end{array}\rb = \lb\begin{array}{cc}
         \alpha &\beta  \\
          \gamma &\delta 
     \end{array}\rb\,.
 \end{equation}
 The covariances can thus be found by using equations (\ref{eq:covlambdaxkx})-(\ref{eq:covlambdaxky}) and (\ref{eq:Covxx})-(\ref{eq:Covxy}) with 
 \begin{equation}
     \begin{array}{ll}
          \alpha = \frac{a}{b}\,, & \beta = -1 \,,\\
      \gamma = \frac{b}{a^2+b^2}\,, & \delta = \frac{a}{a^2+b^2} \,. 
     \end{array}
 \end{equation}

\clearpage

\section{\label{sec:supp_heuristic}Heuristic explanation for the sign of the predictive asymmetry in the general case}
For the unidirectional AR systems we show analytically that the predictive asymmetry is negative in the non-causal direction. In the general case, this behavior may be heuristically understood as follows.

Recall the definition of the predictive asymmetry from  variable $x$ to $y$:

{\small \begin{equation}
\mathbb{A}_{x \to y}(\eta) = \int TE_{x\to y}( \nu) \di \nu -  \int TE_{x\to y}(-\nu) \di \nu \label{eq:supp_asymmetry}
\end{equation}
}

\subsubsection{Unidirectional coupling}Consider first the case of unidirectional coupling $x \to y$. For the simplest possible TE (three-dimensional) analysis, the forward-prediction term in the expression for $\mathbb{A}_{x \to y}$ (eq. \ref{eq:supp_asymmetry}) becomes

{\small
\begin{align}
TE_{x\to y}(\eta) =  \int P(x_t, y_t, y_{t+\eta}) \log_{2}{ \left( \dfrac{P(y_{t+\eta}) |P(y_t), P(x_t)}{P(y_{t+\eta}) |P(y_t)} \right)}, 
\label{eq:te_forwardlags_xtoy_pos}
\end{align}
} 

which measures how much, on average, knowing something about the present of $x$ improves our ability to predict the future of $y$. If $x$ does actually have an influence on $y$, then we expect $TE_{x \to y} > 0$. 

What about the second term? It is not very intuitive to think about backwards prediction. However, after some algebraic manipulation, the backward-prediction term reads:

{\small
\begin{equation}
TE_{x\to y}(-\eta) =  \int P(x_t, y_t, y_{t-\eta}) \log_{2}{ \left( \dfrac{P(x_t) |P(y_t), P(y_{t-\eta})}{P(x_t) |P(y_t)} \right)}. \nonumber
\end{equation}}

The backwards-lag prediction  thus quantifies how well the knowledge about the past of $y$ improves our prediction of the present of $x$, given the present of $y$. One may erroneously conclude that this term should be trivially zero --- that if the dynamical influence is $x \to y$, then neither the past nor present of $y$ should not have a measureable effect on the present of $x$. But if there is dynamical influence $x \to y$, then information about the past of $x(t)$ is encoded in $y(t)$ and therefore $y(t)$ can be considered a proxy for past values of $x$. Including information about $y(t - \eta)$ may thus improve our prediction of the outcome of $x(t)$. If $x \to y$, then we expect to \textit{statistically} quantify some influence $y \to x$ due to $y$ being a proxy for $x$ \footnote[1]{$TE_{x \to y}$ measures a TE-like quantity, except the causal direction flips and the conditioning in the argument of the logarithm occurs only on one variable, not on a mixture of the two variables, as for regular TE (eqs. \ref{eq:te_forwardlags_xtoy_pos} and \ref{eq:te_forwardlags_ytox_pos}).}. 

In summary, $\mathbb{A}_{x \to y}$ compares the direct influence $x_{\text{present}} \to y_{\text{future}}$ resulting from the forcing $x \to y$ with the indirect influence $x_{\text{present}} \to x_{\text{future}}$ arising through the interaction of $x$ with $y$. The key concept of the asymmetry test is that when an underlying coupling $x \to y$ exists, then the latter may be statistically detectable. A reliable causality estimator should be better at detecting direct influences than indirect influences, so we expect $TE_{x \to y}(\eta) > TE_{x \to y}(-\eta)$, and hence
$\mathbb{A}_{x \to y} = TE_{x \to y}(\eta) - TE_{x \to y}(-\eta) > 0$.

In the opposite direction $y \to x$, we are numerically estimating the following integrals

{\small 
\begin{subequations}
\begin{align}
TE_{y\to x}(\eta) &=  \int P(x_t, y_t, y_{t+\eta}) \log_{2}{ \left( \dfrac{P(x_{t+\eta}) |P(x_t), P(y_t)}{P(x_{t+\eta}) |P(x_t)} \right)} \label{eq:te_forwardlags_ytox_pos} \\
TE_{y\to x}(-\eta) &=  \int P(x_t, y_t, y_{t-\eta}) \log_{2}{ \left( \dfrac{P(y_t) |P(x_t), P(x_{t-\eta})}{P(y_t) |P(x_t)} \right) }. \label{eq:te_forwardlags_ytox_neg}
\end{align}
\end{subequations}
} 

The forwards-prediction here quantifies the extent to which having information about the past of $y$ improves our knowledge about the future of $x$. This term also measures an indirect effect of $x$ on it own future through its interaction with $y$. The backwards-prediction  represents the statistical measure of a direct influence $x \to y$ (analogous but not equal to eq. \ref{eq:te_forwardlags_xtoy_pos}). Hence, we expect $TE_{y \to x}(\eta) < TE_{y \to x}(-\eta)$ and hence $\mathbb{A}_{y \to x} < 0$.

\subsubsection{Bidirectional coupling}
For systems that are bidirectionally coupled $x \leftrightarrow y$, the situation is a bit more complicated, but the same basic argument applies. Recall that 

{\normalsize
\begin{align} 
\mathbb{A}_{x \to y} = & TE_{x \to y}(\eta) - TE_{x\to y}(-\eta) = \nonumber\\ 
& \int P(x_t, y_t, y_{t+\eta}) \log_{2}{ \left( \dfrac{P(y_{t+\eta}) |P(y_t), P(x_t)}{P(y_{t+\eta}) |P(y_t)} \right)} - \nonumber\\
& \int P(x_t, y_t, y_{t-\eta}) \log_{2}{ \left( \dfrac{P(x_t) |P(y_t), P(y_{t-\eta})}{P(x_t) |P(y_t)} \right)} \nonumber
\end{align}
}

and

{\small
\begin{align}
\mathbb{A}_{y \to x} = & TE_{y \to x}(\eta) - TE_{y\to x}(-\eta) = \nonumber \\
& \int P(x_t, y_t, y_{t+\eta}) \log_{2}{ \left( \dfrac{P(x_{t+\eta}) |P(x_t), P(y_t)}{P(x_{t+\eta}) |P(x_t)} \right)} - \nonumber \\
& \int P(x_t, y_t, y_{t-\eta}) \log_{2}{ \left( \dfrac{P(y_t) |P(x_t), P(x_{t-\eta})}{P(y_t) |P(x_t)} \right)}. \nonumber
\end{align}
}

What happens if there is a difference in the coupling strengths? The only term that would pick up any change in the coupling strength is in the argument of the logarithm. If there is bidirectional coupling with coupling strengths $c_{x \to y} > c_{y \to x}$, then we expect $TE_{x \to y}(\eta) > TE_{y \to x}(\eta)$, or 

{\scriptsize
\begin{align}
\log_{2}{ \left( \dfrac{P(y_{t+\eta}) |P(y_t), P(x_t)}{P(y_{t+\eta}) |P(y_t)} \right)}  >  \log_{2}{ \left( \dfrac{P(x_{t+\eta}) |P(x_t), P(y_t)}{P(x_{t+\eta}) |P(x_t)} \right)} \nonumber
\end{align}
}

which is the expected from the usual TE. What about the backwards-prediction terms? If $c_{x \to y} > c_{y \to x}$, then 

{\scriptsize
\begin{align}
\log_{2}{ \left( \dfrac{P(x_t) |P(y_t), P(y_{t-\eta})}{P(x_t) |P(y_t)} \right)} <\log_{2}{ \left( \dfrac{P(y_t) |P(x_t), P(x_{t-\eta})}{P(y_t) |P(x_t)} \right)}. \nonumber
\end{align}
}

Thus, for $\mathbb{A}_{x \to y}$, one subtracts  --- relatively speaking --- a smaller indirect effect $x_t \to x_{t + \eta}$ from the direct effect of $x_t \to y_{t + \eta}$. For $\mathbb{A}_{y \to x}$, one subtracts --- relatively speaking --- a larger indirect effect $y_t \to y_{t + \eta}$ from the direct effect of $y_t \to x_{t + \eta}$. The effect of having $c_{xy} > c_{yx}$ is therefore that $\mathbb{A}_{x \to y} > \mathbb{A}_{y \to x}$.

Next, assume the coupling strengths $c_{x \to y}$ and $c_{y \to x}$ are of similar magnitude. Then, $TE_{x \to y}(\eta) \approx TE_{y \to x}(\eta)$ and  $TE_{x \to y}(-\eta) \approx TE_{y \to x}(-\eta)$, so that $TE_{x \to y}(\eta) - TE_{x \to y}(-\eta)  \approx 0$ and $TE_{y \to x}(\eta) - TE_{y \to x}(-\eta) \approx 0$. Hence, we expect $\mathbb{A}_{x \to y} \approx \mathbb{A}_{y \to x} \approx 0$ and the magnitudes of  $\mathbb{A}_{x \to y}$ and $\mathbb{A}_{x \to x}$ to reflect the coupling strengths $c_{x \to y}$ and $c_{y \to x}$.

\subsubsection{No coupling}If $x$ has no influence on $y$, then we expect no detectable influence neither directly from present $x$ values to future $y$ values, nor (because there is no interaction) from the present of $x$ to its own future through its interaction with $y$. Therefore, $\mathbb{A}_{x \to y} \approx \mathbb{A}_{y \to x} \approx 0 $. Simply put, if there is not coupling, then on average none of the predictions involving the other variable will be improved. 

\clearpage

\section{\label{sec:supp_experimental_setup}Experimental setup}

\subsection{\label{sec:supp_Generalized_embedding}Generalized embedding of time series for TE analysis}

Generically denote the time series for the source process $S$ as $S(t)$, and the time series for the target process $T$ as $T(t)$, and $C_i(t)$ as the time series for any conditional processes $C_i$ that might act in tandem with $S$ to influence $T$. To compute (conditional) TE, we need a Generalized embedding \cite{Sauer1991embedology, Deyle2011generalized} incorporating all of these processes.

For convenience, define the state vectors

{\small \begin{align}
T_f^{(k)} &= \{(T(t+\eta_k), \ldots, T(t+\eta_2), T(t+\eta_1))\}, \label{eq:Tf} \\
T_{pp}^{(l)} &= \{ (T(t), T(t-\tau_1), T(t-\tau_2), \ldots, T(t - \tau_{l - 1})) \}, \label{eq:Tpp} \\
S_{pp}^{(m)} &= \{(S(t), S(t-\tau_1), S(t-\tau_2), \ldots, S(t-\tau_{m - 1}))\},\label{eq:Spp}\\
C_{pp}^{(n)} &= \{ (C_1(t), C_1(t-\tau_1), \ldots,  C_2(t), C_2(t-\tau_1) \},\label{eq:Cpp}
\end{align}}

where the state vectors $T_f^{(k)}$ contain $k$ future values of the target variable, $T_{pp}^{(l)}$ contain $l$ present and past values of the target variable, $S_{pp}^{(m)}$ contain $m$ present and past values of the source variable, $C_{pp}^{(n)}$ contain a total of $n$ present and past values of any conditional variable(s). Here, $\tau$ indicates the embedding lag. In real systems, the strategy for choosing $\tau$ depends on the temporal resolution and the auto-correlation function of the time series data. $\eta$ indicates the prediction lag (the lag of the influence the source has on the target). Combining all variables, we have the Generalized embedding 

{
\begin{align}
\mathbb{E} = (T_f^{(k)}, T_{pp}^{(l)}, S_{pp}^{(m)}, C_{pp}^{(n)}), \label{eq:Generalized_embedding}
\end{align}
}

with a total embedding dimension of $k + l + m + n$. Here, only $T_f$ depends on the prediction lag $\eta$, which is to be determined by the analyst; we use multiple negative and positives $\eta$s for computing $\mathbb{A}$.  The remaining variables depend on $\tau$, which 
may be determined from, for example, the minima of the auto-correlation or lagged mutual information function of the time series. For the synthetic examples in this paper, we push the lower limits of time series lengths, so we use $\tau = 1$ to not exclude too many data points. Another reason for choosing $\tau = 1$ is that theoretically it should worsen the performance of the TE method, because it leads to strongly auto-correlated reconstructed states when there is auto-correlation in the time series \cite{Kantz2004nonlinear}. As we shall see, however, this deliberate choice does not diminish the ability of the asymmetry criterion to distinguish directional dynamical influence.

\subsection{\label{sec:transferentropy_def}Transfer entropy}
TE (in bits) from a source variable $S$ to a target variable $T$ with conditioning on variable(s) $C$ is defined as 

{\small
\begin{align}
&TE_{S \rightarrow T|C} = \nonumber \\
&\int_{\mathbb{E}} P(T_f, T_{pp}, S_{pp}, C_{pp}) \log_{2}{\left( \frac{P(T_f | T_{pp}, S_{pp}, C_{pp})}{P(T_f | T_{pp}, C_{pp})}\right)}
\label{eq:te_conditioned}
\end{align}
}

Without conditioning, eq. \ref{eq:te_conditioned} becomes 
{\small
\begin{align}
TE_{S \rightarrow T} = \int_{\mathbb{E}} P(T_f, T_{pp}, S_{pp}) \log_{2}{\left(\frac{P(T_f | T_{pp}, S_{pp})}{P(T_f | T_{pp})}\right)}
\label{eq:te_simple}
\end{align}
}

\subsection{\label{methods:estimating_te}Numerically estimating TE and $\mathbb{A}$}
We have used three different TE estimators to estimate $\mathbb{A}$. The visitation frequency estimator ($TE_{VF}$) \cite{Schreiber2000} computes TE by partitioning the reconstructed state space using a regular binning. The invariant probability over the partition is then estimated by computing how often the orbit  visits each box. From this invariant joint density, we obtain  marginal densities, and TE can be computed using equation \ref{eq:te_simple}. The transfer operator grid estimator ($TE_{TO}$) is also based on a partition of the reconstructed state space using a regular grid. However, the joint probability distribution is computed as the invariant distribution of an  approximation to the transfer operator associated with the system \cite{Diego2018}. 
Lastly, the nearest neighbour based estimator ($TE_{NN}$) uses mutual information (MI) \cite{Cover2012} to compute TE through the identity $TE(S\to T) = MI(T_f,(S_{pp},T_{pp})) - MI(T_f,S_{pp})$. The MI is estimated using the Kraskov estimator \cite{Kraskov2004}. All estimators are available in the CausalityTools.jl Julia software package (\url{https://github.com/kahaaga/CausalityTools.jl}). Finding roughly equivalent results for all three estimators, results presented in the text are generated using the visitation frequency estimator $TE_{VF}$.

For our analyzes, we implement the embedding approach appearing in \cite{Krakovska2018}, in which increasing the dimension of the embedded system implies conditioning on longer sequences of the past of the target variable. Thus, we use time delay embeddings of the same structure i.e., $\mathbb{E}=\left\{(T_f,S_{pp}, T_{pp})\right\}$, where $T_f = (T(t+\eta))$, $S_{pp} = (S(t))$ and $T_{pp} = (T(t), T(t-\tau), \ldots)$.

The TE for the binning-based estimators, $TE_{VF}$ and $TE_{TO}$, is obtained as follows. We consider two different partitions constructed by subdividing each coordinate axis into an integer number of equal-length interval, using two separate partitions. The number of intervals for these partitions are selected according to the following heuristic $n_{b_{min}} = N^{\frac{1}{k+l+m+1}}$ and $ n_{b_{max}} = n_{b_{min}}+1$, roughly following \cite{Krakovska2018}, where $N$ is the number of points in the time series and $k+l+m$ is the embedding dimension. The corresponding absolute bin sizes, $b_{min}$ and $b_{max}$, are then computed for each and the TE is obtained as an average over those two partitions.

For the $TE_{NN}$ estimator, we use the Chebyshev distance metric, and use a different numbers of nearest neighbours for the estimation of each MI term. For all examples, we let $k_1 = 2$ be the number of nearest neighbours used for the highest dimensional MI estimate ($MI(T_f,(S_{pp},T_{pp}))$), and $k_2 = 3$ be the number of nearest neighbours for the lowest dimensional MI estimate.

Because $\mathbb{A}$ is computed as a difference between sums of TE values at symmetric prediction lags, which should not be sensitive to absolute TE values, we do not correct for estimator-intrinsic differences in absolute TE values \cite{Bossomaier2016}. 

\clearpage

\section{\label{sec:supp_test_systems}Test systems with known ground truths}

\subsection{\label{sec:supp_test_systems_logistic_bidir}Bidirectionally coupled logistic maps}

For the main text, we use logistic model for the chaotic population dynamics of two interacting species given by

{
\large
\begin{subequations}
\begin{align}
x(t+1) &= r_1 f_{yx}^{t}(1 - f_{yx}^{t}) \label{eq:coupled_logistic_bidir_x}\\
y(t+1) &= r_2 f_{xy}^{t}(1 - f_{xy}^{t})\label{eq:coupled_logistic_bidir_y} \\
f_{xy}^t &= \dfrac{y(t) + c_{xy}(x(t) + \sigma_{xy} \xi_{xy}^t )}{1 + c_{xy} (1 + \sigma_{xy} )} \label{eq:coupled_logistic_bidir_fxy} \\ 
f_{yx}^t &= \dfrac{x(t) + c_{yx}(y(t) + \sigma_{yx} \xi_{yx}^t )}{1 + c_{yx} (1 + \sigma_{yx} )},
\label{eq:coupled_logistic_bidir_fyx}
\end{align}
\label{eq:logistic_bidir}
\end{subequations}
}

where the coupling strength $c_{xy}$ controls how strongly species $x$ influences species $y$, and vice versa for $c_{yx}$. To simulate time-varying influence of unobserved processes, we use the dynamical noise terms $\xi_{xy}^t \thicksim U(0, 1)$ and $\xi_{yx}^t  \thicksim U(0, 1)$ drawn independently at each time step. If $\sigma_{xy} > 0$, then the influence of $x$ on $y$ is masked by dynamical noise equivalent to $\sigma_{xy} \xi_{xy}^{t}$ at the $t$-th iteration of the map, and vice versa for $\sigma_{yx}$.

\subsection{\label{sec:supp_test_systems_noninteracting}Non-interacting variables forced by common driver}

In this system, two non-interacting variables $x_1$ and $x_2$ are affected by a common external driver $x_3$. All variables have nonlinear deterministic internal dynamics, overprinted by cyclic and stochastic variability, simulating typical paleoclimate time series from the most recent Quaternary period of Earth's history (Fig. \ref{fig:real_data_quaternary}). The coupling between the noninteracting variables and the external forcing is highly nonlinear. 

The model time step is defined as 1 kiloyears, simulating typical Quaternary paleoclimate time series. We randomly draw the periods $\omega_i \thicksim U(20, 100)$, which yields the typical orbital-type dominant frequencies that are pervasive in paleoclimate time series. The signals are also phase-shifted by randomly assigning values to $\phi_i$ (Fig. \ref{fig:real_data_quaternary}). Initial conditions are drawn from a uniform distribution over the unit interval.

{\small
\begin{subequations}
\begin{align}
    x_1(t) &= \alpha_{1} x_1(t-\gamma_{x_1})\left(1 - x_1(t-\gamma_{x_1})^2\right)e^{-x_1(t-\gamma_{x_1})^2} + A_1 \cos{\left( \dfrac{2\pi}{\omega_1}t + \phi_1 \right)} + c_{31} \left(x_3(t - \nu_{31})^2 + \dfrac{\beta_1 x_3(t -\nu_{31})}{1 + e^{-x(t-\nu_{31})}} \right) + \sigma_1 \xi_1(t)\\
    x_2(t) &= \alpha_{i} x_2(t-\gamma_{x_2})\left(1 - x_2(t-\gamma_{x_2})^2\right)e^{-x_2(t-\gamma_{x_2})^2} + A_2 \cos{\left( \dfrac{2\pi}{\omega_2}t + \phi_2 \right)} +c_{32} \left(x_3(t - \nu_{32})^2 + \dfrac{\beta_2 x_3(t -\nu_{32})}{1 + e^{-x(t-\nu_{32})}} \right) + \sigma_2 \xi_2(t)\\
    x_3(t) &= \alpha_{i} x_3(t-\gamma_{x_3})\left(1 - x_3(t-\gamma_{x_3})^2\right)e^{-x_3(t-\gamma_{x_3})^2} + A_3 \cos{\left( \dfrac{2\pi}{\omega_3}t + \phi_3 \right)} + \sigma_3 \xi_3(t)
\end{align}
\label{eq:sys_noninter_ext}
\end{subequations}
}

We generate time series ensembles by drawing parameters randomly from uniform distributions as follows: $\alpha_i \thicksim U(2.5, 4.0)$, $\beta_i \thicksim U(0.2, 0.8)$, and $A_i \thicksim U(0.75, 1.25)$. $\xi_i$ are random uniformly distributed processes drawn independently from $U(0, 1)$ at each time step, where $\sigma_i \thicksim U(0.03, 0.3)$ control the magnitude of the dynamical noise. Additionally, observational noise equivalent to 0.5 the standard deviation of each time series is  added to that time series before analyses. Interaction lags $\nu_{31}$ and $\nu_{32}$ are set to 1, and internal lags $\gamma_{x_i}$ are drawn randomly from the set $\{1, 2, 3, 4\}$.

\subsection{\label{sec:supp_test_systems_var}Vector autoregressive (VAR) processes}

Consider a $p$-dimensional random sequence $\{\vec{x}_t: \, x\in \mathbb{R}^p, t \in \mathbb{Z} \}$. A $p$-dimensional $VAR(k)$ process is given by

\begin{eqnarray}
\vec{x}_t = A_1 \vec{x}_{t-1} + \cdots + A_k \vec{x}_{t-k} + \vec{\epsilon}_t + \vec{d}_t
\end{eqnarray}

\noindent where $\vec{\epsilon}_t$ is a noise sequence drawn from a zero-mean normal distribution with standard deviation $\sigma$ and $\vec{d}_t$ is a deterministic sequence. The coefficient matrices $A_i$ have dimensions $p$-by-$p$, where the entry $c_{kl}^{m}$ is the coefficient controlling the influence of the $k$-th variable on the $l$-th variable at time lag $m$. Interaction strengths between variables are thus governed by the off-diagonal terms of these coefficient matrices. 

For example, consider the following $3$-dimensional AR(2) system with no deterministic sequence:

\begin{equation}
\vec{x}_t = A_1 \vec{x}_{t-1} + A_2 \vec{x}_{t-2} + \vec{\epsilon}_t: \vec{x}_t, \vec{\epsilon}_t \in \mathbb{R}^3
\label{eqn_supp:VARK}
\end{equation}

In matrix notation we have

{\small
\begin{eqnarray}
\vec{x}_t = 
\begin{pmatrix}
x_1(t) \\
x_2(t) \\
x_3(t)
\end{pmatrix} =
\begin{pmatrix}
c_{11}^1 & c_{21}^1 & c_{31}^1 \\
c_{12}^1 & c_{22}^1 & c_{32}^1 \\
c_{13}^1 & c_{23}^1 & c_{33}^1
\end{pmatrix}
\begin{pmatrix}
x_1(t-1) \\
x_2(t-1) \\
x_3(t-1) 
\end{pmatrix} + \nonumber \\
\begin{pmatrix}
c_{11}^2 & c_{21}^2 & c_{31}^2 \\
c_{12}^2 & c_{22}^2 & c_{32}^2 \\
c_{13}^2 & c_{23}^2 & c_{33}^2
\end{pmatrix}
\begin{pmatrix}
x_1(t-2) \\
x_2(t-2) \\
x_3(t-2) \\
\end{pmatrix} + 
\begin{pmatrix}
\epsilon_1(t) \\
\epsilon_2(t) \\
\epsilon_3(t) 
\end{pmatrix}
\end{eqnarray}
}

Stability of the VAR process is ensured if the roots $r_1, r_2, \ldots, r_{np} \in \mathbb{C}$ of the $np$-by-$np$-dimensional companion matrix $A_c$, as defined below, lie inside the unit circle.

\begin{eqnarray}
A_c = \begin{pmatrix}
A_1    & A_2    & \cdots & A_{k-1} & A_k \\
I      & 0      & \cdots & 0       & 0 \\
0      & I      & \cdots & 0       & 0 \\
\vdots & \vdots & \ddots & \vdots  & \vdots \\
0      & 0      & \cdots & I       & 0
\end{pmatrix}
\end{eqnarray}

\noindent Hence, to generate stationary time series from a $VAR(k)$-process with $p$ variables, we assign coefficients such that $|r_i| < 1$ for all $ i \in \{1, \ldots, np\}$.

\subsection{\label{sec:supp_test_systems_noise_normal}Normally distributed noise processes}

To establish a baseline for the significance threshold for the normalized predictive asymmetry test (eq. \ref{eq:normalized_asymmetry_criterion}), we will consider various uncoupled noise processes. 

\begin{eqnarray}
x_t \thicksim N(0, \sigma_x) \\
y_t \thicksim N(0, \sigma_y)
\label{eq:sys_noise_normal}
\end{eqnarray}

\noindent where $N(0, \sigma_x)$ and $N(0, \sigma_y)$ are independent normal distributions with zero mean and standard deviations $\sigma_x$ and $\sigma_y$, and values are drawn independently at each time step.

\subsection{\label{sec:supp_test_systems_noise_uniform}Uniformly distributed noise processes}

Next, we will consider uniformly distributed noise processes

\begin{eqnarray}
x_t \thicksim U(0, 1) \\
y_t \thicksim U(0, 1)
\label{eq:sys_noise_uniform}
\end{eqnarray}

\noindent where $U(0, 1)$ and $U(0, 1)$ are independent uniform distributions with support $[0, 1]$, and values are drawn independently at each time step.

\subsection{\label{sec:supp_test_systems_noise_brownian_uniform}Brownian noise based on uncorrelated uniform noise}

Many observed time series are trended. We simulate this phenomenon by using brownian noise processes

\begin{eqnarray}
x_t = \sum_{i = 1}^{t} x_i, \quad x_i \thicksim U(0, 1) \\
y_t = \sum_{i = 1}^{t} y_i, \quad y_i \thicksim U(0, 1) 
\label{eq:sys_noise_brownian_uniform_based}
\end{eqnarray}

\noindent where $U(0, 1)$ and $U(0, 1)$ are independent uniform distributions with support $[0, 1]$, and values are drawn independently at each time step.

\subsection{\label{sec:supp_test_systems_nontrivially_coupled}Autoregressive systems with periodicity and strongly nonlinear couplings}

This system of unidirectionally chained autoregressive variables was extended from a simpler version in \cite{Peguin-Feissolle1999} and \cite{Chavez2003}, introducing a periodic component and variable parameters, variable internal lags and variable interaction lags. 

Interaction lags are kept constant with $\tau_i \neq \nu_i$ for each instance of the system.  Internal lags $\gamma_i$, as well as $\tau_i$ and $\nu_i$ are selected randomly from the set $\{1, ..., 5\}$ with uniform probability. The $\xi_{i}(t)$ are independent normally distributed dynamical noise processes with zero mean and standard deviations of $\sigma_i$. $\omega_i$ and $\phi_i$ control the period and phase of the periodic component of the $i$-th variable, while $s_i$ scales the magnitudes of the periodic component. The $s_i$ regulate the magnitude of the periodic components of $x_i$ at each time step. The coupling strength between nodes $x_{i-1}$ and $x_i$ in the chain is controlled by the parameter $c_i$. The logistic function responsible for the coupling between adjacent variables $x_{i-1}$ and $x_i$ is parameterized to simulate a wide range of couplings. 

Observational noise equivalent to 20\% of the standard deviation of the respective variable is added to each time series. Parameters are drawn from uniform distributions as specified in the figure texts.

\begin{subequations}
\begin{align}
x_1 &= \alpha_1 + \beta_1 x_1(t - \gamma_1) + \sigma_1 \xi_1(t) + s_1 \cos{\left( \dfrac{2\pi}{\omega_i}t + \phi_i \right)} \label{eq:system_chain_autoregressiveperiodic_stronglynonlinearcoupling_1st}
\\
x_i &= \alpha_i + \beta_i x_i(t- \gamma_i) + \sigma_i \xi_i(t) + s_i \cos{\left( \dfrac{2\pi}{\omega_i}t + \phi_i \right)} + c_i\left(\dfrac{\chi_i - \rho_i x_{i-1}(t - \tau_i)}{1 + e^{-q_ix_{i-1}(t - \nu_i)}}\right)
\label{eq:system_chain_autoregressiveperiodic_stronglynonlinearcoupling_ith}
\end{align}
\label{eq:system_chain_autoregressiveperiodic_stronglynonlinearcoupling}
\end{subequations}

\subsection{Nonlinear systems with linear coupling over multiple forcing lags}

This nonlinear system with linear coupling is modified from  \cite{Chen2004extendedgranger}, but expanding the system to a chain of unidirectionally coupled variables with variable internal lags and interaction lags.

\begin{subequations}
\begin{align}
x_1(t) &= \alpha_{1} x_1(t-\gamma_{1})\left(1 - x_1(t-\gamma_1)^2\right)e^{-x_1(t-\gamma_1)^2} + \beta_{1} x_1(t - \tau_{1}) + \sigma_1 \epsilon_1(t) \\
x_i(t) &= \alpha_{i} x_i(t-\gamma_{i})\left(1 - x_i(t-\gamma_i)^2\right)e^{-x_i(t-\gamma_i)^2} + \beta_{i} x_i(t - \tau_{i}) + c_i x_{i-1}(t - \nu_i) + \sigma_i \epsilon_i(t)
\end{align}
\label{eq:system_chain_nonlinear_chen_linearcoupling}
\end{subequations}

\subsection{Nonlinear systems with nonlinear coupling over multiple forcing lags}

This nonlinear system is also modified from \cite{Chen2004extendedgranger}, but expanding the system to a chain of unidirectionally coupled variables with variable internal lags and interaction lags. Here, the coupling is nonlinear.

\begin{subequations}
\begin{align}
x_1(t) &= \alpha_{1} x_1(t-\gamma_{1})\left(1 - x_1(t-\gamma_1)^2\right)e^{-x_1(t-\gamma_1)^2} + \beta_{1} x_1(t - \tau_{1}) + \sigma_1 \epsilon_1(t) \\
x_i(t) &= \alpha_{i} x_i(t-\gamma_{i})\left(1 - x_i(t-\gamma_i)^2\right)e^{-x_i(t-\gamma_i)^2} + \beta_{i} x_i(t - \tau_{i})  + \sigma_i \epsilon_i(t) + c_i x_{i-1}(t - \nu_i)^2
\end{align}
\label{eq:system_chain_nonlinear_chen_nonlinearcoupling}
\end{subequations}

\subsection{Nonlinear systems with periodic component, linear coupling}

This system is modified from \cite{Chen2004extendedgranger}, but expanding the system to a chain of unidirectionally coupled variables with variable internal lags and interaction lags. Cyclic components have also been introducing to the signals, where $\omega_i$ and $\phi_i$ controls the period and phase of the periodic component of the $i$-th variable. The interaction between adjacent nodes in the chain is linear, and the coupling strength is controlled by the parameter $c_i$.

\begin{subequations}
\begin{align}
x_1(t) &= \alpha_{1} x_1(t-\gamma_{1})\left(1 - x_1(t-\gamma_1)^2\right)e^{-x_1(t-\gamma_1)^2} + \beta_{1} x_1(t - \tau_{1}) + \cos{\left( \dfrac{2\pi}{\omega_1}t + \phi_1 \right)} + \sigma_1 \epsilon_1(t) \\
x_i(t) &= \alpha_{i} x_i(t-\gamma_{i})\left(1 - x_i(t-\gamma_i)^2\right)e^{-x_i(t-\gamma_i)^2} + \beta_{i} x_i(t - \tau_{i}) + c_i x_{i-1}(t - \nu_i) + \cos{\left( \dfrac{2\pi}{\omega_i}t + \phi_i \right)} + \sigma_i \epsilon_i(t)
\end{align}
\label{eq:system_chain_chen_nonlinearperiodic_linearcoupling}
\end{subequations}

\subsection{\label{sec:supp_test_systems_logistic_chain}Unidirectional chain of logistic maps with variable internal lags and forcing lags}

The following system of difference equations describes a $K$-dimensional system of logistic maps. Its interaction network is characterised by unidirectional coupling between adjacent nodes. The first map is independent, while for the remaining $K-1$ maps, the map $k$ is affected by itself and the $(k-1)$-th map. 

$$
\large
\begin{subequations}
\begin{aligned}
x^{(1)}_{t} &= r_1 f_{1}(1 - f_{1}) \label{eq:system_chain_logistic_chain_variablelags_a}\\
x^{(k)}_{t} &= r_k f_{k-1}^{k} \left(1 - f_{k-1}^{k} \right) \label{eq:system_chain_logistic_chain_variablelags_b} \\
f_{1} &= x^{(1)}_{t - \gamma_1}  \label{eq:system_chain_logistic_chain_variablelags_c}\\
f_{k-1}^{k} &= \dfrac{x^{(k)}_{t - \gamma_k} + c_{k-1}^{k} \left( x^{(k-1)}_{t - \tau_{k-1}^{k}} + \sigma_{k-1}^{k} \xi^{k} \right)}{1 +  c_{k-1}^{k} \left(1 + \sigma_{k-1}^{k} \right)}  \label{eq:system_chain_logistic_chain_variablelags_d}
\end{aligned}
\label{eq:system_chain_logistic_chain_variablelags}
\end{subequations}
$$

Here, $x^{(k)}_{t}$ is the value of the $k$-th variable at time $t$ and $x^{(k-1)}_{t}$ is the value of the $(k-1)$-th variable at time $t$. The strength of the unidirectional forcing from $x^{(k-1)}$ to $x^{(k)}$ is controlled by $c_{k-1}^{k}$. However, the influence from $x^{(k-1)}$ to $x^{(k)}$ is also masked by dynamical noise. The average magnitude of this noise is given by $\sigma_{k-1}^{k} \in [0, 1]$ (given as a percentage of the allowed range of values, so a relatively low value should be chosen to not completely obscure the signals). The noise, $\xi^{k}$, is dynamical noise drawn independently from a uniform distribution $U(0, 1)$ independently at every time step (masking the influence of $x^{(k-1)}$ on $x^{(k)}$), and then scaled by $\sigma_{k-1}^{k}$.

For the lags, $\tau_{k-1}^{k} \in \{1, 2, ..., K_\tau\}$ is the time lag of the influence from $x^{(k-1)}$ to $x^{(k)}$, and $\gamma_k \in  \{1, 2, ..., K_\gamma \}$ is the time lag of the influence from $x^{(k)}$ on itself, chosen randomly from the allowed lags for each variable $k$. 

For every realization of the system, initial conditions and parameters $r_j$ are randomized over uniform distributions $U(0.0, 1.0)$ and $U(3.86, 3.9)$, respectively. Observational noise equivalent to 20\% of the standard deviation of the respective variable is added to each time series. The dynamical noise is set to $\sigma = 0.05$ for all interactions.

\subsection{\label{sec:supp_test_systems_henon_chain_unidir}Unidirectional chain of Henon maps}

Consider a $K$-dimensional system consisting of $K$ unidirectionally coupled Henon maps given by

{\begin{equation}
X_i(t) = 
\begin{cases}
    a - X_i(t-1)^2 + b X_i(t-2), & \text{for } i = 1 \\
    a - 0.5C \left[X_{i-1}(t-1) + X_i(t-1)\right] + (1-C)X_i(t-1)^2 + b X_i(t-2), & \text{for } i > 1
  \end{cases}, 
  \label{eq:system_chain_henonchain}
\end{equation}
}

where $X_i$ is the $i$-th map, $a = 1.4$ and $b = 0.3$ and $C$ is the coupling strength from variable $i$ to $i+1$. For every realization, initial conditions are drawn from uniform distributions over $[0.0, 1.0]$. Observational noise equivalent to 20\% of the standard deviation of the respective variable is added to each time series.

\subsection{\label{sec:supp_test_systems_rossler_lorenz_unidir}Rössler-Lorenz system}

Here, we use two coupled Rössler and Lorenz systems, where the Rössler subsystem unidirectionally drives the Lorenz subsystem.

\begin{subequations}

\begin{align}
\dot x_1 &= a_1(x_2 + x_3) \\
\dot x_2 &= a_2(x_1 + 0.2x_2) \\
\dot x_3 &= a_2(0.2 + x_3(x_1 - a_3)) \\
\dot y_1 &= b_1(y_2 - y_1) \\
\dot y_2 &= y_1(b_2 - y_3) - y_2 +c_{xy}(x_2)^2 \\
\dot y_3 &= y_1 y_2 - b_3y_3
\end{align}
\label{eq:system_rosslerlorenz}
\end{subequations}

with the coupling constant $c_{xy} \geq 0$.

\subsection{\label{sec:supp_test_systems_bidirectional_nonlinear_periodic}Bidirectional nonlinear system}

This system is also modified from \cite{Chen2004extendedgranger}, but noise and cyclic components have also been introducing to the signals, where $\omega_i$ and $\phi_i$ controls the period and phase of the periodic component of the $i$-th variable. The interaction between adjacent nodes in the chain is linear, and the coupling strength is controlled by the parameter $c_i$.

{
\small
\begin{subequations}
\begin{align}
x_1(t) &= \alpha_{1} x_1(t-\gamma_{x_1})\left(1 - x_1(t-\gamma_{x_1})^2\right)e^{-x_1(t-\gamma_{x_1})^2} + \beta_{1} x_1(t - \tau_{x_1}) + c_{21} \sin(x_2(t - \nu_1))+ \sigma_1 \epsilon_1(t) + \cos{\left( \dfrac{2\pi}{\omega_1}t + \phi_1 \right)}x_1(t-\tau_1)\\
x_2(t) &= \alpha_{2} x_i(t-\gamma_{x_2})\left(1 - x_2(t-\gamma_{x_2})^2\right)e^{-x_2(t-\gamma_{x_2})^2} + \beta_{2} x_i(t - \tau_{x_2}) + c_{12} \sin(x_1(t - \nu_2)) + \sigma_2 \epsilon_2(t) + \cos{\left( \dfrac{2\pi}{\omega_2}t + \phi_2 \right)}x_2(t-\tau_2)
\end{align}
\label{eq:system_bidirectional_nonlinear_periodic}
\end{subequations}
}

We generate time series ensembles by drawing parameters randomly from uniform distributions as follows: $\alpha_i \thicksim U(3.0, 3.6)$,$\beta_i \thicksim U(0.2, 0.8)$, $\omega_i \thicksim U(5, 20)$, and $\phi_i \thicksim U(0, 2\pi)$. $\xi_i$ are random uniformly distributed processes drawn independently from $U(0, 1)$ at each time step, where $\sigma_i = 0.5$ control the magnitude of the dynamical noise. Additionally, observational noise equivalent to 0.5 the standard deviation of each time series is  added to that time series before analyses. Interaction lags $\nu_{31}$ and $\nu_{32}$ are set to 1, while internal lags $\gamma_{x_i}$ are drawn randomly from the set $\{1, 2\}$ and internal lags $\tau_{x_i}$ are set to 1.  

\subsection{Nonlinear system without dynamical noise and periodicity}

This system is also modified from \cite{Chen2004extendedgranger}, but contains no dynamical noise or periodicity.

\begin{subequations}
\begin{align}
    x(t+1) & = a_1 x(t-\tau_{x_1}) \lb 1-x(t-\tau_{x_1})^2\rb e^{-x(t-\tau_{x_1})^2} + a_2 x(t-\tau_{x_2})\,,\\
    y(t+1) & = b_1 y(t-\tau_{y_1}) \lb 1-y(t-\tau_{y_1})^2\rb e^{-y(t-\tau_{y_2})^2} + b_2 y(t-\tau_{y_2})+c_{xy} x(t-\tau_{c_{xy}})^2\,
\end{align}
\label{eq:nonlinear_withoutdynamicalnoise_withoutperiodicity}
\end{subequations}

\clearpage

\section{\label{sec_supp:significance_test}System-specific and estimator-specific significance test}

\subsection{\label{sec_supp_significance_test_demonstration_of_criterion}Statistical robustness when no coupling is present}

\begin{figure*}[h]
\centering
\includegraphics[width=1.0\linewidth]{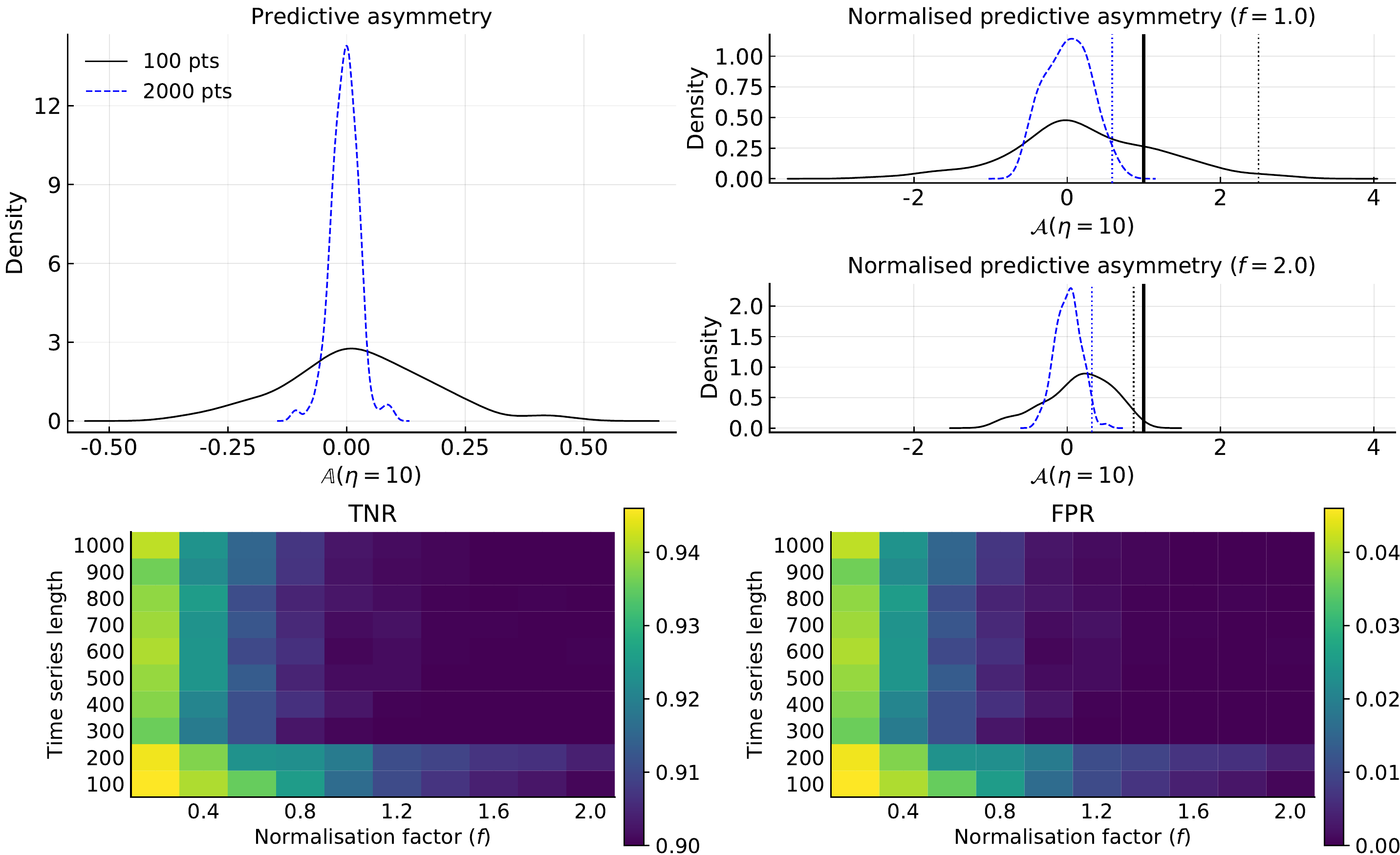}
\caption{Predictive asymmetry $\mathbb{A}$ (eq. \ref{eq:asymmetry}), and normalized predictive asymmetry $\mathcal{A}^f$ (eq. \ref{eq:normalized_asymmetry_criterion}) for varying normalization factor $f$ for a model of two uncoupled uniform-noise processes $x_t$ and $y_t$ (eq. \ref{eq:sys_noise_uniform}). Vertical, dotted lines in the upper panels indicate the 99th percentiles for $\mathcal{A}^f$. Heatmaps show the statistical robustness of $\mathcal{A}^f$ (expressed by TNR and FPR) to time series length and varying $f$. Density plots and values in each heatmap cell are computed over 300 independent pairs of time series, using a fixed maximum prediction lag $\eta = 10$.
Generalized embeddings were constructed with $k = 1$, $m = 1$ and varying  $l$. The latter, along with its reconstruction delay, were optimised optimised using the false first nearest neighbors method \cite{Krakovska2015f1nn}, with the optimal delay estimated using the first zero-crossing of the auto-correlation function of the target time series. 
}
\label{fig_supp:significance_test_uniform}
\end{figure*}

\begin{figure*}[h]
\centering
\includegraphics[width=1.0\linewidth]{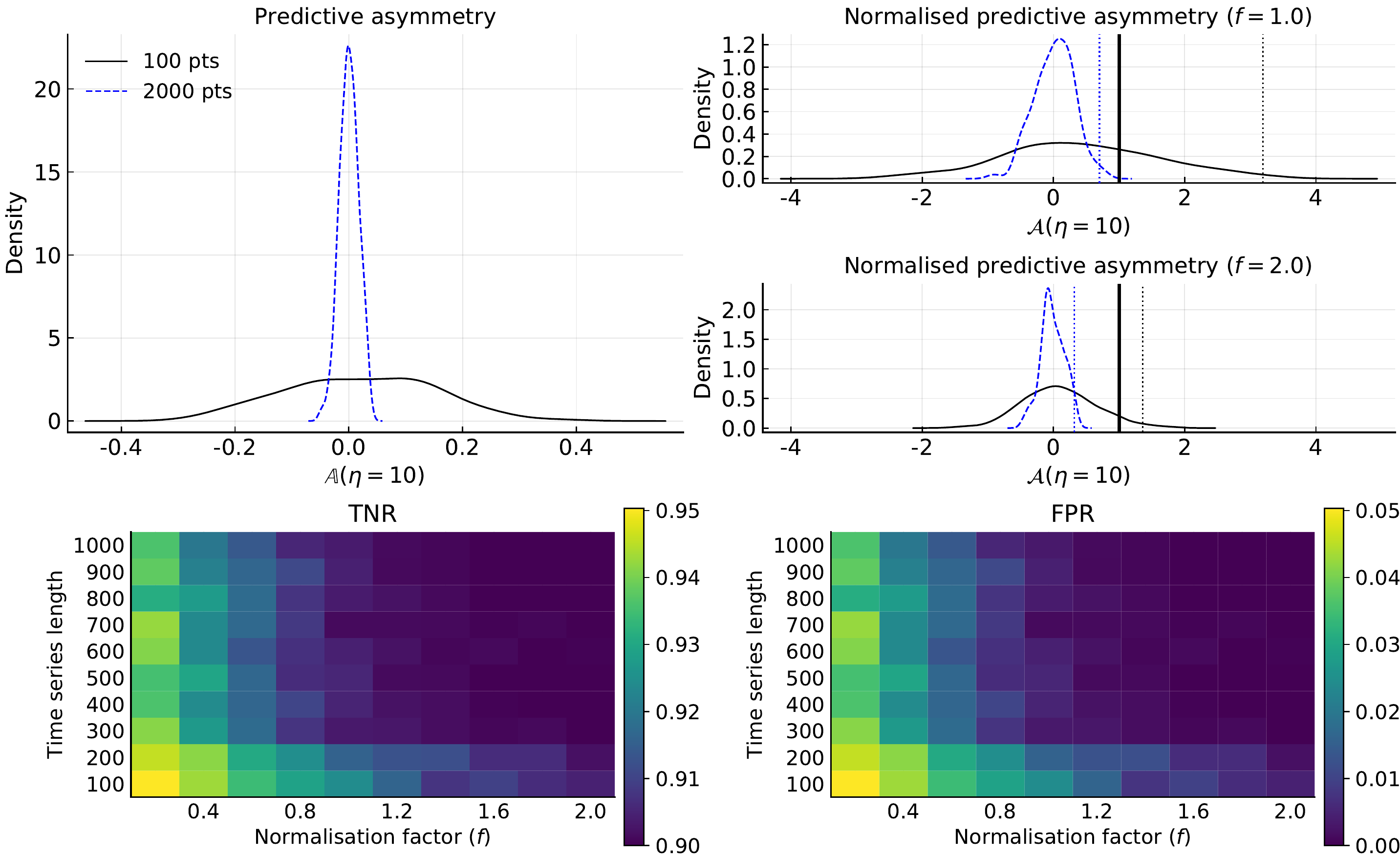}
\caption{Predictive asymmetry $\mathbb{A}$ (eq. \ref{eq:asymmetry}), and normalized predictive asymmetry $\mathcal{A}^f$ (eq. \ref{eq:normalized_asymmetry_criterion}) for varying normalization factor $f$ for a model of two uncoupled noise (normally distributed) processes $x_t$ and $y_t$ (eq. \ref{eq:sys_noise_normal}). Vertical, dotted lines in the upper panels indicate the 99th percentiles for $\mathcal{A}^f$, were computed for fixed maximum prediction lag $\eta = 10$. Heatmaps show the statistical robustness of $\mathcal{A}^f$ (expressed by TNR and FPR) to time series length and varying $f$.  Density plots and values in each heatmap cell are computed over 300 independent pairs of time series, using a fixed maximum prediction lag $\eta = 10$.
Generalized embeddings were constructed with $k = 1$, $m = 1$ and varying  $l$. The latter, along with its reconstruction delay, were optimised optimised using the false first nearest neighbors method \cite{Krakovska2015f1nn}, with the optimal delay estimated using the first zero-crossing of the auto-correlation function of the target time series. 
}
\label{fig_supp:significance_test_normal}
\end{figure*}

\begin{figure*}[h]
\centering
\includegraphics[width=1.0\linewidth]{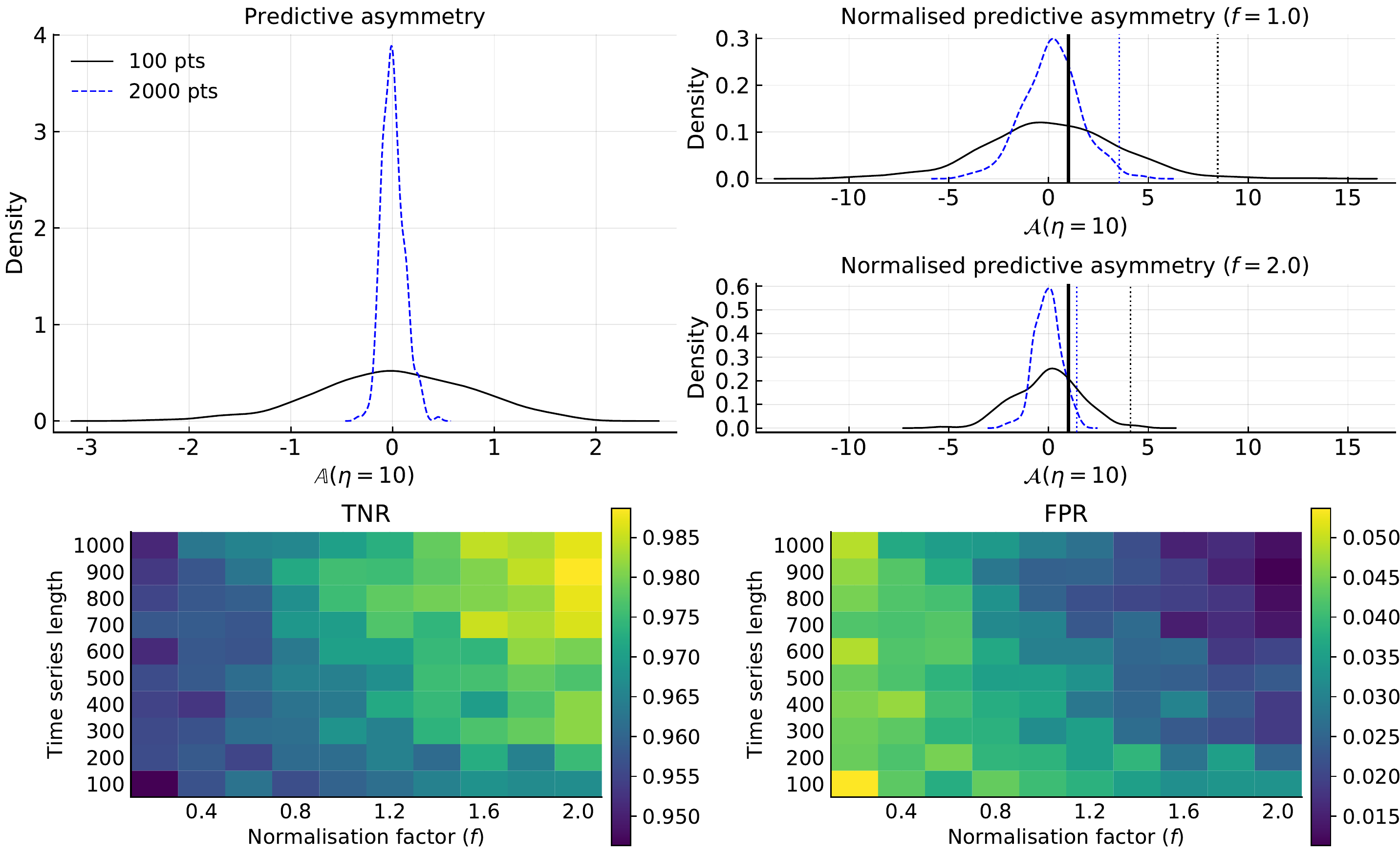}
\caption{Predictive asymmetry $\mathbb{A}$ (eq. \ref{eq:asymmetry}), and normalized predictive asymmetry $\mathcal{A}^f$ (eq. \ref{eq:normalized_asymmetry_criterion}) for varying normalization factor $f$ for two uncoupled brownian noise processes (eq. \ref{eq:sys_noise_brownian_uniform_based}). Vertical, dotted lines in the upper panels indicate the 99th percentiles for $\mathcal{A}^f$. Heatmaps show the statistical robustness of $\mathcal{A}^f$ (expressed by TNR and FPR) to time series length and varying $f$. Density plots and values in each heatmap cell are computed over 300 independent pairs of time series, using a fixed maximum prediction lag $\eta = 10$.
Generalized embeddings were constructed with $k = 1$, $m = 1$ and $l = 1$. 
}
\label{fig_supp:significance_test_brownian}
\end{figure*}

\begin{figure*}[h]
\centering
\includegraphics[width=1.0\linewidth]{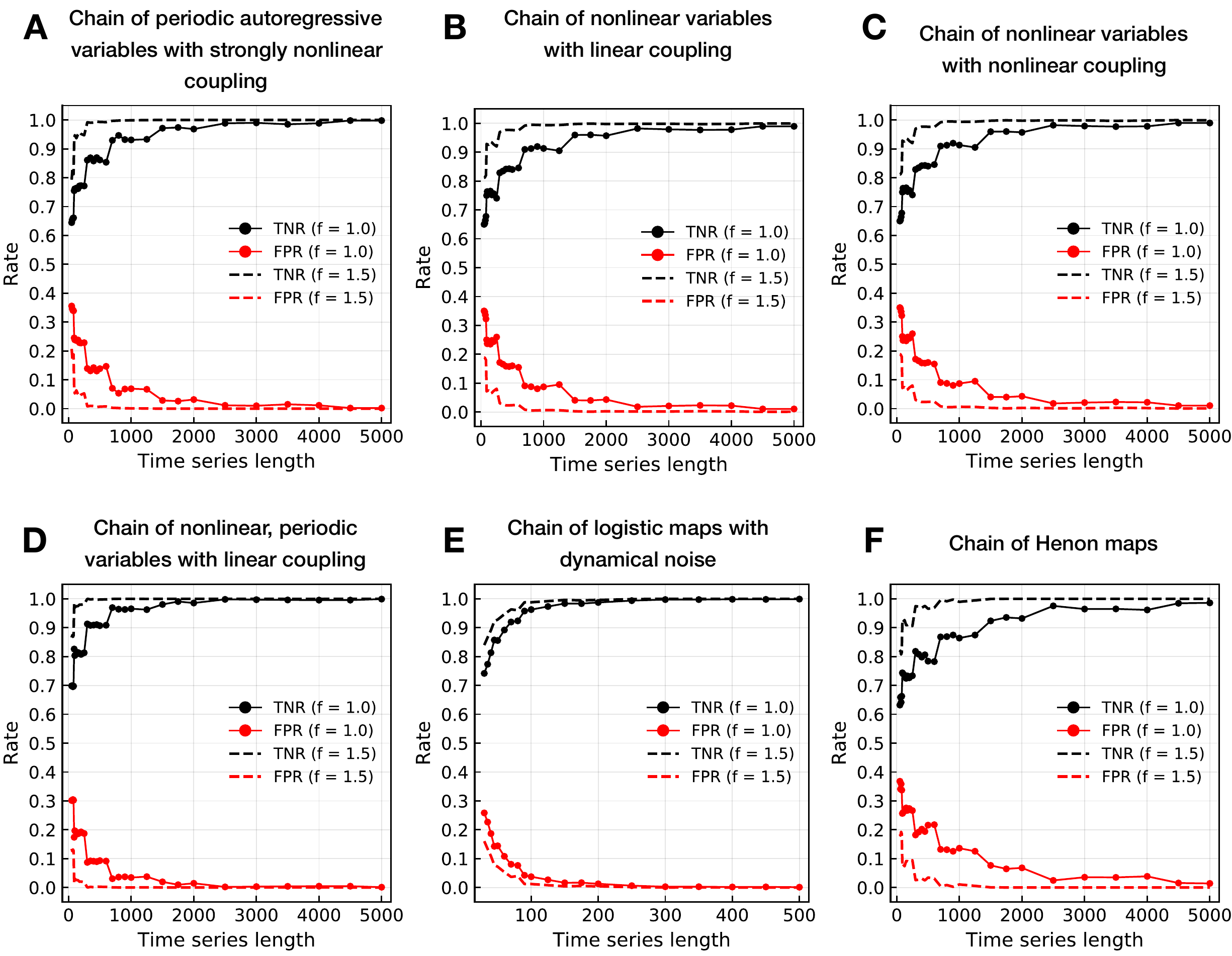}
\caption{Statistical robustness of the normalized predictive asymmetry causality criterion $\mathcal{A}^f$ (eq. \ref{eq:normalized_asymmetry_criterion}) for various systems, varying time series length, and varying $f$. Here, we show the ability of the test to correctly reject interactions when there are none, as expressed by true negative and false positive rates. TNR and FPR rates are computed over 1000 independent pairs of time series for each time series length, using a variable maximum prediction lag $\eta = 10 + \Gamma - 1$, where $\Gamma$ is the maximum internal/interaction delay for that particular system realization. Generalized embeddings were constructed with $k = 1$, $m = 1$ and $l = 1$. A: Periodic autoregressive variables with strongly nonlinear coupling (eq. \ref{eq:system_chain_autoregressiveperiodic_stronglynonlinearcoupling}); B: Nonlinear system with linear coupling (eq. \ref{eq:system_chain_nonlinear_chen_linearcoupling}); C: Nonlinear system with nonlinear coupling (eq. \ref{eq:system_chain_nonlinear_chen_nonlinearcoupling}); D: Nonlinear system with periodic component and linear coupling (eq. \ref{eq:system_chain_chen_nonlinearperiodic_linearcoupling}); E: Logistic map system with dynamical noise, variable interaction lags, variable internal lags, and dynamical noise (eq. \ref{eq:system_chain_logistic_chain_variablelags}); F: Henon map (eq. \ref{eq:system_chain_henonchain}).
}
\label{fig_supp:statistical_robustness_TNR_FPR_compilation}
\end{figure*}

\begin{figure*}[h]
\centering
\includegraphics[width=0.5\linewidth]{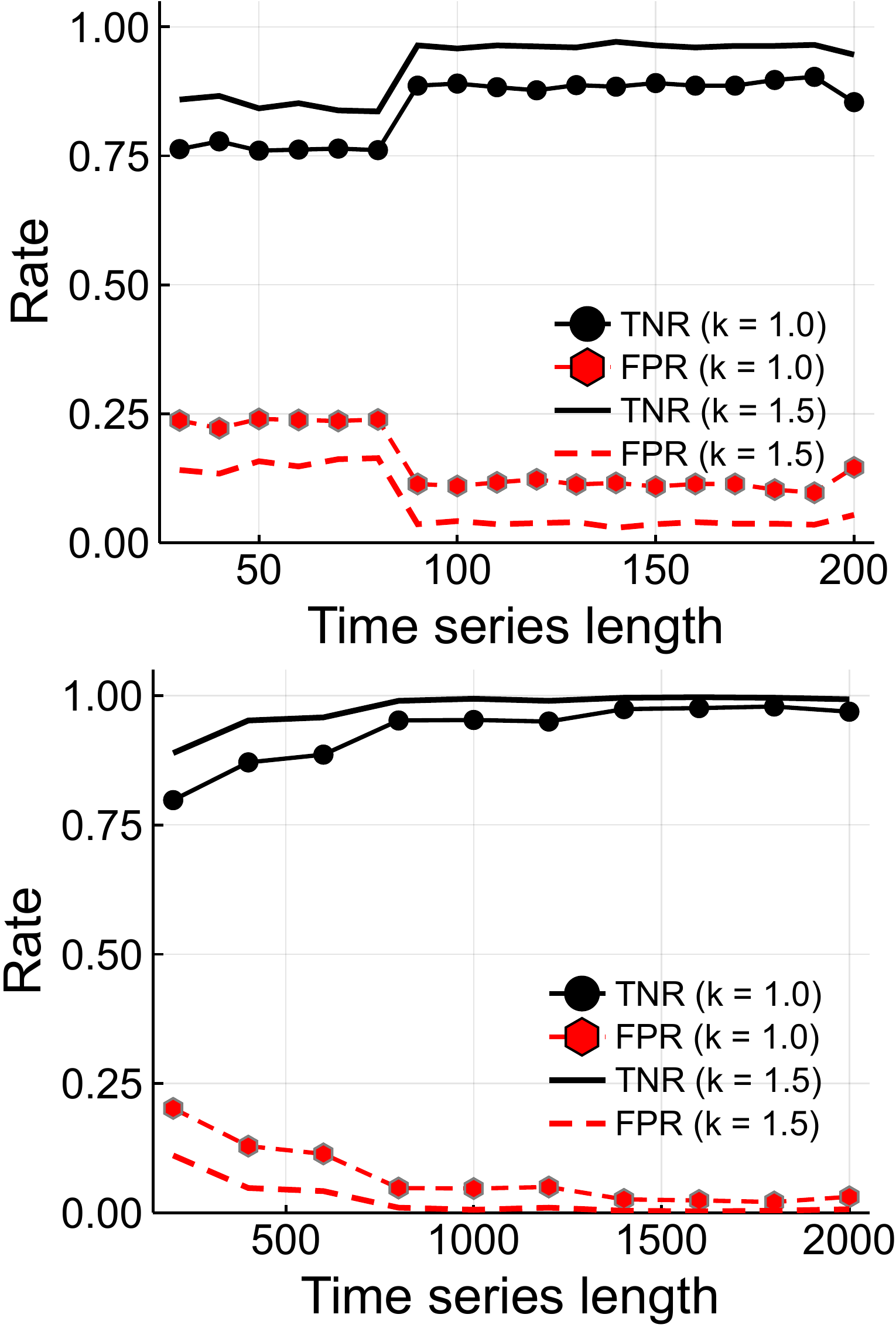}
\caption{TNR and FPR for the test (eq. \ref{eq:normalized_asymmetry_criterion}) for non-coupled autoregressive systems. Lines with and without points show rates when $\mathbb{A}$ is compared to 1.0 and 1.5 times the average TE, respectively (i.e. $f=1.0$ and $f=1.5$ in eq. \ref{eq:normalized_asymmetry_criterion}). 
\\ \\
\textit{Upper panel}: TNR and FPR in the limit of very short times. The improved performance around time series length 90 correspond to when the number of subdivisions along each axis changes due to the partition heuristic. For the particular case of no coupling, we find that better TNR and FPR are achieved by finer partitions, but when coupling exists, sticking to the partition heuristic yields better performance. For each time series length, $\mathbb{A}$ is computed on 500 unique $2$-dimensional $VAR(k)$ systems with no coupling, where the order $k$ is randomly chosen from the set $\{1, 2, \ldots, 5\}$ (eq. \ref{eqn_supp:VARK}) for each system. For each system, the standard deviations for the error terms are drawn from uniform distributions on $[0.95, 1.05]$. Each variable affects itself at exactly one time lag, which is also chosen randomly from $\{1, 2, \ldots, 5\}$. The diagonal terms of the relevant coefficient matrices $A_i$ are independently drawn from a uniform distribution on $[0.1, 0.9]$, while off-diagonal terms of the $A_i$ are set to zero, yielding no coupling.
\\ \\
\textit{Lower panel}: TNR and FPR for longer time series. For each time series length, $\mathbb{A}$ is computed on 1000 unique $2$-dimensional $VAR(k)$ systems with no coupling and maximum order $k$,  randomly chosen from the set $\{1, 2, \ldots, 20\}$ (eq. \ref{eqn_supp:VARK}) for each system. The coefficient matrices are generated as for the short time series. 
}
\label{fig_supp:TNR_FPR_nocoupling_VAR}
\end{figure*}

By relating the $\mathbb{A}$ to some fraction $f$ of the system-specific and estimator-specific empirical TE, $\mathcal{A}^f$ can be used as a criterion for statistical significance: $\mathcal{A}^f > 1$ indicates the presence of directional coupling, while $\mathcal{A}^f <= 1$ rejects coupling. First, we demonstrate the difference between the predictive asymmetry $\mathbb{A}$ (eq. \ref{eq:asymmetry}) and its normalized counterpart $\mathcal{A}^f$ (eq. \ref{eq:normalized_asymmetry_criterion}), and how the value of $f$ affects the ability of the test to reject coupling for uncoupled systems (Figs. \ref{fig_supp:significance_test_uniform}, \ref{fig_supp:significance_test_normal}, \ref{fig_supp:significance_test_brownian}, 
\ref{fig_supp:statistical_robustness_TNR_FPR_compilation}, 
\ref{fig_supp:TNR_FPR_nocoupling_VAR}). 
Because there are no true positives when there is no coupling, only the TNR and FPR are meaningful summary statistics to use in this context. To determine the ability of the test to correctly classify absence of coupling, we hence compute TNR and FPR for multiple realizations of the following systems.

\begin{itemize}
    \item Noise time series with uniformly distributed noise (Fig. \ref{fig_supp:significance_test_uniform})
    \item Noise time series with normally distributed noise (Fig. \ref{fig_supp:significance_test_normal})
    \item Brownian noise time series (Fig. \ref{fig_supp:significance_test_brownian}).
    \item Chain of periodic autoregressive variables with strongly nonlinear coupling (Fig. \ref{fig_supp:statistical_robustness_TNR_FPR_compilation}).
    \item Chain of nonlinear variables (Fig. \ref{fig_supp:statistical_robustness_TNR_FPR_compilation}).
    \item Another chain of nonlinear variables (Fig. \ref{fig_supp:statistical_robustness_TNR_FPR_compilation}).
    \item Chain of nonlinear, periodic variables with linear coupling (Fig. \ref{fig_supp:statistical_robustness_TNR_FPR_compilation}).
    \item Chain of logistic maps with dynamical noise (Fig. \ref{fig_supp:statistical_robustness_TNR_FPR_compilation}).
    \item Chain of Henon maps (Fig. \ref{fig_supp:statistical_robustness_TNR_FPR_compilation}).
    \item Non-coupled autoregressive systems of maximum order $k=5$ for very short time series (Fig. \ref{fig_supp:TNR_FPR_nocoupling_VAR}, upper panel).
    \item Non-coupled autoregressive systems of maximum order $k=20$ for longer time series (Fig. \ref{fig_supp:TNR_FPR_nocoupling_VAR}, lower panel).
\end{itemize}

We find that a normalization factor $f = 1$ (i.e. normalizing to the mean TE) is a good-trade off between statistical robustness (here: the ability to reject coupling when there is none) and sensitivity to time series length. For a given time series length, higher $f$ reduces the number of false positives.  Moreover, for all the tested systems, the ability of the test to reject coupling when there is none approaches perfect for sufficient time series length.

\clearpage

\subsection{\label{sec_supp:significance_test_unidircoupling}Statistical robustness for systems with unidirectional coupling}

For systems with unidirectional coupling, we use the following single-valued statistics to evaluate the performance of the test: accuracy, sensitivity, TPR, TNR, FPR and FNR, and for some systems PPV (positive predictive value), NPP (negative predictive value) and the F1 score. We computed these statistics as a function of coupling strength and time series length for the  systems listed below. We find that, provided sufficient coupling strength and long enough time series, the performance approaches perfect across all statistical performance measures for all the tested systems.

\begin{figure*}[h]
\centering
\includegraphics[width=1.0\linewidth]{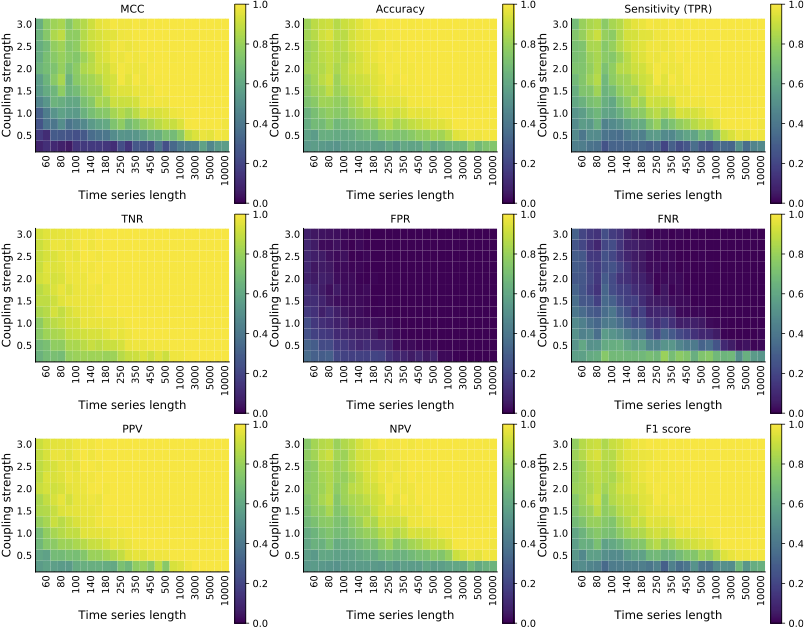}
\caption{
Statistical robustness of the normalized predictive asymmetry causality criterion $\mathcal{A}^{f=1.0}$ (eq. \ref{eq:normalized_asymmetry_criterion}) for chained periodic autoregressive systems with strongly nonlinear coupling, where the systems have variable internal lags, variable interaction lags, dynamical noise and observational noise (eq. \ref{eq:system_chain_autoregressiveperiodic_stronglynonlinearcoupling}). In each heat map cell (for each combination of coupling strength and time series length) the statistical measures are computed over 300 independent realizations of the system with randomized initial conditions and randomized parameters.
}
\label{fig_supp:statistical_robustness_bigpanel_autoregressiveperiodic_stronglynonlinearcoupling}
\end{figure*}

\begin{figure*}[h]
\centering
\includegraphics[width=1.0\linewidth]{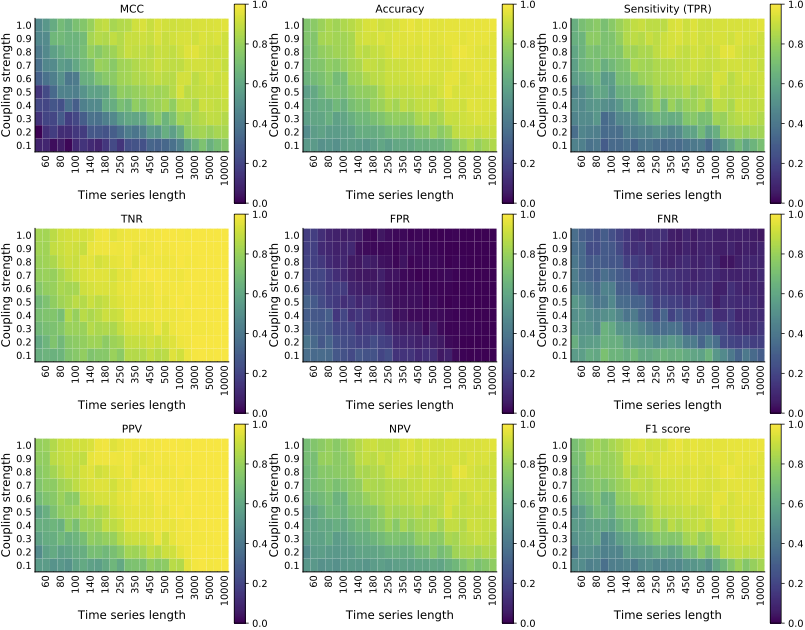}
\caption{
Statistical robustness of the normalized predictive asymmetry causality criterion $\mathcal{A}^{f=1.0}$ (eq. \ref{eq:normalized_asymmetry_criterion}) for a chained nonlinear system with linear coupling, where the systems have variable internal lags, variable interaction lags, dynamical noise and observational noise (eq. \ref{eq:system_chain_chen_nonlinearperiodic_linearcoupling}). In each heat map cell (for each combination of coupling strength and time series length) the statistical measures are computed over 300 independent realizations of the system with randomized initial conditions and randomized parameters.
}
\label{fig_supp:statistical_robustness_bigpanel_nonlinear_linearcoupling}
\end{figure*}

\begin{figure*}[h]
\centering
\includegraphics[width=1.0\linewidth]{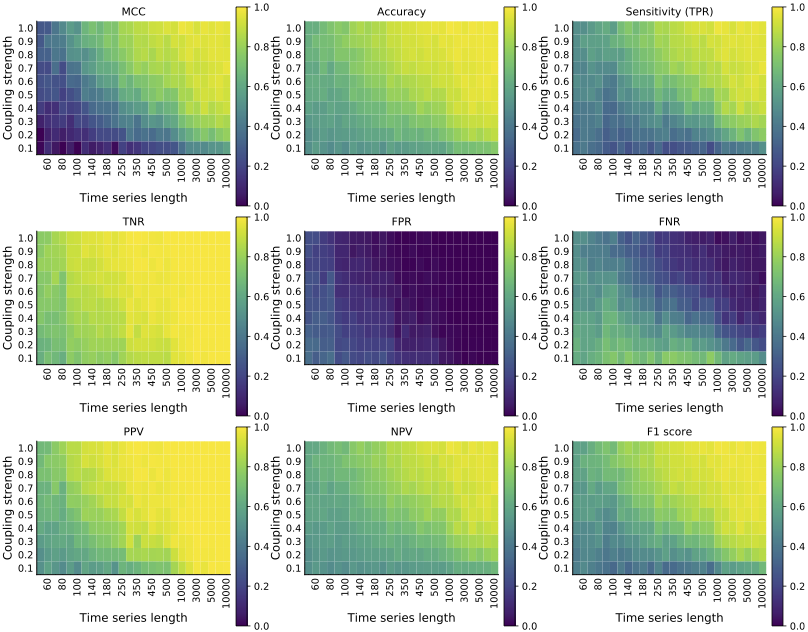}
\caption{
Statistical robustness of the normalized predictive asymmetry causality criterion $\mathcal{A}^{f=1.0}$  (eq. \ref{eq:normalized_asymmetry_criterion}) for a chained nonlinear system with nonlinear coupling, where the systems have variable internal lags, variable interaction lags, dynamical noise and observational noise (eq. \ref{eq:system_chain_nonlinear_chen_nonlinearcoupling}). In each heat map cell (for each combination of coupling strength and time series length) the statistical measures are computed over 300 independent realizations of the system with randomized initial conditions and randomized parameters.
}
\label{fig_supp:statistical_robustness_bigpanel_nonlinear_nonlinearcoupling}
\end{figure*}

\begin{figure*}[h]
\centering
\includegraphics[width=1.0\linewidth]{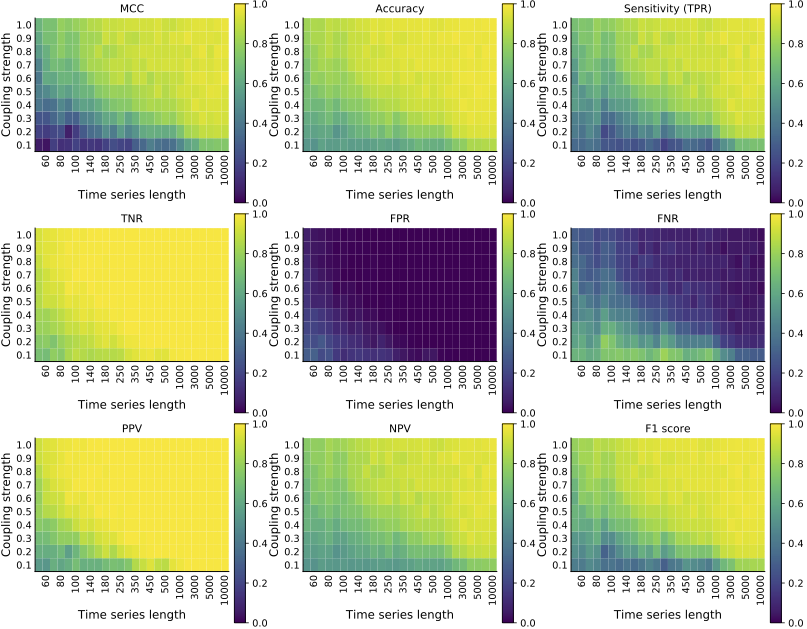}
\caption{
Statistical robustness of the normalized predictive asymmetry causality criterion $\mathcal{A}^{f=1.0}$  (eq. \ref{eq:normalized_asymmetry_criterion}) for a chained periodic and nonlinear with linear coupling, where the systems have variable internal lags, variable interaction lags, dynamical noise and observational noise (eq. \ref{eq:system_chain_chen_nonlinearperiodic_linearcoupling}). In each heat map cell (for each combination of coupling strength and time series length) the statistical measures are computed over 300 independent realizations of the system with randomized initial conditions and randomized parameters.
}
\label{fig_supp:statistical_robustness_bigpanel_nonlinearperiodic_linearcoupling}
\end{figure*}

\begin{figure*}[h]
\centering
\includegraphics[width=1.0\linewidth]{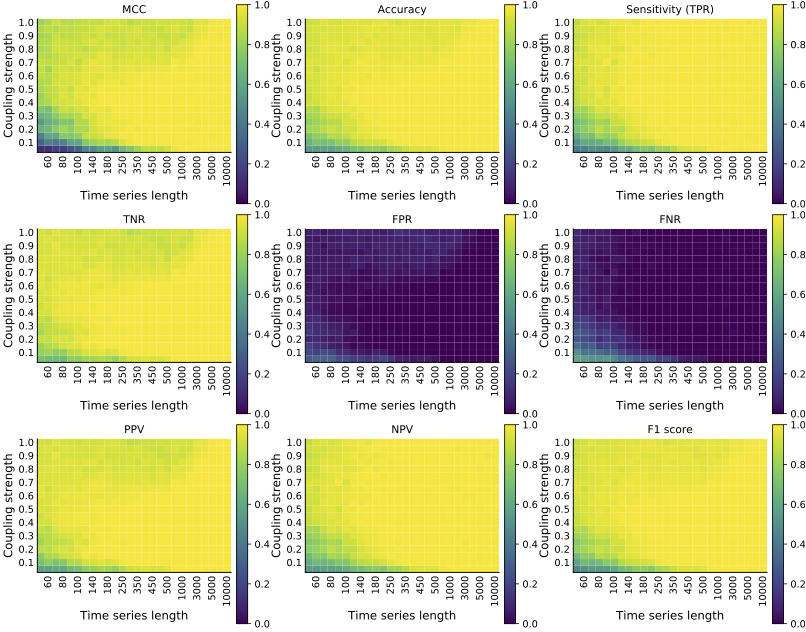}
\caption{
Statistical robustness of the normalized predictive asymmetry causality criterion $\mathcal{A}^{f=1.0}$  (eq. \ref{eq:normalized_asymmetry_criterion}) for chained logistic map systems with variable internal lags, variable interaction lags, dynamical noise and observational noise (eq. \ref{eq:system_chain_logistic_chain_variablelags}). In each heat map cell (for each combination of coupling strength and time series length) the statistical measures are computed over 300 independent realizations of the system with randomized initial conditions on the unit interval and randomized parameters in the chaotic regime ($r_i \thicksim U(3.86, 3.9)$). The dynamical noise level is set to $\sigma = 0.05$. Observational noise equivalent to 0.3 times the standard deviation of each time series is added to the respective time series before analysis. 
Lags $\tau_k$ and $\gamma_k$ are drawn with uniform probability over the set $\{1, 2, \ldots, K_\tau\}$ and $\{1, 2, \ldots, K_\gamma\}$ with $K_\tau = K_\gamma = 5$ For the computation of $\mathcal{A}$, the prediction lag $\eta_{max}$ is set to $10 + \max(K_\tau, K_\gamma) - 1$ (varies between realizations), and embedding parameters are kept constant at $k=l=m=1$. 
}
\label{fig_supp:statistical_robustness_bigpanel_logisticchain}
\end{figure*}

\begin{figure*}[h]
\centering
\includegraphics[width=1.0\linewidth]{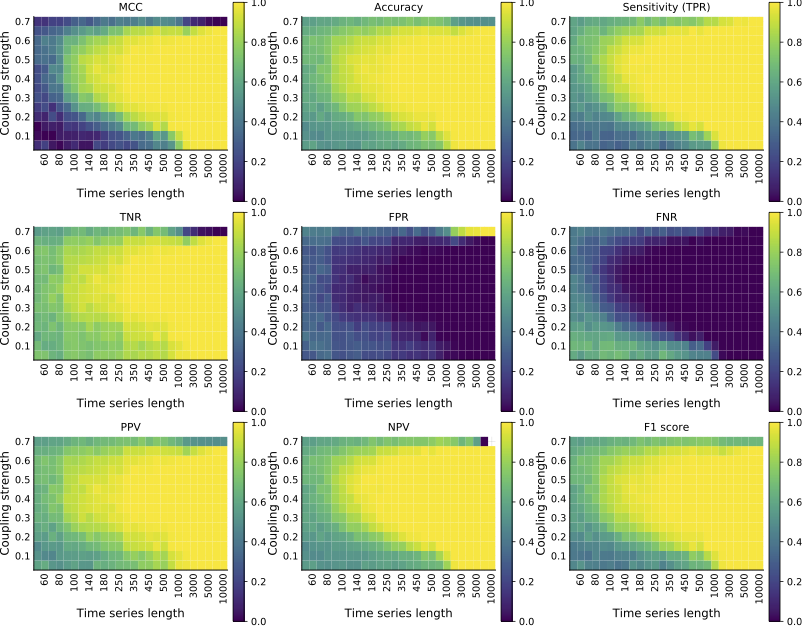}
\caption{
Statistical robustness of the normalized predictive asymmetry causality criterion $\mathcal{A}^{f=1.0}$  (eq. \ref{eq:normalized_asymmetry_criterion}) for a chained Henon map system (eq. \ref{eq:system_chain_henonchain}). In each heat map cell (for each combination of coupling strength and time series length) the statistical measures are computed over 300 independent realizations of the system with randomized initial conditions. Synchronization occurs for coupling strengths around $0.7$ and higher.
}
\label{fig_supp:statistical_robustness_bigpanel_henonchain}
\end{figure*}

\begin{figure*}[h]
\centering
\includegraphics[width=1.0\linewidth]{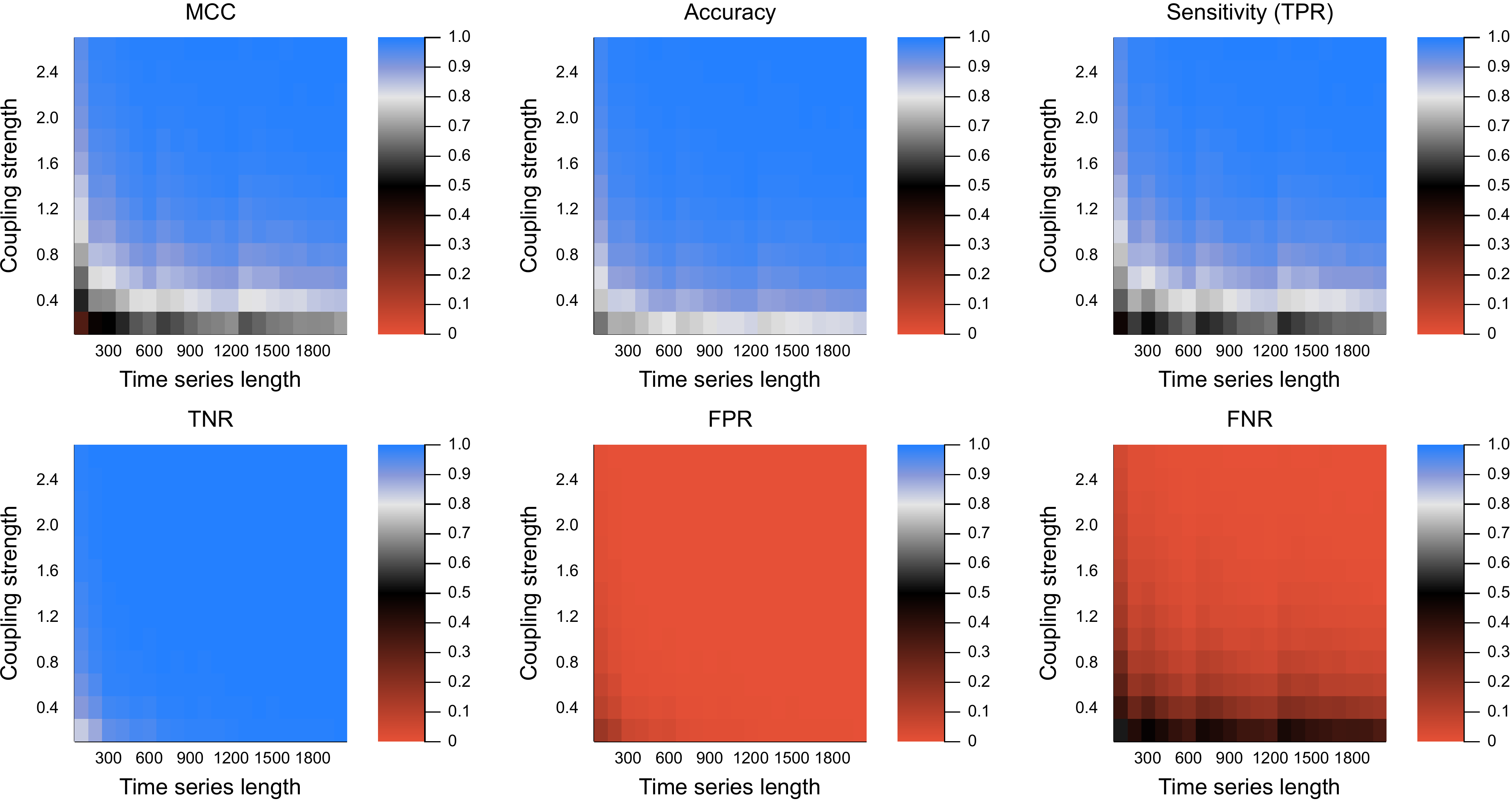}
\caption{Statistical robustness of the test (eq. \ref{eq:normalized_asymmetry_criterion}) for unidirectionally coupled autoregressive systems. For each combination of time series length $L \in \{L_1, L_2, \ldots, L_{n_{L}}\} = \{100, 200, \ldots, 2000\}$, and coupling strength interval $C \in \{C_1, C_2, \ldots, C_{n_{C}}\} = \{[0.0, 0.2], [0.2, 0.4], \ldots, [2.4, 2.6]\}$, $\mathcal{A}$  (eq. \ref{eq:normalized_asymmetry_criterion})  is computed on 5000 $2$-dimensional $VAR(k)$ systems (eq. \ref{eqn_supp:VARK}) with unique coefficient matrices and standard deviations for the noise distributions, and with the model order $k$ randomly chosen from the set $\{1, 2, \ldots, 20\}$ for each system. 
\\ \\
Diagonal terms of the relevant coefficient matrices $A_i$ are independently drawn from a uniform distribution on $[0.1, 0.9]$, and each variable affects itself at exactly one lag (i.e. its coefficient appears in only one of the $k$ coefficient matrices). For a particular coupling strength interval $C_m$, off-diagonal terms giving rise to the coupling generated randomly from a uniform distribution on $[\min{(C_m)}, \max{(C_m)}]$, i.e. the uppermost row in the heatmaps are generated with coupling strengths ranging from $2.4$ to $2.6$. Coupling terms are generated such that there is only unidirectional coupling (i.e. for $p=2$ variables, one off-diagonal term is zero and the other is nonzero in the coefficient matrix). Standard deviations for the error terms are drawn from uniform distributions on $[0.95, 1.05]$, and are drawn independently for each variable.
\\ \\
Generalized embeddings were constructed with $k = l = m = 1$, and $\mathcal{A}$ was computed at $\eta_{max} = 10$ with $f=1.0$. The color scheme is such that light gray corresponds to a rate of 0.8 and black to a rate of 0.5. 
}
\label{fig_supp:VAR_cl_long_maxorder20}
\end{figure*}

\begin{itemize}
    \item Chain of periodic autoregressive variables with strongly nonlinear coupling (Fig. \ref{fig_supp:statistical_robustness_bigpanel_autoregressiveperiodic_stronglynonlinearcoupling}).
    \item Chain of nonlinear variables with linear coupling (Fig. \ref{fig_supp:statistical_robustness_bigpanel_nonlinear_linearcoupling}).
    \item Chain of nonlinear variables with nonlinear coupling (Fig. \ref{fig_supp:statistical_robustness_bigpanel_nonlinear_nonlinearcoupling}).
    \item Chain of nonlinear, periodic variables with linear coupling (Fig. \ref{fig_supp:statistical_robustness_bigpanel_nonlinearperiodic_linearcoupling}).
    \item Chain of logistic maps with dynamical noise (Fig. \ref{fig_supp:statistical_robustness_bigpanel_logisticchain}).
    \item Chain of Henon maps (Fig. \ref{fig_supp:statistical_robustness_bigpanel_henonchain}).
    \item Unidirectionally coupled autoregressive systems of maximum order $k=20$ for longer time series (Fig. \ref{fig_supp:VAR_cl_long_maxorder20}).
\end{itemize}

\clearpage

\subsection{\label{sec_supp:significance_test_bidircoupling}Statistical robustness for systems with bidirectional coupling}

Here, we use sensitivity (TPR) and FNR to characterise the statistical robustness of the normalized predictive asymmetry test for bidirectional systems. 

\begin{figure*}[h]
\centering
\includegraphics[width=1.0\linewidth]{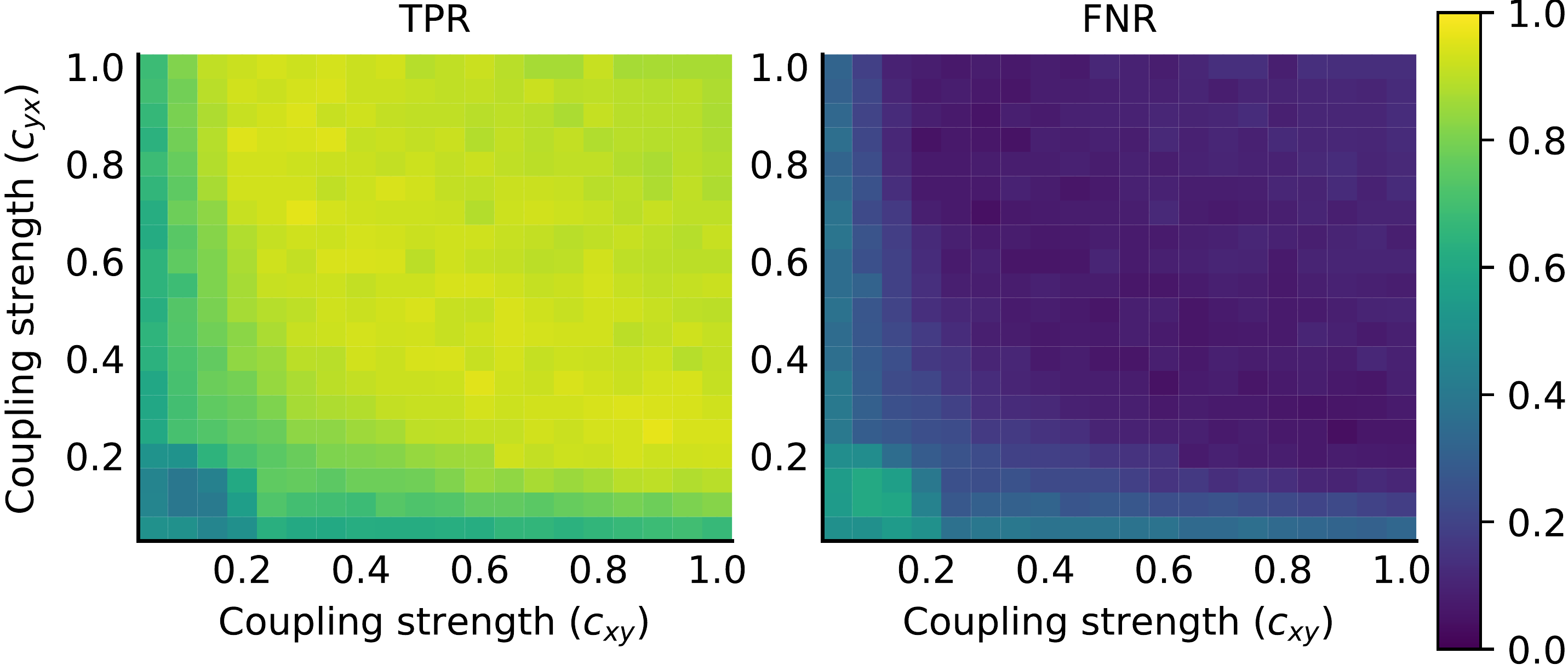}
\caption{Statistical robustness of the test (eq. \ref{eq:normalized_asymmetry_criterion}) for a 2-dimensional system of bidirectionally coupled logistic maps (eq. \ref{eq:logistic_bidir}).  For each combination of coupling strengths $c_{xy} \in \{C_{xy}^1, C_{xy}^2, \ldots, C_{xy}^{n_{C_{xy}}}\} = \{0.05, 0.1, \ldots, 1.0\}$, and $C_{yx} \in \{C_{yx}^1, C_{yx}^2, \ldots, C_{yx}^{n_{C_{yx}}}\} = \{0.05, 0.1, \ldots, 1.0\}$, 
$\mathcal{A}$  is computed on 300 unique realizations of (eq. \ref{eq:logistic_bidir}) with parameters as described in section \ref{sec:supp_test_systems_logistic_bidir}, using time series consisting of 300 observations. By comparing the sign of $\mathcal{A}$ with the known interactions for each of the systems, we then compute $n_{C_{xy}}n_{C_{yx}}$ different confusion matrices, and from those, sensitivity (TPR) and FNR for each combination of $c_{yx}$ and $c_{xy}$.
Generalized embeddings were constructed with $k = l = m = 1$, and $\mathcal{A}$ was computed at $\eta_{max} = 10$ with $f=1.0$. The color scheme is such that light gray corresponds to a rate of 0.8 and black to a rate of 0.5. 
}
\label{fig_supp:heatmaps_CL_bidirectional_logistic}
\end{figure*}

For an ensemble of realizations of the bidirectional logistic map system from the main text, we computed TPR and FNR as a function of coupling strengths in both directions for fixed time series length (Fig. \ref{fig_supp:heatmaps_CL_bidirectional_logistic}). For this system, even for short time series (here 300 observations), the test consistently detects the bidirectional coupling across all but the lowest coupling strengths (for the most part, TPR > 0.8). For higher coupling strengths, the variables become partially synchronized (but not completely due to the dynamical noise), so the detection rates weaken slightly.

\begin{figure*}[h]
\centering
\includegraphics[width=1.0\linewidth]{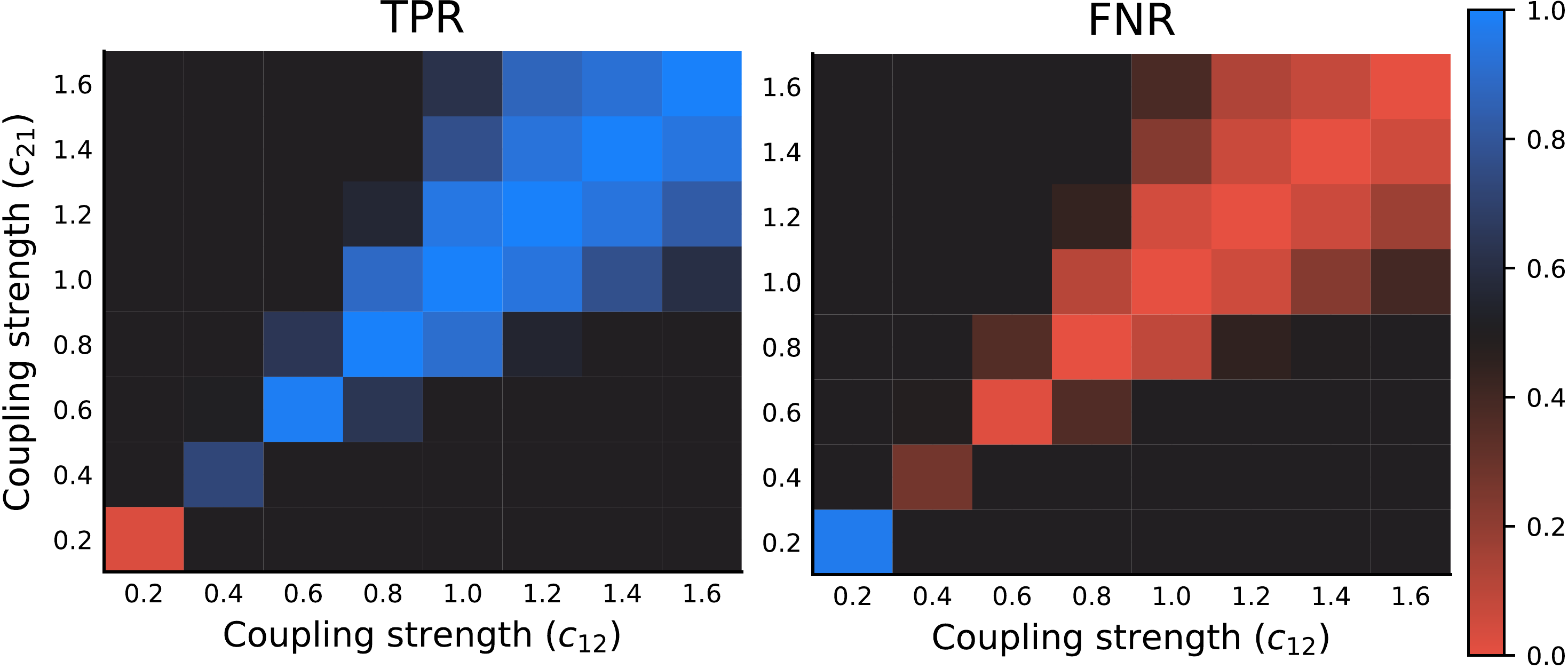}
\caption{Statistical robustness of the test (eq. \ref{eq:normalized_asymmetry_criterion}) for a bidirectionally coupled nonlinear 2-dimensional system (eq. \ref{eq:system_bidirectional_nonlinear_periodic}).  For each combination of coupling strengths $c_{12} \in \{C_{12}^1, C_{12}^2, \ldots, C_{12}^{n_{C_{12}}}\} = \{0.2, 0.4, \ldots, 1.6\}$, and $C_{21} \in \{C_{21}^1, C_{21}^2, \ldots, C_{21}^{n_{C_{21}}}\} = \{0.2, 0.4, \ldots, 1.6\}$, 
$\mathcal{A}$  is computed on 300 unique realizations of (eq. \ref{eq:system_bidirectional_nonlinear_periodic}) with parameters as described in section \ref{sec:supp_test_systems_bidirectional_nonlinear_periodic}, using time series consisting of 20000 observations. By comparing the sign of $\mathcal{A}$ with the known interactions for each of the systems, we then compute $n_{C_{12}}n_{C_{21}}$ different confusion matrices, and from those, sensitivity (TPR) and FNR for each combination of $c_{21}$ and $c_{12}$.
Generalized embeddings were constructed with $k = l = m = 1$, and $\mathcal{A}$ was computed at $\eta_{max} = 15$ with $f=1.0$. The color scheme is such that black corresponds to a rate of 0.5. 
}
\label{fig_supp:heatmaps_CL_bidirectional_nonlinear}
\end{figure*}

We also explored another bidirectional system similar to the common-cause scenario in the main text. This system consists of two bidirectionally coupled nonlinear variables with periodicity and dynamical noise, where the coupling is also nonlinear (eq. \ref{eq:system_bidirectional_nonlinear_periodic}). For this system, we find that longer time series are needed to consistently detect the bidirectional relationship between the variables (Fig. \ref{fig_supp:heatmaps_CL_bidirectional_nonlinear}). If coupling strengths in both directions are non-vanishing and roughly equal, then the bidirectional relationship is detected most of the time (TPR > 0.9). If the coupling in one direction is much stronger in one direction than in the other direction, then the system will appear unidirectional in the eyes of the predictive asymmetry test (\ref{fig_supp:characteristic_asymmetries_heatmap_bidirectional_nonlinear_periodic}; black heat map cells away from the diagonal in Fig. \ref{fig_supp:heatmaps_CL_bidirectional_nonlinear}). This happens because the predictive asymmetry in the direction of the strongest forcing becomes positive, while in the direction of the weakest direction, the predictive asymmetry becomes negative (\ref{fig_supp:characteristic_asymmetries_heatmap_bidirectional_nonlinear_periodic}), essentially rendering the detectable relationship unidirectional.  Stronger mutual coupling increases the deviation between relative coupling strengths that can be tolerated before the bidirectional relationship starts to appear unidirectional (wider blue areas around the diagonal for higher coupling strengths in Fig. \ref{fig_supp:heatmaps_CL_bidirectional_nonlinear}). 

\clearpage

% Appendix:
% Characteristic asymmetries
\section{\label{sec:characteristic_asymmetries}Characteristic predictive asymmetries}

Here, we further demonstrate the typical asymmetries obtained for the cases  unidirectional coupling and bidirectional coupling discussed in the main text. We consider the cases of unidirectional coupling and bidirectional coupling separately, visualizing the asymmetries using two types of plots: (1) Line plots with error bars of $\mathcal{A}(\eta)$ versus $\eta$ at a fixed time series length $L$ and coupling strength $C$, and (2) Heat maps with average $\mathcal{A}$ over multiple configurations of time series lengths and coupling strengths.

For all the tested systems, the ensemble median of the predictive asymmetry converges to values around zero with increasing variability for higher prediction lags (not shown here). 

\subsection{\label{sec:characteristic_asymmetries_unidir_chains}Causal chains}

\begin{figure*}[h]
\centering
\includegraphics[width=1.0\linewidth]{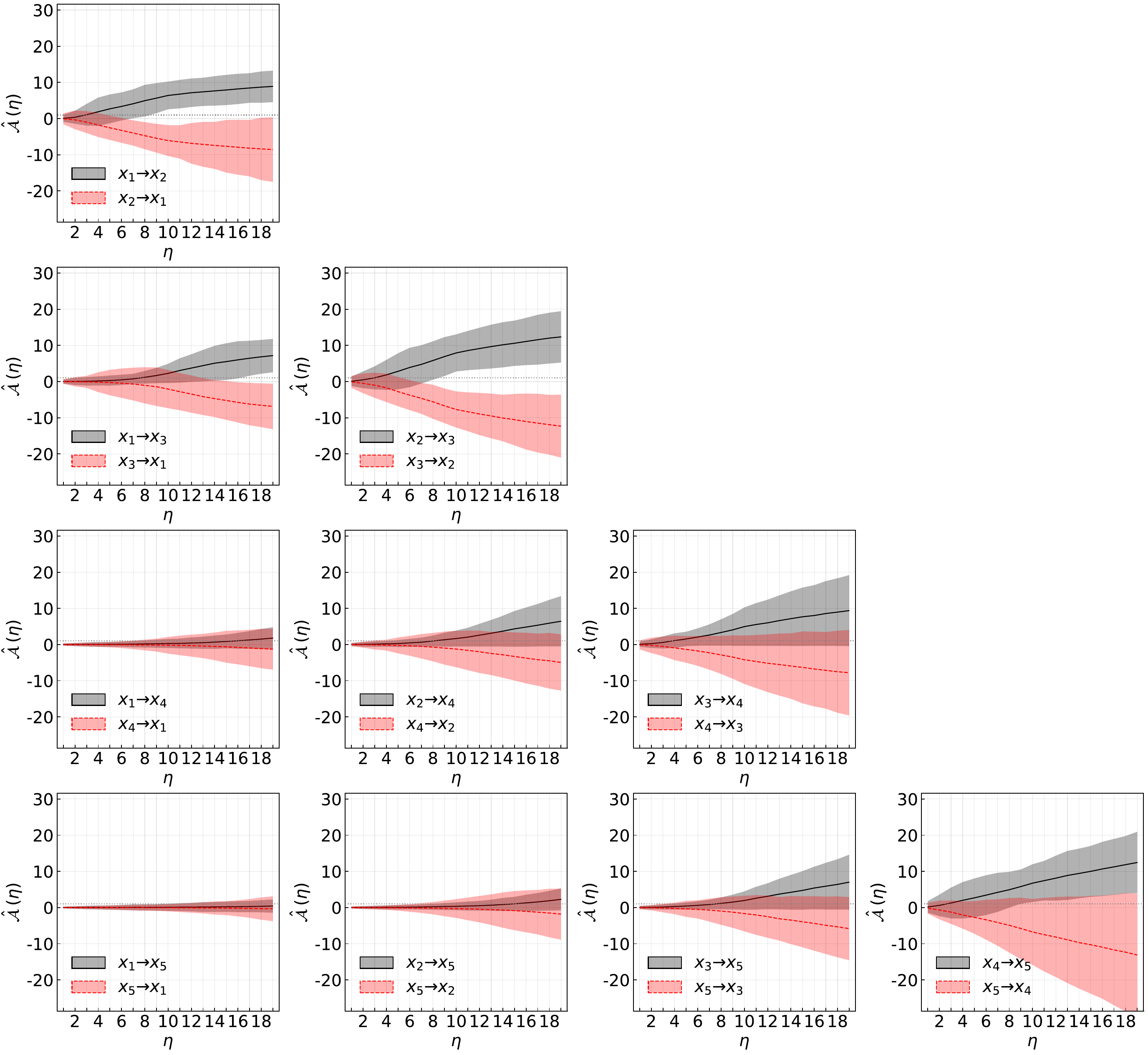}
\caption{
Statistical robustness of the normalized predictive asymmetry causality criterion $\mathcal{A}^{f=1.0}$ (eq. \ref{eq:normalized_asymmetry_criterion}) for systems of chained periodic autoregressive variables which interact in a strongly nonlinear manner, and where the systems have variable internal lags, variable interaction lags, and observational noise (eq. \ref{eq:system_chain_autoregressiveperiodic_stronglynonlinearcoupling}).
}
\label{fig_supp:causal_chain_bigpanel_autoregressiveperiodic_stronglynonlinearcoupling}
\end{figure*}

\begin{figure*}[h]
\centering
\includegraphics[width=1.0\linewidth]{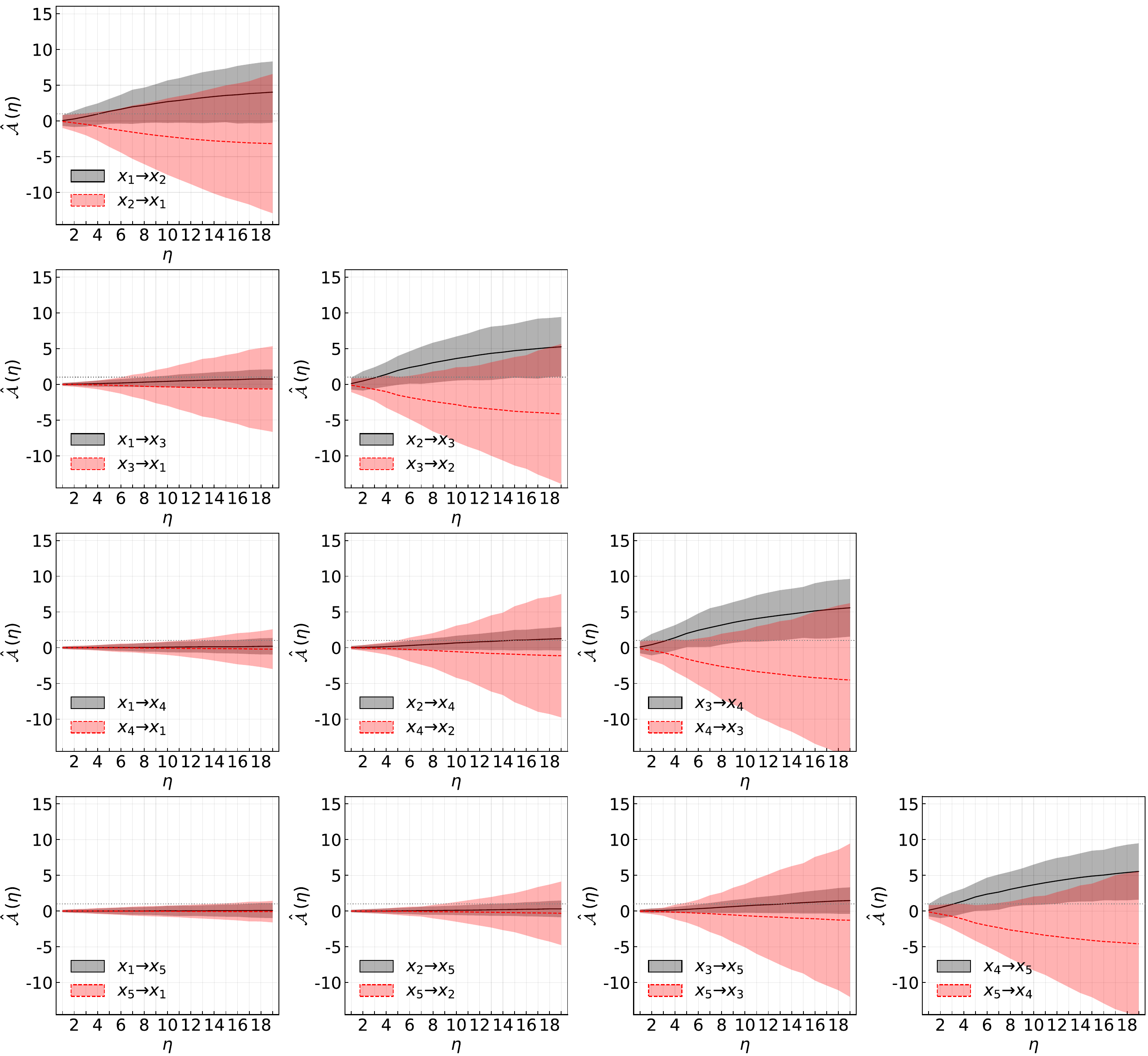}
\caption{
Statistical robustness of the normalized predictive asymmetry causality criterion $\mathcal{A}^{f=1.0}$  (eq. \ref{eq:normalized_asymmetry_criterion}) for a chained nonlinear system with linear coupling, where the systems have variable internal lags, variable interaction lags, dynamical noise and observational noise (eq. \ref{eq:system_chain_nonlinear_chen_linearcoupling}).
}
\label{fig_supp:causal_chain_bigpanel_nonlinear_linearcoupling}
\end{figure*}

\begin{figure*}[h]
\centering
\includegraphics[width=1.0\linewidth]{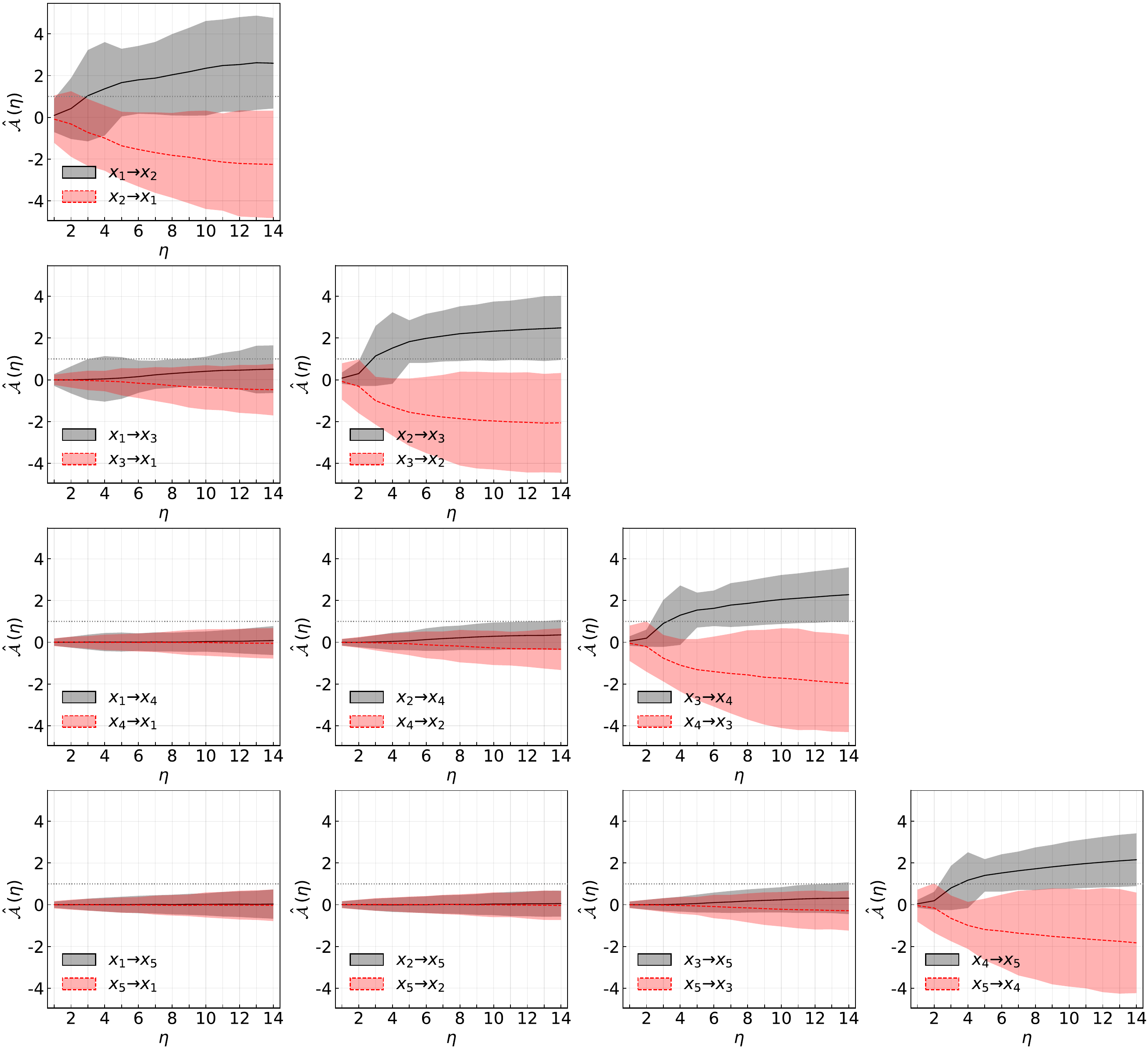}
\caption{
Statistical robustness of the normalized predictive asymmetry causality criterion $\mathcal{A}^{f=1.0}$  (eq. \ref{eq:normalized_asymmetry_criterion}) for a chained nonlinear system with nonlinear coupling, where the systems have variable internal lags, variable interaction lags, dynamical noise and observational noise (eq. \ref{eq:system_chain_nonlinear_chen_nonlinearcoupling}).
}
\label{fig_supp:causal_chain_bigpanel_nonlinear_nonlinearcoupling}
\end{figure*}

\begin{figure*}[h]
\centering
\includegraphics[width=1.0\linewidth]{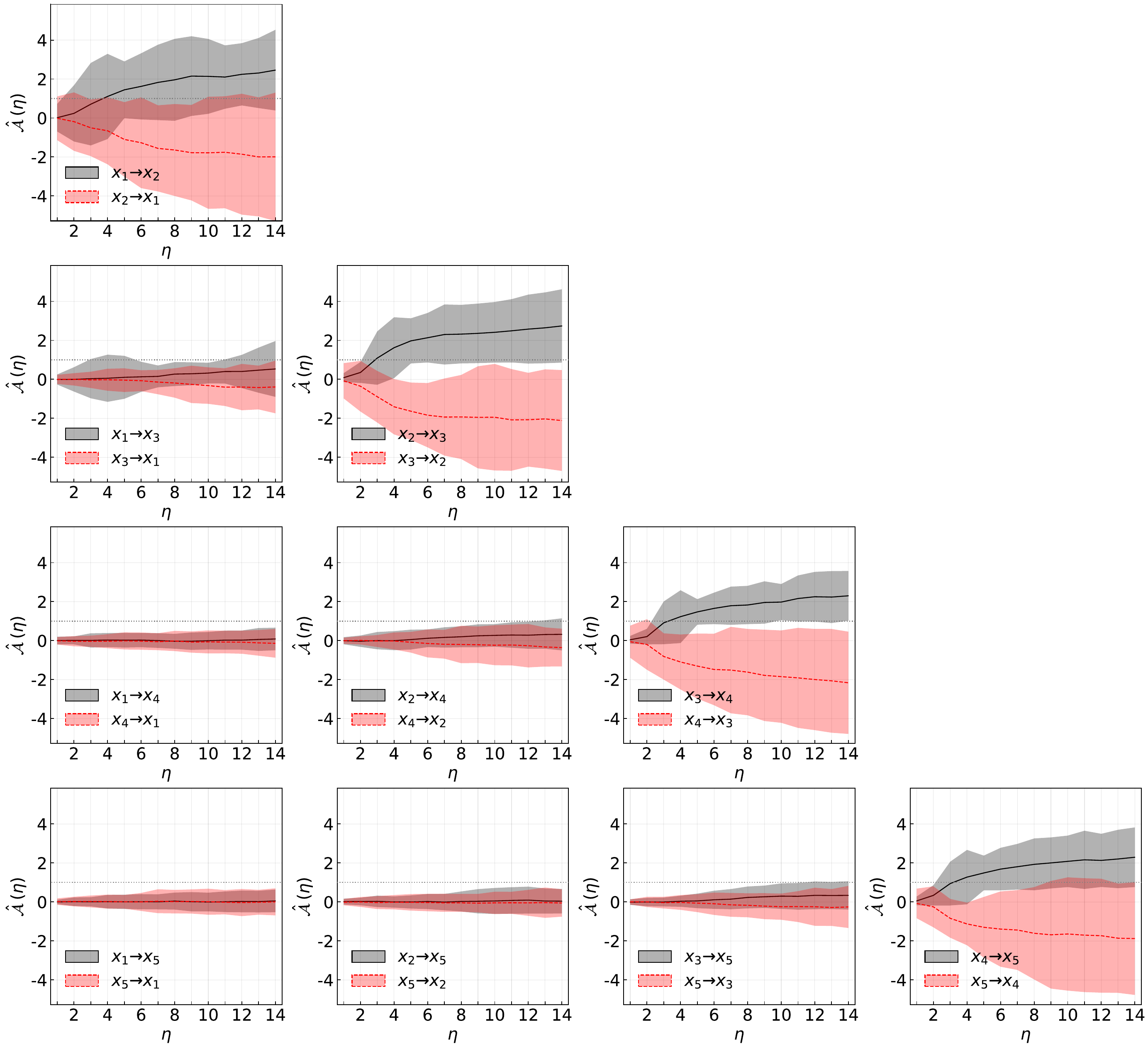}
\caption{
Statistical robustness of the normalized predictive asymmetry causality criterion $\mathcal{A}^{f=1.0}$  (eq. \ref{eq:normalized_asymmetry_criterion}) for a chained periodic and nonlinear with linear coupling, where the systems have variable internal lags, variable interaction lags, dynamical noise and observational noise (eq. \ref{eq:system_chain_chen_nonlinearperiodic_linearcoupling}).
}
\label{fig_supp:causal_chain_bigpanel_nonlinearperiodic_linearcoupling}
\end{figure*}

\begin{figure*}[h]
\centering
\includegraphics[width=1.0\linewidth]{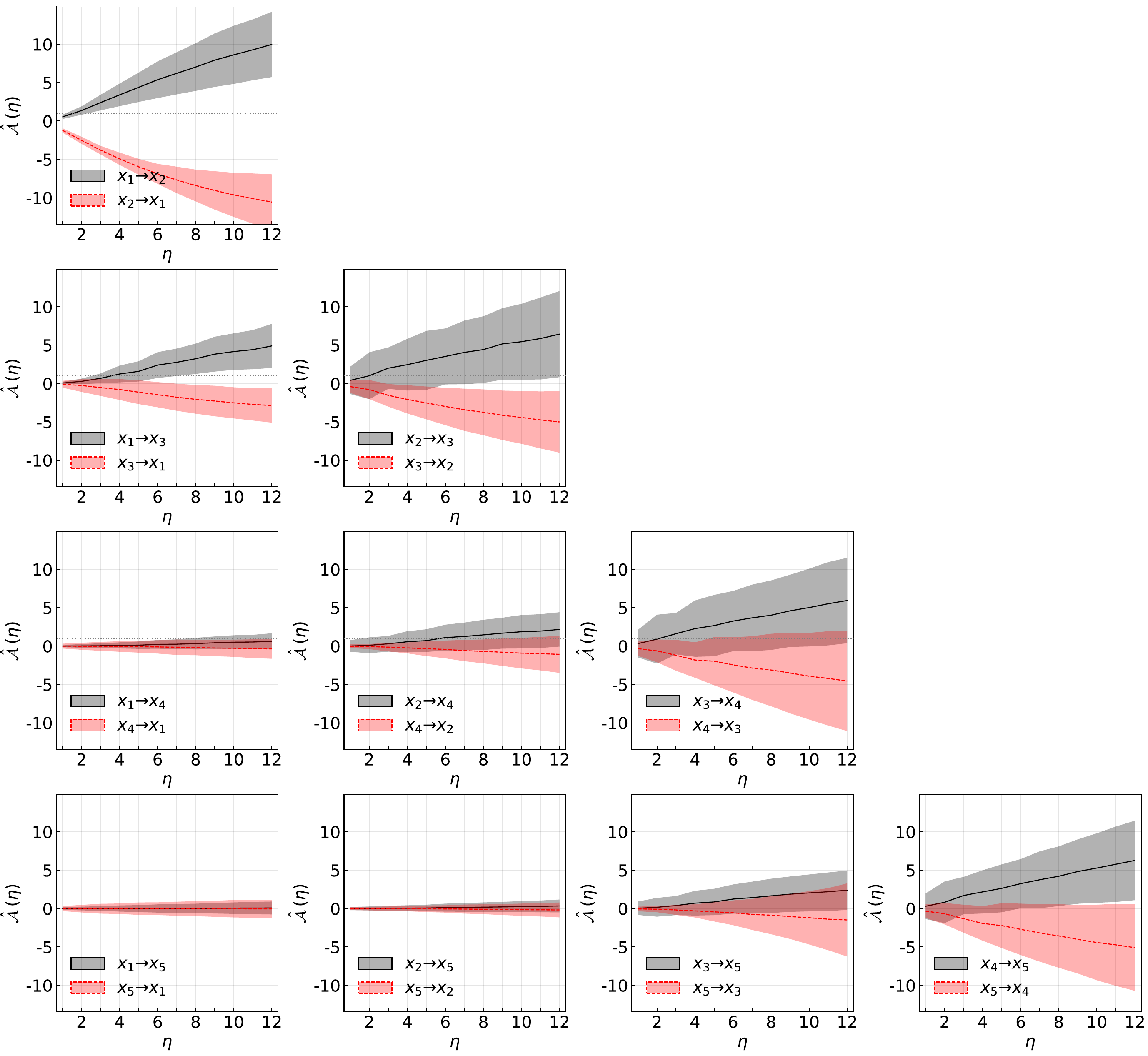}
\caption{
Statistical robustness of the normalized predictive asymmetry causality criterion $\mathcal{A}^{f=1.0}$  (eq. \ref{eq:normalized_asymmetry_criterion}) for chained systems of logistic maps, where the systems have variable internal lags, variable interaction lags, dynamical noise and observational noise (eq. \ref{eq:system_chain_logistic_chain_variablelags}).
}
\label{fig_supp:causal_chain_bigpanel_logisticchain}
\end{figure*}

\begin{figure*}[h]
\centering
\includegraphics[width=1.0\linewidth]{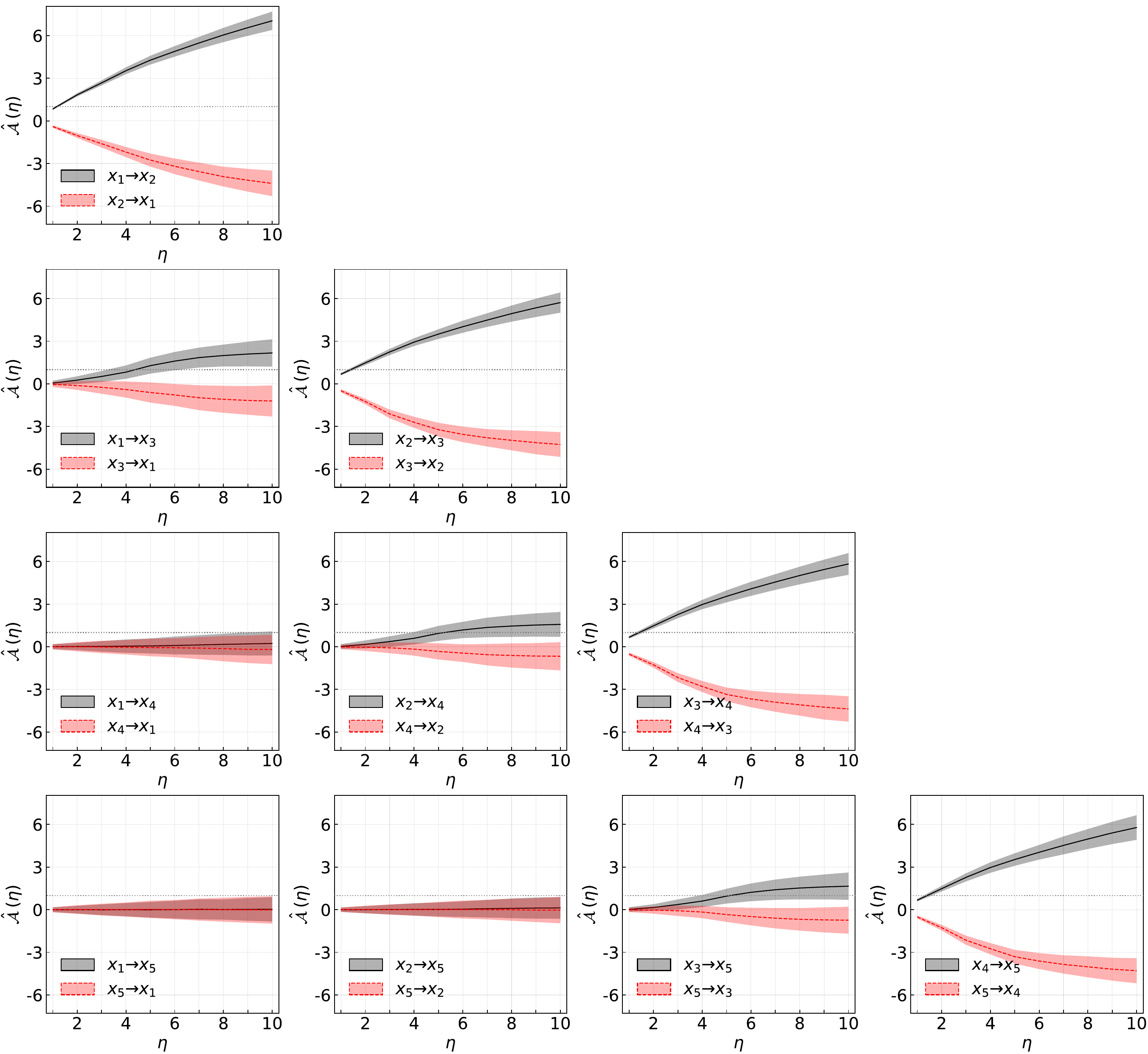}
\caption{
Statistical robustness of the normalized predictive asymmetry causality criterion $\mathcal{A}^{f=1.0}$ (eq. \ref{eq:normalized_asymmetry_criterion}) for chained systems of Henon maps with observational noise (eq. \ref{eq:system_chain_henonchain}).
}
\label{fig_supp:causal_chain_bigpanel_henonchain}
\end{figure*}

For unidirectional causal chains, the normalized predictive asymmetry is best at detecting adjacent links. Indirect links are also detectable, but predictive asymmetries decrease in absolute magnitude with an increasing number of intermediate links (Figs. 
\ref{fig_supp:causal_chain_bigpanel_autoregressiveperiodic_stronglynonlinearcoupling},
\ref{fig_supp:causal_chain_bigpanel_nonlinear_linearcoupling},
\ref{fig_supp:causal_chain_bigpanel_nonlinear_nonlinearcoupling},
\ref{fig_supp:causal_chain_bigpanel_nonlinearperiodic_linearcoupling},
\ref{fig_supp:causal_chain_bigpanel_logisticchain},
\ref{fig_supp:causal_chain_bigpanel_henonchain}). The exact number of intermediate links that are detectable varies between systems. For our example systems, the predictive asymmetry gets indistinguishable from non-coupled systems after after two-to-three intermediate links. 

\begin{itemize}
    \item Chain of periodic autoregressive variables with strongly nonlinear coupling (Fig. \ref{fig_supp:causal_chain_bigpanel_autoregressiveperiodic_stronglynonlinearcoupling}).
    \item Chain of nonlinear variables with linear coupling (Fig. \ref{fig_supp:causal_chain_bigpanel_nonlinear_linearcoupling}).
    \item Chain of nonlinear variables with nonlinear coupling (Fig. \ref{fig_supp:causal_chain_bigpanel_nonlinear_nonlinearcoupling}).
    \item Chain of nonlinear, periodic variables with linear coupling (Fig. \ref{fig_supp:causal_chain_bigpanel_nonlinearperiodic_linearcoupling}).
    \item Chain of logistic maps with dynamical noise (Fig. \ref{fig_supp:causal_chain_bigpanel_logisticchain}).
    \item Chain of Henon maps (Fig. \ref{fig_supp:causal_chain_bigpanel_henonchain}).
\end{itemize}

\clearpage

\clearpage

\subsection{\label{sec:characteristic_asymmetries_unidircoupling}Average magnitude of $\mathcal{A}$ for systems with unidirectional coupling}

Here, we corroborate the statement that on average, systems with unidirectional coupling $\mathcal{A}^f > 0$ in the direction where dynamical coupling exists, and that $\mathcal{A}^f <= 0$ in the direction where the is dynamical no coupling. We illustrate this by heat maps of average $\mathcal{A}^f$ across ensembles of realizations of different systems, as a function of time series length and coupling strength (Figs. 
\ref{fig_supp:statistical_robustness_bigpanel_autoregressiveperiodic_stronglynonlinearcoupling},
\ref{fig_supp:statistical_robustness_bigpanel_nonlinear_linearcoupling},
\ref{fig_supp:statistical_robustness_bigpanel_nonlinear_nonlinearcoupling},
\ref{fig_supp:statistical_robustness_bigpanel_nonlinearperiodic_linearcoupling},
\ref{fig_supp:statistical_robustness_bigpanel_logisticchain},
\ref{fig_supp:statistical_robustness_bigpanel_henonchain}).

\begin{itemize}
    \item Chain of periodic autoregressive variables with strongly nonlinear coupling (Fig. \ref{fig_supp:statistical_robustness_bigpanel_autoregressiveperiodic_stronglynonlinearcoupling}).
    \item Chain of nonlinear variables with linear coupling (Fig. \ref{fig_supp:statistical_robustness_bigpanel_nonlinear_linearcoupling}).
    \item Chain of nonlinear variables with nonlinear coupling (Fig. \ref{fig_supp:statistical_robustness_bigpanel_nonlinear_nonlinearcoupling}).
    \item Chain of nonlinear, periodic variables with linear coupling (Fig. \ref{fig_supp:statistical_robustness_bigpanel_nonlinearperiodic_linearcoupling}).
    \item Chain of logistic maps with dynamical noise (Fig. \ref{fig_supp:statistical_robustness_bigpanel_logisticchain}).
    \item Chain of Henon maps (Fig. \ref{fig_supp:statistical_robustness_bigpanel_henonchain}).
\end{itemize}

\begin{figure*}[h]
\centering
\includegraphics[width=0.7\linewidth]{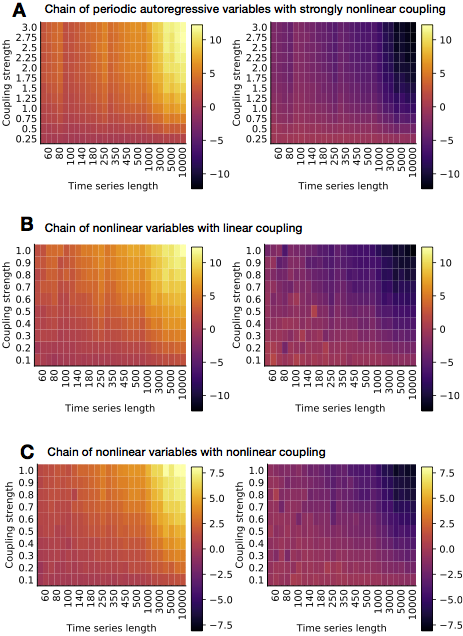}
\caption{
Average magnitude of the normalized predictive asymmetry causality criterion $\mathcal{A}^{f=1.0}$ for various systems, with variable variable internal lags, variable interaction lags, dynamical noise and observational noise. In each heat map cell (for each combination of coupling strength and time series length) the average magnitude is computed over 300 independent realizations of the system with randomized initial conditions and randomized parameters. A: Periodic autoregressive systems with strongly nonlinear coupling (eq. \ref{eq:system_chain_autoregressiveperiodic_stronglynonlinearcoupling}; B: Nonlinear systems with linear coupling (eq. \ref{eq:system_chain_nonlinear_chen_linearcoupling}); C: Nonlinear systems with nonlinear coupling (eq. \ref{eq:system_chain_nonlinear_chen_nonlinearcoupling}).
}
\label{fig_supp:average_Amagnitude_ABC}
\end{figure*}

\begin{figure*}[h]
\centering
\includegraphics[width=0.7\linewidth]{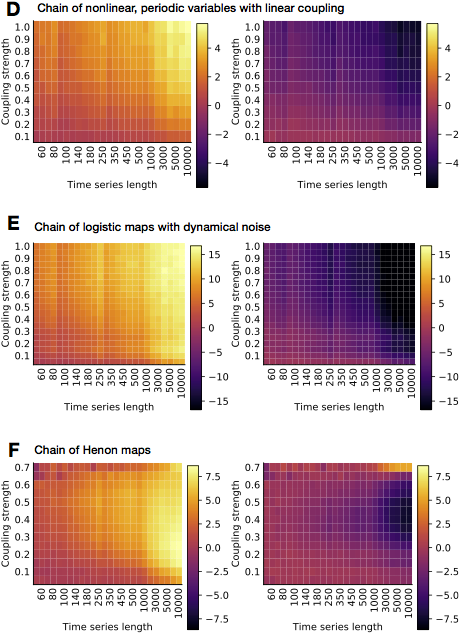}
\caption{
Continued from Fig. \ref{fig_supp:average_Amagnitude_ABC}. Average magnitude of the normalized predictive asymmetry causality criterion $\mathcal{A}^{f=1.0}$ for various systems, with variable variable internal lags, variable interaction lags, dynamical noise and observational noise. In each heat map cell (for each combination of coupling strength and time series length) the average magnitude is computed over 300 independent realizations of the system with randomized initial conditions and randomized parameters. D: Nonlinear periodic systems with linear coupling (eq. \ref{eq:system_chain_chen_nonlinearperiodic_linearcoupling}); E: Chain of logistic maps with variable interaction lags, and variable internal lags, dynamical noise, and observational noise (eq. \ref{eq:system_chain_logistic_chain_variablelags}); F: Chain of Henon maps with observational noise (eq. \ref{eq:system_chain_henonchain}).
}
\label{fig_supp:average_Amagnitude_DEF}
\end{figure*}

\clearpage

\subsection{\label{sec:characteristic_asymmetries_bidircoupling}Average magnitude of $\mathcal{A}$ for systems with bidirectional coupling}

\begin{figure*}[h]
\centering
\includegraphics[width=1.0\linewidth]{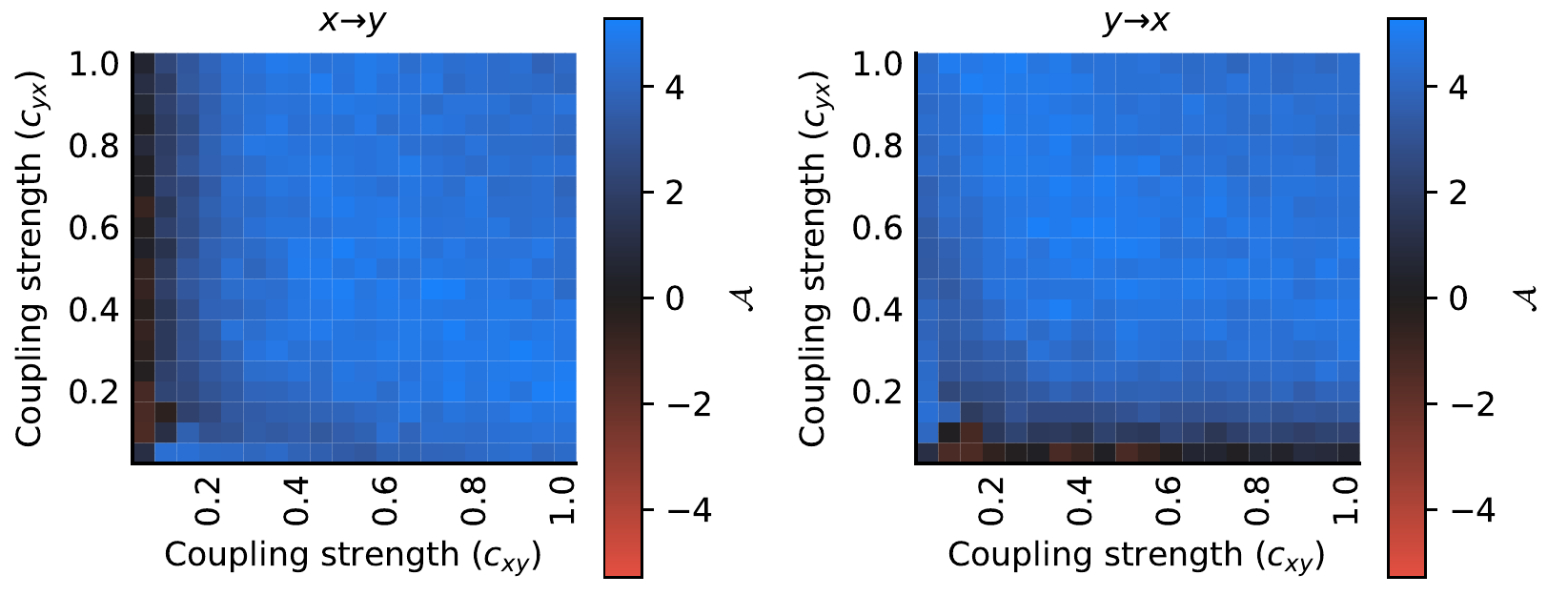}
\caption{normalized predictive asymmetry (eq. \ref{eq:normalized_asymmetry_criterion}) for a 2-dimensional system of bidirectionally coupled logistic maps (eq. \ref{eq:logistic_bidir}). For each combination of coupling strengths $c_{xy} \in \{C_{xy}^1, C_{xy}^2, \ldots, C_{xy}^{n_{C_{xy}}}\} = \{0.05, 0.1, \ldots, 1.0\}$, and $C_{yx} \in \{C_{yx}^1, C_{yx}^2, \ldots, C_{yx}^{n_{C_{yx}}}\} = \{0.05, 0.1, \ldots, 1.0\}$, 
$\mathcal{A}$  is computed on 300 unique realizations of (eq. \ref{eq:logistic_bidir}) with parameters as described in section \ref{sec:supp_test_systems_logistic_bidir}, using time series consisting of 300 observations. The value in each cell is the mean $\mathcal{A}$ over the 300 realizations.
Generalized embeddings were constructed with $k = l = m = 1$, and $\mathcal{A}$ was computed at $\eta_{max} = 10$ with $f=1.0$. 
}
\label{fig_supp:characteristic_asymmetries_heatmap_logistic_bidir}
\end{figure*}

\begin{figure*}[h]
\centering
\includegraphics[width=1.0\linewidth]{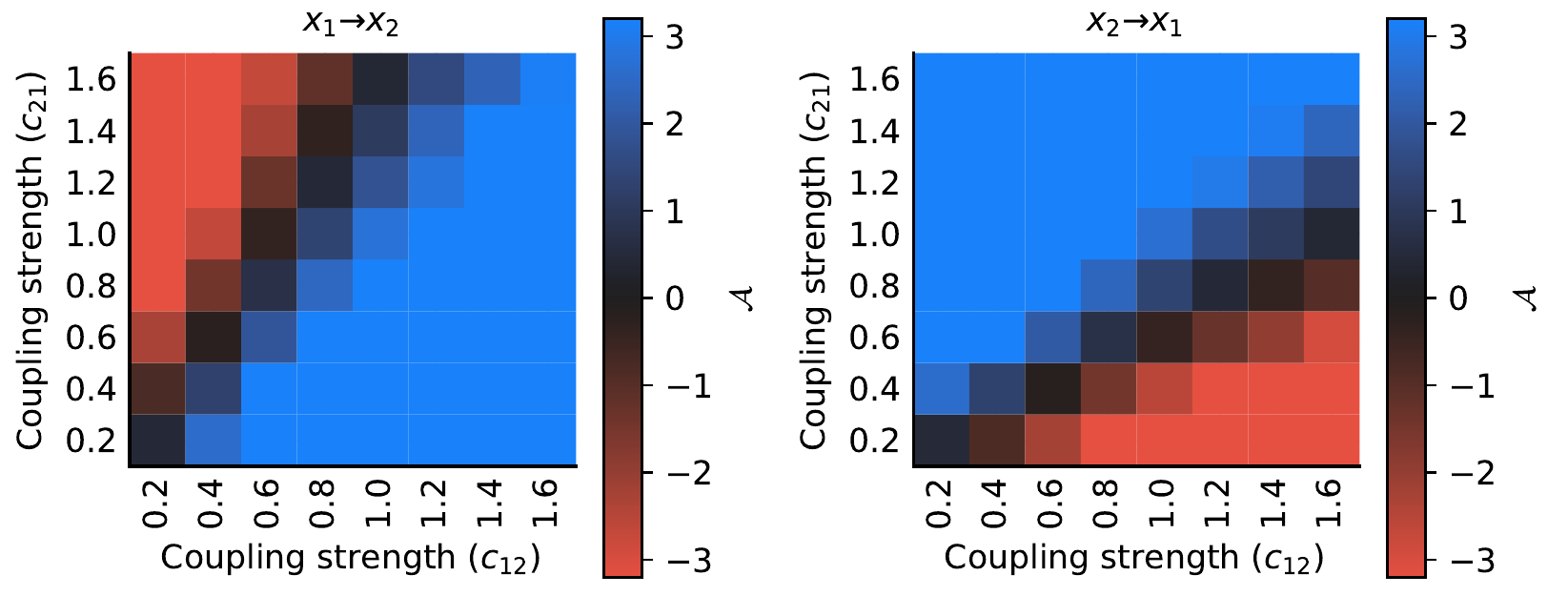}
\caption{normalized predictive asymmetry (eq. \ref{eq:normalized_asymmetry_criterion}) for a nonlinear and bidirectionally coupled 2-dimensional system (eq. \ref{eq:system_bidirectional_nonlinear_periodic}). For each combination of coupling strengths $c_{xy} \in \{C_{xy}^1, C_{xy}^2, \ldots, C_{xy}^{n_{C_{xy}}}\} = \{0.05, 0.1, \ldots, 0.7\}$, and $C_{yx} \in \{C_{yx}^1, C_{yx}^2, \ldots, C_{yx}^{n_{C_{yx}}}\} = \{0.05, 0.1, \ldots, 0.7\}$, 
$\mathcal{A}$  is computed on 300 unique realizations of (eq. \ref{eq:logistic_bidir}) with parameters as described in section \ref{sec:supp_test_systems_logistic_bidir}, using time series consisting of 20000 observations. The value in each cell is the mean $\mathcal{A}$ over the 300 realizations.
Generalized embeddings were constructed with $k = l = m = 1$, and $\mathcal{A}$ was computed at $\eta_{max} = 10$ with $f=1.0$. 
}
\label{fig_supp:characteristic_asymmetries_heatmap_bidirectional_nonlinear_periodic}
\end{figure*}

Here, we demonstrate the median magnitude of $\mathcal{A}^f$ over varying $c_{xy}$ and $c_{yx}$, for fixed time series length for the bidirectional logistic map system from the main text (Fig. \ref{fig_supp:characteristic_asymmetries_heatmap_logistic_bidir}) and a bidirectional nonlinear system with periodicity and dynamical noise (Fig. \ref{fig_supp:characteristic_asymmetries_heatmap_bidirectional_nonlinear_periodic}). We find that for the bidirectional logistic maps, the ensemble median $\mathcal{A}$ is positive in both directions, thus capturing the underlying bidirectional coupling, even for short time series (here 300 observations). For the second bidirectional nonlinear system, the system appears bidirectional if coupling strengths are roughly equal. However, if the relative coupling strengths are different, then the system appears unidirectional (positive $\mathcal{A}$ in the direction of the strongest forcing, and negative $\mathcal{A}$ in the direction of the weakest forcing).

\clearpage 

\section{Application to real datasets\label{sec_supp:application_to_real_data}}

In this appendix, we apply the ensemble sub-sampling approach described in section \ref{sec:applications_real_data} in the main manuscript to infer interaction networks from time series where ground truths are known. We analyze the following pairs of time series from the cause-effect pair database of Mooji et al.'s cause-effect pair database \cite{Mooij2016} (\url{https://webdav.tuebingen.mpg.de/cause-effect/}, accessed January 15th, 2020).

\begin{figure*}[h]
\centering
\includegraphics[width=1.0\linewidth]{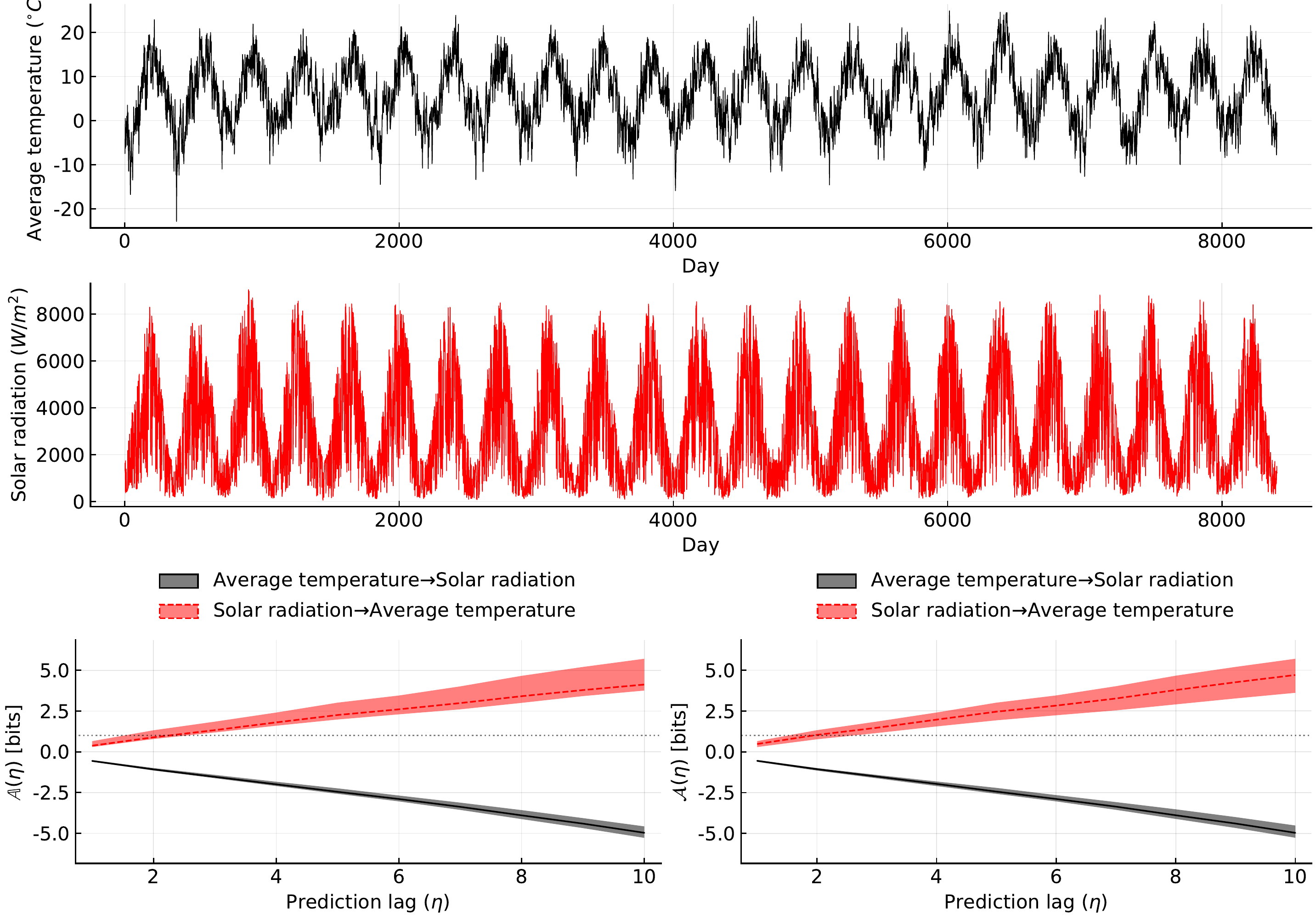}
\caption{Predictive asymmetries (lower left panel) and normalized predictive asymmetries (lower right panel) for time series of solar radiation and average air temperature in Furtwangen, Black Forest, Germany between January 1, 1985 and December 31, 2008 (pair 0077 of the cause-effect pair database \cite{Mooij2016}). Data were provided by Bernward Janzing and processed by Dominik Janzing. Generalized embeddings were constructed with $k = l = m = 1$. Lines and ribbons are the median and 99 percentile confidence intervals for the sample statistic over 50 randomly selected contiguous sub-segments of the time series, where subsegments have lengths ranging from 75\% to 100\% of the total number of observations.  The significance threshold (eq. \ref{eq:normalized_asymmetry_criterion} with $f = 1.0$) is indicated by the dotted horisontal line.
}
\label{fig_supp:real_data_solarradiation_avgtemp}
\end{figure*}

\begin{figure*}[h]
\centering
\includegraphics[width=1.0\linewidth]{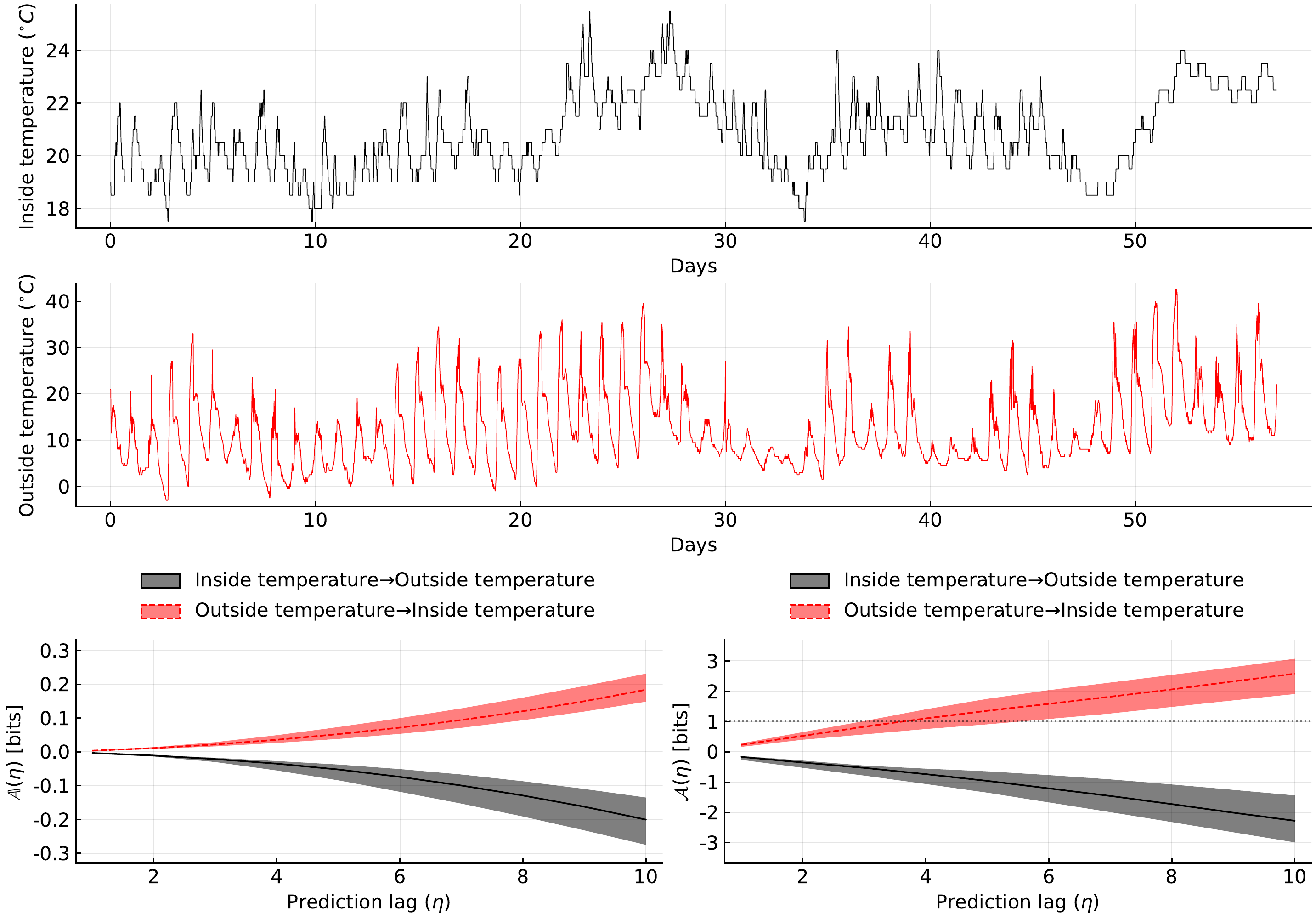}
\caption{Predictive asymmetries (lower left panel) and normalized predictive asymmetries (lower right panel) for time series of inside room temperature and outside temperature (pair 0069 of the cause-effect pair database \cite{Mooij2016}). Data were provided by Joris M. Mooij. Generalized embeddings were constructed with $k = l = m = 1$. Lines and ribbons are the median and 99 percentile confidence intervals for the sample statistic over 50 randomly selected contiguous sub-segments of the time series, where subsegments have lengths ranging from 75\% to 100\% of the total number of observations.  The significance threshold (eq. \ref{eq:normalized_asymmetry_criterion} with $f = 1.0$) is indicated by the dotted horisontal line.
}
\label{fig_supp:real_data_insidetemp_outsidetemp}
\end{figure*}

\begin{figure*}[h]
\centering
\includegraphics[width=1.0\linewidth]{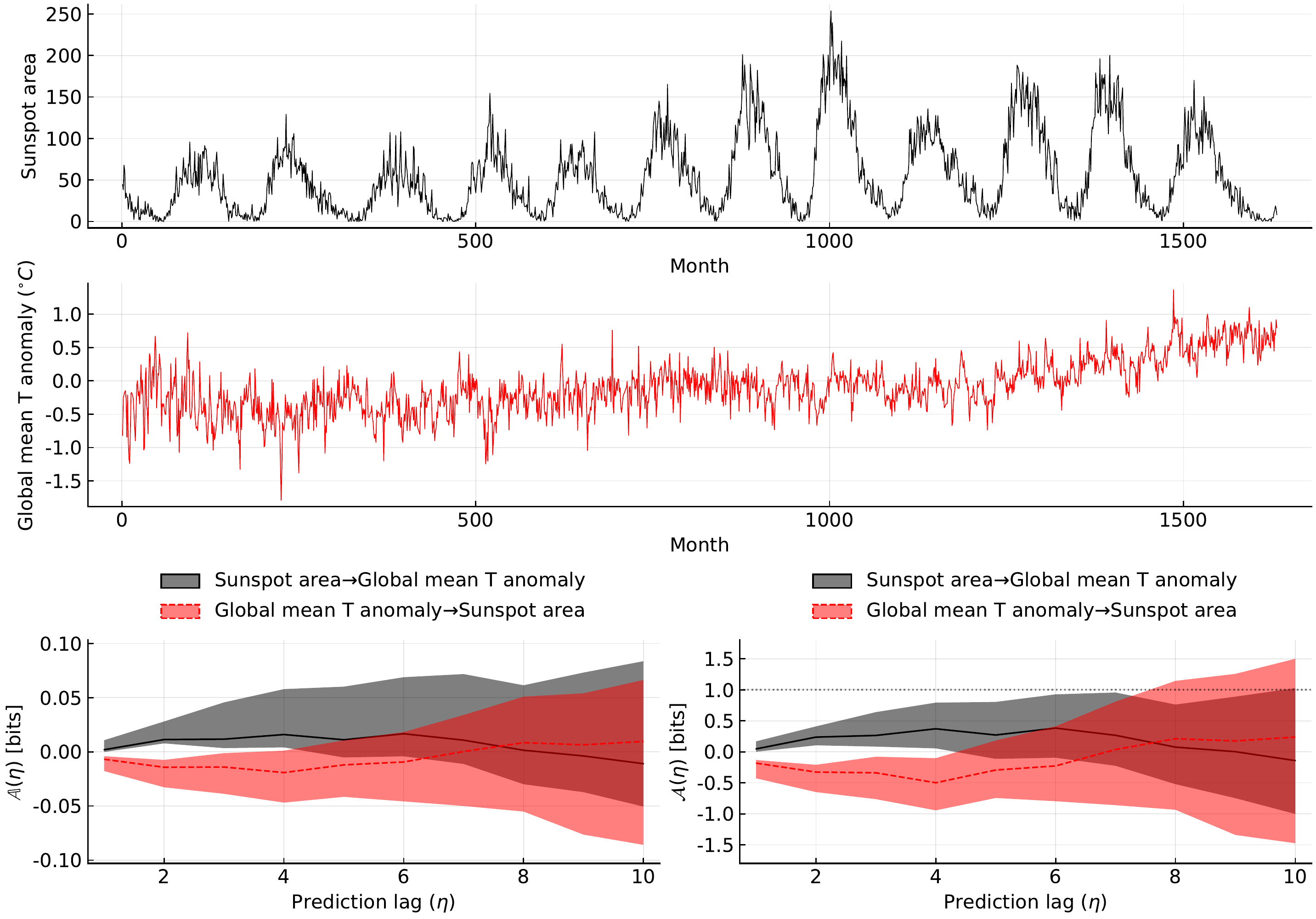}
\caption{Predictive asymmetries (lower left panel) and normalized predictive asymmetries (lower right panel) for time series of sunspot area and global temperature anomalies (deviations from 1961-1990). This is pair 0072 of the cause-effect pair database \cite{Mooij2016}. Generalized embeddings were constructed with $k = l = m = 1$. Lines and ribbons are the median and 99 percentile confidence intervals for the sample statistic over 50 randomly selected contiguous sub-segments of the time series, where subsegments have lengths ranging from 75\% to 100\% of the total number of observations.  The significance threshold (eq. \ref{eq:normalized_asymmetry_criterion} with $f = 1.0$) is indicated by the dotted horisontal line.
}
\label{fig_supp:real_data_sunspotarea_globaTanomaly}
\end{figure*}

\begin{itemize}
    \item Dataset 1: Solar radiation ($W/m^2)$ vs average air temperature ($^\circ C$) at the same location in Furtwangen, Black Forest, Germany between January 1, 1985 and December 31, 2008 (Fig. \ref{fig_supp:real_data_solarradiation_avgtemp}). This is pair 0077 of the cause-effect pair database \cite{Mooij2016}. Ground truth: $\textnormal{solar radiation} \to \textnormal{average temperature}$.
    
    \item Dataset 2: Inside room temperature ($^\circ C$) vs. outside temperature ($^\circ C$) (Fig. \ref{fig_supp:real_data_insidetemp_outsidetemp}). This is pair 0069 of the cause-effect pair database \cite{Mooij2016}. Ground truth: $\textnormal{Outside temperature}$ $\to$ $\textnormal{inside temperature}$. 
    
    \item Dataset 3: Average precipitation ($mm/day$) vs. average runoff ($mm/day$) for 438 river catchments in the US. This is pair 0093 of the cause-effect pair database \cite{Mooij2016}. Ground truth: $\textnormal{Precipitation}$ $\to$ $\textnormal{run-off}$. 
    
    \item Dataset 4: Sunspot area vs. global temperature anomalies (deviations from 1961-1990) (Fig. \ref{fig_supp:real_data_sunspotarea_globaTanomaly}). This is pair 0072 of the cause-effect pair database \cite{Mooij2016}. Ground truth: unclear. If there is any coupling, it is from sunspot area to global temperature.
\end{itemize}

Our test correct infers a statistically significant causal influence in the correct direction for all these datasets. One exception occurs for the sunspot-temperature data, where neither direction is significant. This may be because the putative coupling is too weak to detect with so little data, or because there is no coupling at all. Nevertheless, the dynamical coupling between sunspots and temperature on Earth is disputed.

% SOLAR RADIATION VS. AVERAGE TEMP IN GERMANY

\newpage 

% INSIDE TEMP VS. OUTSIDE TEMP

% INSIDE TEMP VS. OUTSIDE TEMP
\begin{figure*}[h]
\centering
\includegraphics[width=1.0\linewidth]{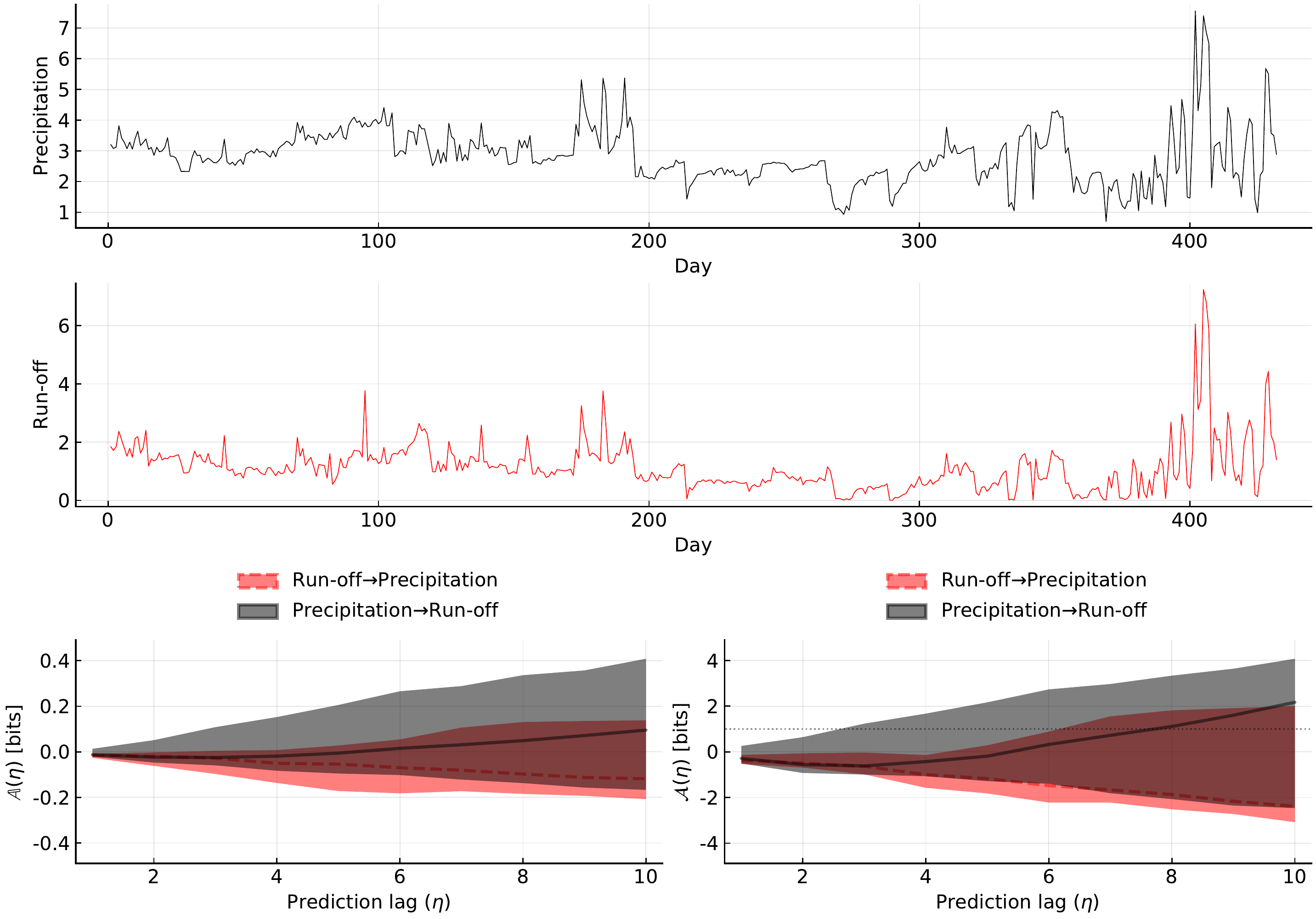}
\caption{Predictive asymmetries (lower left panel) and normalized predictive asymmetries (lower right panel) for time series of average precipitation and run-off for 438 river catchments in the US (pair 0093 of the cause-effect pair database \cite{Mooij2016}). Generalized embeddings were constructed with $k = l = m = 1$. Lines and ribbons are the median and 99 percentile confidence intervals for the sample statistic over 50 randomly selected contiguous sub-segments of the time series, where subsegments have lengths ranging from 75\% to 100\% of the total number of observations.  The significance threshold (eq. \ref{eq:normalized_asymmetry_criterion} with $f = 1.0$) is indicated by the dotted horisontal line.
}
\label{fig_supp:real_data_precipitation_runoff}
\end{figure*}

% SUNSPOT AREA vs. GLOBAL TEMP

\clearpage
\newpage

%%% Alternative Pleistocene analysis using Grant SL
\section{\label{sec_supp:pleistocene_grant}Predictive asymmetry analysis of Late Pleistocene paleoclimate records}
Here we repeat the analysis of the causal interactions among key climate system components in the Late Pleistocene using a different sea level (ice volume) record that is chronologically independent of orbital parameters \citep{grant2014sea}.

\begin{figure*}[h]
\centering\includegraphics[width=0.6\linewidth]{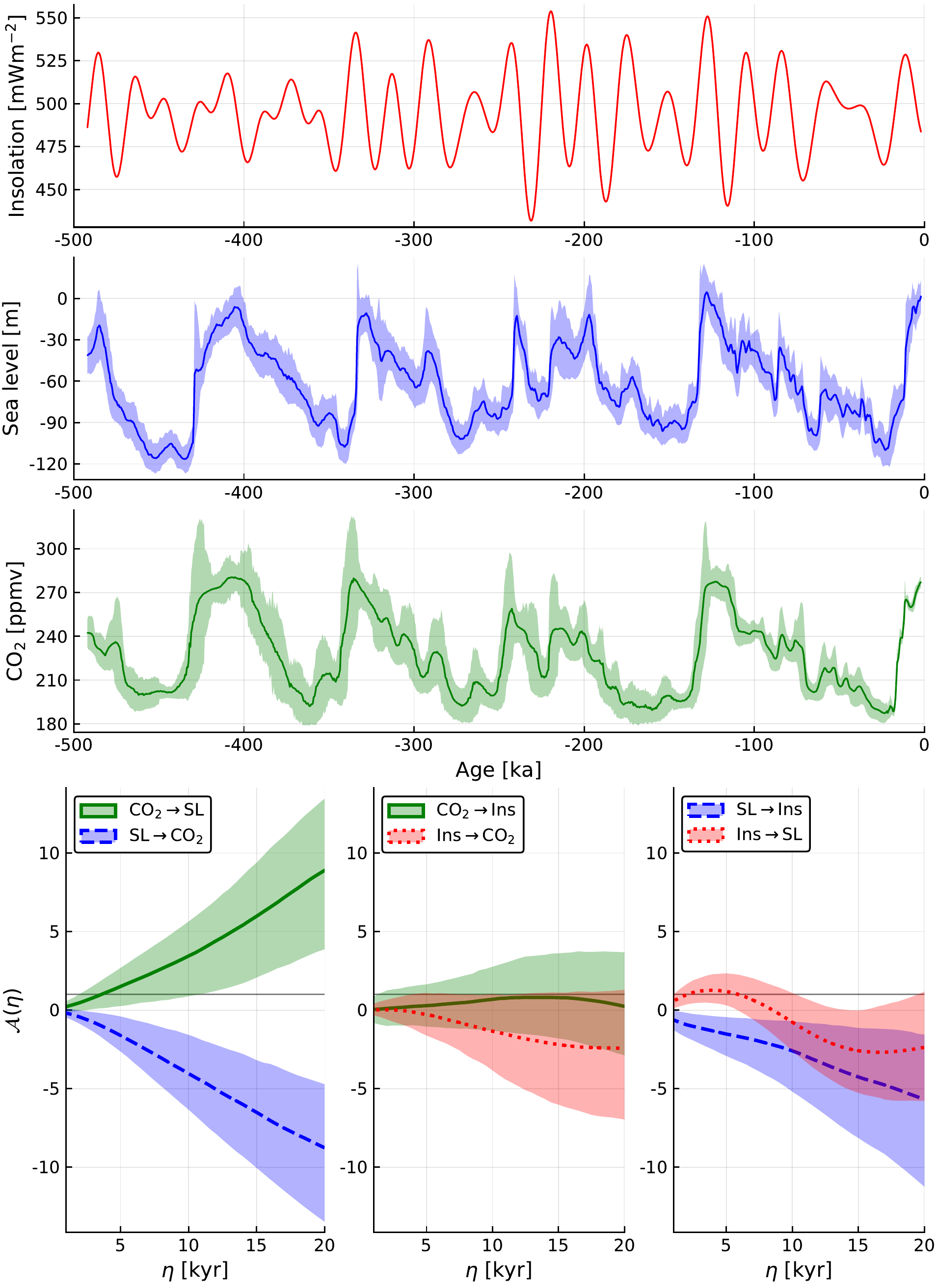}
\caption{Predictive asymmetry analysis of key climate variables over the last 500 kyr. (A) The Laskar 2004 \cite{Laskar2004long} solution for June 21 insolation at 65$^{\circ}$N.  (B) Sea level estimates for the Red Sea \cite{grant2014sea}. Values are medians and 95\% confidence ribbon representing uncertainty in both sea level estimates and ages. (C) Composite ice core record of atmospheric CO$_2$ \cite{bereiter2015revision}, with AICC2012 age model uncertainty \cite{bazin2013optimized,veres2013antarctic}. Values are medians and  95\% confidence ribbon representing uncertainty in both CO$_2$ values and ages.  Uncertainties in both sea level and CO$_2$ were computed by Monte Carlo resampling in 500-yr bins using the UncertainData.jl Julia package \cite{Haaga2019uncertaindata}. (D-F) Mean normalized predictive asymmetry $\overline{\mathcal{A}}(\eta)$ with $f=1$, computed over 1,000 randomly positioned segments, each of length ranging from 370 to 500 kyr. Ribbons represent 95\% confidence intervals from resampling within uncertainties across the ensemble of segments. 
}
\label{fig_supp:pleistocene_grant}
\end{figure*}

%%%%%%%%%%%%%%%%%%%

\clearpage

\bibliography{references}% Produces the bibliography via BibTeX.

\begin{thebibliography}{61}%
\makeatletter
\providecommand \@ifxundefined [1]{%
 \@ifx{#1\undefined}
}%
\providecommand \@ifnum [1]{%
 \ifnum #1\expandafter \@firstoftwo
 \else \expandafter \@secondoftwo
 \fi
}%
\providecommand \@ifx [1]{%
 \ifx #1\expandafter \@firstoftwo
 \else \expandafter \@secondoftwo
 \fi
}%
\providecommand \natexlab [1]{#1}%
\providecommand \enquote  [1]{``#1''}%
\providecommand \bibnamefont  [1]{#1}%
\providecommand \bibfnamefont [1]{#1}%
\providecommand \citenamefont [1]{#1}%
\providecommand \href@noop [0]{\@secondoftwo}%
\providecommand \href [0]{\begingroup \@sanitize@url \@href}%
\providecommand \@href[1]{\@@startlink{#1}\@@href}%
\providecommand \@@href[1]{\endgroup#1\@@endlink}%
\providecommand \@sanitize@url [0]{\catcode `\\12\catcode `\$12\catcode
  `\&12\catcode `\#12\catcode `\^12\catcode `\_12\catcode `\%12\relax}%
\providecommand \@@startlink[1]{}%
\providecommand \@@endlink[0]{}%
\providecommand \url  [0]{\begingroup\@sanitize@url \@url }%
\providecommand \@url [1]{\endgroup\@href {#1}{\urlprefix }}%
\providecommand \urlprefix  [0]{URL }%
\providecommand \Eprint [0]{\href }%
\providecommand \doibase [0]{https://doi.org/}%
\providecommand \selectlanguage [0]{\@gobble}%
\providecommand \bibinfo  [0]{\@secondoftwo}%
\providecommand \bibfield  [0]{\@secondoftwo}%
\providecommand \translation [1]{[#1]}%
\providecommand \BibitemOpen [0]{}%
\providecommand \bibitemStop [0]{}%
\providecommand \bibitemNoStop [0]{.\EOS\space}%
\providecommand \EOS [0]{\spacefactor3000\relax}%
\providecommand \BibitemShut  [1]{\csname bibitem#1\endcsname}%
\let\auto@bib@innerbib\@empty
%</preamble>
\bibitem [{\citenamefont {Granger}(1969)}]{Granger1969}%
  \BibitemOpen
  \bibfield  {author} {\bibinfo {author} {\bibfnamefont {C.~W.}\ \bibnamefont
  {Granger}},\ }\bibfield  {title} {\bibinfo {title} {Investigating causal
  relations by econometric models and cross-spectral methods},\ }\href@noop {}
  {\bibfield  {journal} {\bibinfo  {journal} {Econometrica: Journal of the
  Econometric Society}\ ,\ \bibinfo {pages} {424}} (\bibinfo {year}
  {1969})}\BibitemShut {NoStop}%
\bibitem [{\citenamefont {Chen}\ \emph {et~al.}(2004)\citenamefont {Chen},
  \citenamefont {Rangarajan}, \citenamefont {Feng},\ and\ \citenamefont
  {Ding}}]{Chen2004extendedgranger}%
  \BibitemOpen
  \bibfield  {author} {\bibinfo {author} {\bibfnamefont {Y.}~\bibnamefont
  {Chen}}, \bibinfo {author} {\bibfnamefont {G.}~\bibnamefont {Rangarajan}},
  \bibinfo {author} {\bibfnamefont {J.}~\bibnamefont {Feng}},\ and\ \bibinfo
  {author} {\bibfnamefont {M.}~\bibnamefont {Ding}},\ }\bibfield  {title}
  {\bibinfo {title} {Analyzing multiple nonlinear time series with extended
  granger causality},\ }\href@noop {} {\bibfield  {journal} {\bibinfo
  {journal} {Physics Letters A}\ }\textbf {\bibinfo {volume} {324}},\ \bibinfo
  {pages} {26} (\bibinfo {year} {2004})}\BibitemShut {NoStop}%
\bibitem [{\citenamefont {Marinazzo}\ \emph {et~al.}(2008)\citenamefont
  {Marinazzo}, \citenamefont {Pellicoro},\ and\ \citenamefont
  {Stramaglia}}]{Marinazzo2008kernelgranger}%
  \BibitemOpen
  \bibfield  {author} {\bibinfo {author} {\bibfnamefont {D.}~\bibnamefont
  {Marinazzo}}, \bibinfo {author} {\bibfnamefont {M.}~\bibnamefont
  {Pellicoro}},\ and\ \bibinfo {author} {\bibfnamefont {S.}~\bibnamefont
  {Stramaglia}},\ }\bibfield  {title} {\bibinfo {title} {Kernel method for
  nonlinear granger causality},\ }\href@noop {} {\bibfield  {journal} {\bibinfo
   {journal} {Physical Review Letters}\ }\textbf {\bibinfo {volume} {100}},\
  \bibinfo {pages} {144103} (\bibinfo {year} {2008})}\BibitemShut {NoStop}%
\bibitem [{\citenamefont {Schreiber}(2000)}]{Schreiber2000}%
  \BibitemOpen
  \bibfield  {author} {\bibinfo {author} {\bibfnamefont {T.}~\bibnamefont
  {Schreiber}},\ }\bibfield  {title} {\bibinfo {title} {{Measuring information
  transfer}},\ }\bibfield  {journal} {\bibinfo  {journal} {Physical Review
  Letters}\ }\href {https://doi.org/10.1103/PhysRevLett.85.461}
  {10.1103/PhysRevLett.85.461} (\bibinfo {year} {2000}),\ \Eprint
  {https://arxiv.org/abs/0001042v1} {arXiv:0001042v1 [nlin]} \BibitemShut
  {NoStop}%
\bibitem [{\citenamefont {Palu\ifmmode~\check{s}\else \v{s}\fi{}}\ and\
  \citenamefont {Vejmelka}(2007)}]{Palus2007cmi}%
  \BibitemOpen
  \bibfield  {author} {\bibinfo {author} {\bibfnamefont {M.}~\bibnamefont
  {Palu\ifmmode~\check{s}\else \v{s}\fi{}}}\ and\ \bibinfo {author}
  {\bibfnamefont {M.}~\bibnamefont {Vejmelka}},\ }\bibfield  {title} {\bibinfo
  {title} {Directionality of coupling from bivariate time series: How to avoid
  false causalities and missed connections},\ }\href
  {https://doi.org/10.1103/PhysRevE.75.056211} {\bibfield  {journal} {\bibinfo
  {journal} {Physical Review E}\ }\textbf {\bibinfo {volume} {75}},\ \bibinfo
  {pages} {056211} (\bibinfo {year} {2007})}\BibitemShut {NoStop}%
\bibitem [{\citenamefont {Rulkov}\ \emph {et~al.}(1995)\citenamefont {Rulkov},
  \citenamefont {Sushchik}, \citenamefont {Tsimring},\ and\ \citenamefont
  {Abarbanel}}]{Rulkov1995generalizedsync_closeness}%
  \BibitemOpen
  \bibfield  {author} {\bibinfo {author} {\bibfnamefont {N.~F.}\ \bibnamefont
  {Rulkov}}, \bibinfo {author} {\bibfnamefont {M.~M.}\ \bibnamefont
  {Sushchik}}, \bibinfo {author} {\bibfnamefont {L.~S.}\ \bibnamefont
  {Tsimring}},\ and\ \bibinfo {author} {\bibfnamefont {H.~D.}\ \bibnamefont
  {Abarbanel}},\ }\bibfield  {title} {\bibinfo {title} {Generalized
  synchronization of chaos in directionally coupled chaotic systems},\
  }\href@noop {} {\bibfield  {journal} {\bibinfo  {journal} {Physical Review
  E}\ }\textbf {\bibinfo {volume} {51}},\ \bibinfo {pages} {980} (\bibinfo
  {year} {1995})}\BibitemShut {NoStop}%
\bibitem [{\citenamefont {Schiff}\ \emph {et~al.}(1996)\citenamefont {Schiff},
  \citenamefont {So}, \citenamefont {Chang}, \citenamefont {Burke},\ and\
  \citenamefont {Sauer}}]{Schiff1996mutualprediction}%
  \BibitemOpen
  \bibfield  {author} {\bibinfo {author} {\bibfnamefont {S.~J.}\ \bibnamefont
  {Schiff}}, \bibinfo {author} {\bibfnamefont {P.}~\bibnamefont {So}}, \bibinfo
  {author} {\bibfnamefont {T.}~\bibnamefont {Chang}}, \bibinfo {author}
  {\bibfnamefont {R.~E.}\ \bibnamefont {Burke}},\ and\ \bibinfo {author}
  {\bibfnamefont {T.}~\bibnamefont {Sauer}},\ }\bibfield  {title} {\bibinfo
  {title} {Detecting dynamical interdependence and generalized synchrony
  through mutual prediction in a neural ensemble},\ }\href
  {https://doi.org/10.1103/PhysRevE.54.6708} {\bibfield  {journal} {\bibinfo
  {journal} {Physical Review E}\ }\textbf {\bibinfo {volume} {54}},\ \bibinfo
  {pages} {6708} (\bibinfo {year} {1996})}\BibitemShut {NoStop}%
\bibitem [{\citenamefont {Arnhold}\ \emph {et~al.}(1999)\citenamefont
  {Arnhold}, \citenamefont {Grassberger}, \citenamefont {Lehnertz},\ and\
  \citenamefont {Elger}}]{Arnhold1999robustinterdependences}%
  \BibitemOpen
  \bibfield  {author} {\bibinfo {author} {\bibfnamefont {J.}~\bibnamefont
  {Arnhold}}, \bibinfo {author} {\bibfnamefont {P.}~\bibnamefont
  {Grassberger}}, \bibinfo {author} {\bibfnamefont {K.}~\bibnamefont
  {Lehnertz}},\ and\ \bibinfo {author} {\bibfnamefont {C.~E.}\ \bibnamefont
  {Elger}},\ }\bibfield  {title} {\bibinfo {title} {A robust method for
  detecting interdependences: application to intracranially recorded eeg},\
  }\href@noop {} {\bibfield  {journal} {\bibinfo  {journal} {Physica D:
  Nonlinear Phenomena}\ }\textbf {\bibinfo {volume} {134}},\ \bibinfo {pages}
  {419} (\bibinfo {year} {1999})}\BibitemShut {NoStop}%
\bibitem [{\citenamefont {Quiroga}\ \emph {et~al.}(2002)\citenamefont
  {Quiroga}, \citenamefont {Kraskov}, \citenamefont {Kreuz},\ and\
  \citenamefont {Grassberger}}]{Quiroga2002synchronization}%
  \BibitemOpen
  \bibfield  {author} {\bibinfo {author} {\bibfnamefont {R.~Q.}\ \bibnamefont
  {Quiroga}}, \bibinfo {author} {\bibfnamefont {A.}~\bibnamefont {Kraskov}},
  \bibinfo {author} {\bibfnamefont {T.}~\bibnamefont {Kreuz}},\ and\ \bibinfo
  {author} {\bibfnamefont {P.}~\bibnamefont {Grassberger}},\ }\bibfield
  {title} {\bibinfo {title} {Performance of different synchronization measures
  in real data: a case study on electroencephalographic signals},\ }\href@noop
  {} {\bibfield  {journal} {\bibinfo  {journal} {Physical Review E}\ }\textbf
  {\bibinfo {volume} {65}},\ \bibinfo {pages} {041903} (\bibinfo {year}
  {2002})}\BibitemShut {NoStop}%
\bibitem [{\citenamefont {Chicharro}\ and\ \citenamefont
  {Andrzejak}(2009)}]{Chicharro2009directionrankstats}%
  \BibitemOpen
  \bibfield  {author} {\bibinfo {author} {\bibfnamefont {D.}~\bibnamefont
  {Chicharro}}\ and\ \bibinfo {author} {\bibfnamefont {R.~G.}\ \bibnamefont
  {Andrzejak}},\ }\bibfield  {title} {\bibinfo {title} {Reliable detection of
  directional couplings using rank statistics},\ }\href@noop {} {\bibfield
  {journal} {\bibinfo  {journal} {Physical Review E}\ }\textbf {\bibinfo
  {volume} {80}},\ \bibinfo {pages} {026217} (\bibinfo {year}
  {2009})}\BibitemShut {NoStop}%
\bibitem [{\citenamefont {Sugihara}\ \emph {et~al.}(2012)\citenamefont
  {Sugihara}, \citenamefont {May}, \citenamefont {Ye}, \citenamefont {Hsieh},
  \citenamefont {Deyle}, \citenamefont {Fogarty},\ and\ \citenamefont
  {Munch}}]{Sugihara2012}%
  \BibitemOpen
  \bibfield  {author} {\bibinfo {author} {\bibfnamefont {G.}~\bibnamefont
  {Sugihara}}, \bibinfo {author} {\bibfnamefont {R.}~\bibnamefont {May}},
  \bibinfo {author} {\bibfnamefont {H.}~\bibnamefont {Ye}}, \bibinfo {author}
  {\bibfnamefont {C.~H.}\ \bibnamefont {Hsieh}}, \bibinfo {author}
  {\bibfnamefont {E.}~\bibnamefont {Deyle}}, \bibinfo {author} {\bibfnamefont
  {M.}~\bibnamefont {Fogarty}},\ and\ \bibinfo {author} {\bibfnamefont
  {S.}~\bibnamefont {Munch}},\ }\bibfield  {title} {\bibinfo {title}
  {{Detecting causality in complex ecosystems}},\ }\bibfield  {journal}
  {\bibinfo  {journal} {Science}\ }\href
  {https://doi.org/10.1126/science.1227079} {10.1126/science.1227079} (\bibinfo
  {year} {2012}),\ \Eprint {https://arxiv.org/abs/0711.2729} {arXiv:0711.2729}
  \BibitemShut {NoStop}%
\bibitem [{\citenamefont {Wiesenfeldt}\ \emph {et~al.}(2001)\citenamefont
  {Wiesenfeldt}, \citenamefont {Parlitz},\ and\ \citenamefont
  {Lauterborn}}]{Wiesenfeldt2001mixedstateanalysis}%
  \BibitemOpen
  \bibfield  {author} {\bibinfo {author} {\bibfnamefont {M.}~\bibnamefont
  {Wiesenfeldt}}, \bibinfo {author} {\bibfnamefont {U.}~\bibnamefont
  {Parlitz}},\ and\ \bibinfo {author} {\bibfnamefont {W.}~\bibnamefont
  {Lauterborn}},\ }\bibfield  {title} {\bibinfo {title} {Mixed state analysis
  of multivariate time series},\ }\href@noop {} {\bibfield  {journal} {\bibinfo
   {journal} {International Journal of Bifurcation and Chaos}\ }\textbf
  {\bibinfo {volume} {11}},\ \bibinfo {pages} {2217} (\bibinfo {year}
  {2001})}\BibitemShut {NoStop}%
\bibitem [{\citenamefont {Feldmann}\ and\ \citenamefont
  {Bhattacharya}(2004)}]{Feldmann2004predictabilityimprovement}%
  \BibitemOpen
  \bibfield  {author} {\bibinfo {author} {\bibfnamefont {U.}~\bibnamefont
  {Feldmann}}\ and\ \bibinfo {author} {\bibfnamefont {J.}~\bibnamefont
  {Bhattacharya}},\ }\bibfield  {title} {\bibinfo {title} {Predictability
  improvement as an asymmetrical measure of interdependence in bivariate time
  series},\ }\href@noop {} {\bibfield  {journal} {\bibinfo  {journal}
  {International Journal of Bifurcation and Chaos}\ }\textbf {\bibinfo {volume}
  {14}},\ \bibinfo {pages} {505} (\bibinfo {year} {2004})}\BibitemShut
  {NoStop}%
\bibitem [{\citenamefont {Krakovsk{\'a}}\ and\ \citenamefont
  {Hanzely}(2016)}]{Krakovska2016mixredprediction}%
  \BibitemOpen
  \bibfield  {author} {\bibinfo {author} {\bibfnamefont {A.}~\bibnamefont
  {Krakovsk{\'a}}}\ and\ \bibinfo {author} {\bibfnamefont {F.}~\bibnamefont
  {Hanzely}},\ }\bibfield  {title} {\bibinfo {title} {Testing for causality in
  reconstructed state spaces by an optimized mixed prediction method},\
  }\href@noop {} {\bibfield  {journal} {\bibinfo  {journal} {Physical Review
  E}\ }\textbf {\bibinfo {volume} {94}},\ \bibinfo {pages} {052203} (\bibinfo
  {year} {2016})}\BibitemShut {NoStop}%
\bibitem [{\citenamefont {Liang}(2013)}]{Liang2013information}%
  \BibitemOpen
  \bibfield  {author} {\bibinfo {author} {\bibfnamefont {X.}~\bibnamefont
  {Liang}},\ }\bibfield  {title} {\bibinfo {title} {The liang-kleeman
  information flow: Theory and applications},\ }\href@noop {} {\bibfield
  {journal} {\bibinfo  {journal} {Entropy}\ }\textbf {\bibinfo {volume} {15}},\
  \bibinfo {pages} {327} (\bibinfo {year} {2013})}\BibitemShut {NoStop}%
\bibitem [{\citenamefont {McCracken}\ and\ \citenamefont
  {Weigel}(2016)}]{McCracken2016}%
  \BibitemOpen
  \bibfield  {author} {\bibinfo {author} {\bibfnamefont {J.~M.}\ \bibnamefont
  {McCracken}}\ and\ \bibinfo {author} {\bibfnamefont {R.~S.}\ \bibnamefont
  {Weigel}},\ }\bibfield  {title} {\bibinfo {title} {Nonparametric causal
  inference for bivariate time series},\ }\href@noop {} {\bibfield  {journal}
  {\bibinfo  {journal} {Physical Review E}\ }\textbf {\bibinfo {volume} {93}},\
  \bibinfo {pages} {022207} (\bibinfo {year} {2016})}\BibitemShut {NoStop}%
\bibitem [{\citenamefont {Ye}\ \emph {et~al.}(2015)\citenamefont {Ye},
  \citenamefont {Deyle}, \citenamefont {Gilarranz},\ and\ \citenamefont
  {Sugihara}}]{Ye2015}%
  \BibitemOpen
  \bibfield  {author} {\bibinfo {author} {\bibfnamefont {H.}~\bibnamefont
  {Ye}}, \bibinfo {author} {\bibfnamefont {E.~R.}\ \bibnamefont {Deyle}},
  \bibinfo {author} {\bibfnamefont {L.~J.}\ \bibnamefont {Gilarranz}},\ and\
  \bibinfo {author} {\bibfnamefont {G.}~\bibnamefont {Sugihara}},\ }\bibfield
  {title} {\bibinfo {title} {{Distinguishing time-delayed causal interactions
  using convergent cross mapping}},\ }\bibfield  {journal} {\bibinfo  {journal}
  {Scientific Reports}\ }\href {https://doi.org/10.1038/srep14750}
  {10.1038/srep14750} (\bibinfo {year} {2015})\BibitemShut {NoStop}%
\bibitem [{\citenamefont {Krakovsk{\'{a}}}\ \emph {et~al.}(2018)\citenamefont
  {Krakovsk{\'{a}}}, \citenamefont {Jakub{\'{i}}k}, \citenamefont
  {Chvostekov{\'{a}}}, \citenamefont {Coufal}, \citenamefont {Jajcay},\ and\
  \citenamefont {Palu{\v{s}}}}]{Krakovska2018}%
  \BibitemOpen
  \bibfield  {author} {\bibinfo {author} {\bibfnamefont {A.}~\bibnamefont
  {Krakovsk{\'{a}}}}, \bibinfo {author} {\bibfnamefont {J.}~\bibnamefont
  {Jakub{\'{i}}k}}, \bibinfo {author} {\bibfnamefont {M.}~\bibnamefont
  {Chvostekov{\'{a}}}}, \bibinfo {author} {\bibfnamefont {D.}~\bibnamefont
  {Coufal}}, \bibinfo {author} {\bibfnamefont {N.}~\bibnamefont {Jajcay}},\
  and\ \bibinfo {author} {\bibfnamefont {M.}~\bibnamefont {Palu{\v{s}}}},\
  }\bibfield  {title} {\bibinfo {title} {{Comparison of six methods for the
  detection of causality in a bivariate time series}},\ }\href
  {https://doi.org/10.1103/PhysRevE.97.042207} {\bibfield  {journal} {\bibinfo
  {journal} {Physical Review E}\ }\textbf {\bibinfo {volume} {97}},\ \bibinfo
  {pages} {042207} (\bibinfo {year} {2018})}\BibitemShut {NoStop}%
\bibitem [{\citenamefont {Smirnov}(2013)}]{Smirnov2013spurious}%
  \BibitemOpen
  \bibfield  {author} {\bibinfo {author} {\bibfnamefont {D.~A.}\ \bibnamefont
  {Smirnov}},\ }\bibfield  {title} {\bibinfo {title} {Spurious causalities with
  transfer entropy},\ }\href@noop {} {\bibfield  {journal} {\bibinfo  {journal}
  {Physical Review E}\ }\textbf {\bibinfo {volume} {87}},\ \bibinfo {pages}
  {042917} (\bibinfo {year} {2013})}\BibitemShut {NoStop}%
\bibitem [{\citenamefont {Theiler}\ \emph {et~al.}(1992)\citenamefont
  {Theiler}, \citenamefont {Eubank}, \citenamefont {Longtin}, \citenamefont
  {Galdrikian},\ and\ \citenamefont {{Doyne Farmer}}}]{Theiler1992}%
  \BibitemOpen
  \bibfield  {author} {\bibinfo {author} {\bibfnamefont {J.}~\bibnamefont
  {Theiler}}, \bibinfo {author} {\bibfnamefont {S.}~\bibnamefont {Eubank}},
  \bibinfo {author} {\bibfnamefont {A.}~\bibnamefont {Longtin}}, \bibinfo
  {author} {\bibfnamefont {B.}~\bibnamefont {Galdrikian}},\ and\ \bibinfo
  {author} {\bibfnamefont {J.}~\bibnamefont {{Doyne Farmer}}},\ }\bibfield
  {title} {\bibinfo {title} {{Testing for nonlinearity in time series: the
  method of surrogate data}},\ }\bibfield  {journal} {\bibinfo  {journal}
  {Physica D: Nonlinear Phenomena}\ }\href
  {https://doi.org/10.1016/0167-2789(92)90102-S} {10.1016/0167-2789(92)90102-S}
  (\bibinfo {year} {1992}),\ \Eprint {https://arxiv.org/abs/9909037}
  {arXiv:9909037 [chao-dyn]} \BibitemShut {NoStop}%
\bibitem [{\citenamefont {Lancaster}\ \emph {et~al.}(2018)\citenamefont
  {Lancaster}, \citenamefont {Iatsenko}, \citenamefont {Pidde}, \citenamefont
  {Ticcinelli},\ and\ \citenamefont {Stefanovska}}]{Lancaster2018}%
  \BibitemOpen
  \bibfield  {author} {\bibinfo {author} {\bibfnamefont {G.}~\bibnamefont
  {Lancaster}}, \bibinfo {author} {\bibfnamefont {D.}~\bibnamefont {Iatsenko}},
  \bibinfo {author} {\bibfnamefont {A.}~\bibnamefont {Pidde}}, \bibinfo
  {author} {\bibfnamefont {V.}~\bibnamefont {Ticcinelli}},\ and\ \bibinfo
  {author} {\bibfnamefont {A.}~\bibnamefont {Stefanovska}},\ }\bibfield
  {title} {\bibinfo {title} {{Surrogate data for hypothesis testing of physical
  systems}},\ }\bibfield  {journal} {\bibinfo  {journal} {Physics Reports}\
  }\href {https://doi.org/10.1016/j.physrep.2018.06.001}
  {10.1016/j.physrep.2018.06.001} (\bibinfo {year} {2018})\BibitemShut
  {NoStop}%
\bibitem [{\citenamefont {James}\ \emph {et~al.}(2016)\citenamefont {James},
  \citenamefont {Barnett},\ and\ \citenamefont
  {Crutchfield}}]{James2016information}%
  \BibitemOpen
  \bibfield  {author} {\bibinfo {author} {\bibfnamefont {R.~G.}\ \bibnamefont
  {James}}, \bibinfo {author} {\bibfnamefont {N.}~\bibnamefont {Barnett}},\
  and\ \bibinfo {author} {\bibfnamefont {J.~P.}\ \bibnamefont {Crutchfield}},\
  }\bibfield  {title} {\bibinfo {title} {Information flows? a critique of
  transfer entropies},\ }\href@noop {} {\bibfield  {journal} {\bibinfo
  {journal} {Physical {R}eview {L}etters}\ }\textbf {\bibinfo {volume} {116}},\
  \bibinfo {pages} {238701} (\bibinfo {year} {2016})}\BibitemShut {NoStop}%
\bibitem [{\citenamefont {Hahs}\ and\ \citenamefont {Pethel}(2013)}]{Hahs2013}%
  \BibitemOpen
  \bibfield  {author} {\bibinfo {author} {\bibfnamefont {D.~W.}\ \bibnamefont
  {Hahs}}\ and\ \bibinfo {author} {\bibfnamefont {S.~D.}\ \bibnamefont
  {Pethel}},\ }\bibfield  {title} {\bibinfo {title} {{Transfer entropy for
  coupled autoregressive processes}},\ }\href
  {https://doi.org/10.3390/e15030767} {\bibfield  {journal} {\bibinfo
  {journal} {Entropy}\ }\textbf {\bibinfo {volume} {15}},\ \bibinfo {pages}
  {767} (\bibinfo {year} {2013})}\BibitemShut {NoStop}%
\bibitem [{\citenamefont {Barnett}\ \emph {et~al.}(2009)\citenamefont
  {Barnett}, \citenamefont {Barrett},\ and\ \citenamefont
  {Seth}}]{Barnett2009}%
  \BibitemOpen
  \bibfield  {author} {\bibinfo {author} {\bibfnamefont {L.}~\bibnamefont
  {Barnett}}, \bibinfo {author} {\bibfnamefont {A.~B.}\ \bibnamefont
  {Barrett}},\ and\ \bibinfo {author} {\bibfnamefont {A.~K.}\ \bibnamefont
  {Seth}},\ }\bibfield  {title} {\bibinfo {title} {Granger causality and
  transfer entropy are equivalent for gaussian variables},\ }\href
  {https://doi.org/10.1103/PhysRevLett.103.238701} {\bibfield  {journal}
  {\bibinfo  {journal} {Phys. Rev. Lett.}\ }\textbf {\bibinfo {volume} {103}},\
  \bibinfo {pages} {238701} (\bibinfo {year} {2009})}\BibitemShut {NoStop}%
\bibitem [{\citenamefont {Hannisdal}(2011)}]{hannisdal2011non}%
  \BibitemOpen
  \bibfield  {author} {\bibinfo {author} {\bibfnamefont {B.}~\bibnamefont
  {Hannisdal}},\ }\bibfield  {title} {\bibinfo {title} {Non-parametric
  inference of causal interactions from geological records},\ }\href@noop {}
  {\bibfield  {journal} {\bibinfo  {journal} {American Journal of Science}\
  }\textbf {\bibinfo {volume} {311}},\ \bibinfo {pages} {315} (\bibinfo {year}
  {2011})}\BibitemShut {NoStop}%
\bibitem [{\citenamefont {Diego}\ \emph {et~al.}(2018)\citenamefont {Diego},
  \citenamefont {Haaga},\ and\ \citenamefont {Hannisdal}}]{Diego2018}%
  \BibitemOpen
  \bibfield  {author} {\bibinfo {author} {\bibfnamefont {D.}~\bibnamefont
  {Diego}}, \bibinfo {author} {\bibfnamefont {K.~A.}\ \bibnamefont {Haaga}},\
  and\ \bibinfo {author} {\bibfnamefont {B.}~\bibnamefont {Hannisdal}},\
  }\bibfield  {title} {\bibinfo {title} {{Transfer entropy computation using
  the Perron-Frobenius operator}},\ }\href {https://arxiv.org/abs/1811.01677}
  {\  (\bibinfo {year} {2018})},\ \Eprint {https://arxiv.org/abs/1811.01677}
  {arXiv:1811.01677} \BibitemShut {NoStop}%
\bibitem [{\citenamefont {Matthews}(1975)}]{Matthews1975corr}%
  \BibitemOpen
  \bibfield  {author} {\bibinfo {author} {\bibfnamefont {B.~W.}\ \bibnamefont
  {Matthews}},\ }\bibfield  {title} {\bibinfo {title} {Comparison of the
  predicted and observed secondary structure of t4 phage lysozyme},\
  }\href@noop {} {\bibfield  {journal} {\bibinfo  {journal} {Biochimica et
  Biophysica Acta (BBA)-Protein Structure}\ }\textbf {\bibinfo {volume}
  {405}},\ \bibinfo {pages} {442} (\bibinfo {year} {1975})}\BibitemShut
  {NoStop}%
\bibitem [{\citenamefont {Chicco}(2017)}]{Chicco2017tenmachinetips}%
  \BibitemOpen
  \bibfield  {author} {\bibinfo {author} {\bibfnamefont {D.}~\bibnamefont
  {Chicco}},\ }\bibfield  {title} {\bibinfo {title} {Ten quick tips for machine
  learning in computational biology},\ }\href@noop {} {\bibfield  {journal}
  {\bibinfo  {journal} {BioData mining}\ }\textbf {\bibinfo {volume} {10}},\
  \bibinfo {pages} {35} (\bibinfo {year} {2017})}\BibitemShut {NoStop}%
\bibitem [{\citenamefont {Haaga}(2019)}]{Haaga2019uncertaindata}%
  \BibitemOpen
  \bibfield  {author} {\bibinfo {author} {\bibfnamefont {K.}~\bibnamefont
  {Haaga}},\ }\bibfield  {title} {\bibinfo {title} {Uncertaindata. jl: a julia
  package for working with measurements and datasets with uncertainties.},\
  }\href@noop {} {\bibfield  {journal} {\bibinfo  {journal} {Journal of Open
  Source Software}\ }\textbf {\bibinfo {volume} {4}},\ \bibinfo {pages} {1666}
  (\bibinfo {year} {2019})}\BibitemShut {NoStop}%
\bibitem [{\citenamefont {Esmark}(1824)}]{esmark1824bidrag}%
  \BibitemOpen
  \bibfield  {author} {\bibinfo {author} {\bibfnamefont {J.}~\bibnamefont
  {Esmark}},\ }\bibfield  {title} {\bibinfo {title} {Bidrag til vor jordklodes
  historie},\ }\href@noop {} {\bibfield  {journal} {\bibinfo  {journal}
  {Magazin for Naturvidenskaberne}\ }\textbf {\bibinfo {volume} {2}},\ \bibinfo
  {pages} {28} (\bibinfo {year} {1824})}\BibitemShut {NoStop}%
\bibitem [{\citenamefont {Croll}(1875)}]{croll1875climate}%
  \BibitemOpen
  \bibfield  {author} {\bibinfo {author} {\bibfnamefont {J.}~\bibnamefont
  {Croll}},\ }\bibfield  {title} {\bibinfo {title} {Climate and time},\
  }\href@noop {} {\bibfield  {journal} {\bibinfo  {journal} {Nature}\ }\textbf
  {\bibinfo {volume} {12}},\ \bibinfo {pages} {329} (\bibinfo {year}
  {1875})}\BibitemShut {NoStop}%
\bibitem [{\citenamefont {Milankovitch}(1941)}]{Milankovitch1941}%
  \BibitemOpen
  \bibfield  {author} {\bibinfo {author} {\bibfnamefont {M.}~\bibnamefont
  {Milankovitch}},\ }\href@noop {} {\emph {\bibinfo {title} {{Kanon der
  Erdebestrahlung und seine Anwendung auf das Eiszeitenproblem}}}}\ (\bibinfo
  {publisher} {K{\"o}niglich Serbische Akademie},\ \bibinfo {year}
  {1941})\BibitemShut {NoStop}%
\bibitem [{\citenamefont {Hays}\ \emph {et~al.}(1976)\citenamefont {Hays},
  \citenamefont {Imbrie},\ and\ \citenamefont {Shackleton}}]{Hays1976}%
  \BibitemOpen
  \bibfield  {author} {\bibinfo {author} {\bibfnamefont {J.~D.}\ \bibnamefont
  {Hays}}, \bibinfo {author} {\bibfnamefont {J.}~\bibnamefont {Imbrie}},\ and\
  \bibinfo {author} {\bibfnamefont {N.~J.}\ \bibnamefont {Shackleton}},\
  }\bibfield  {title} {\bibinfo {title} {{Variations in the Earth's orbit:
  pacemaker of the ice ages}},\ }\href@noop {} {\bibfield  {journal} {\bibinfo
  {journal} {{Science}}\ }\textbf {\bibinfo {volume} {194}},\ \bibinfo {pages}
  {1121} (\bibinfo {year} {1976})}\BibitemShut {NoStop}%
\bibitem [{\citenamefont {Pisias}\ and\ \citenamefont
  {Moore~Jr}(1981)}]{pisias1981evolution}%
  \BibitemOpen
  \bibfield  {author} {\bibinfo {author} {\bibfnamefont {N.~G.}\ \bibnamefont
  {Pisias}}\ and\ \bibinfo {author} {\bibfnamefont {T.}~\bibnamefont
  {Moore~Jr}},\ }\bibfield  {title} {\bibinfo {title} {The evolution of
  pleistocene climate: a time series approach},\ }\href@noop {} {\bibfield
  {journal} {\bibinfo  {journal} {Earth and Planetary Science Letters}\
  }\textbf {\bibinfo {volume} {52}},\ \bibinfo {pages} {450} (\bibinfo {year}
  {1981})}\BibitemShut {NoStop}%
\bibitem [{\citenamefont {Raymo}(1997)}]{Raymo1997}%
  \BibitemOpen
  \bibfield  {author} {\bibinfo {author} {\bibfnamefont {M.}~\bibnamefont
  {Raymo}},\ }\bibfield  {title} {\bibinfo {title} {{The timing of major
  climate terminations}},\ }\href@noop {} {\bibfield  {journal} {\bibinfo
  {journal} {Paleoceanography}\ }\textbf {\bibinfo {volume} {12}},\ \bibinfo
  {pages} {577} (\bibinfo {year} {1997})}\BibitemShut {NoStop}%
\bibitem [{\citenamefont {Tzedakis}\ \emph {et~al.}(2017)\citenamefont
  {Tzedakis}, \citenamefont {Crucifix}, \citenamefont {Mitsui},\ and\
  \citenamefont {Wolff}}]{tzedakis2017simple}%
  \BibitemOpen
  \bibfield  {author} {\bibinfo {author} {\bibfnamefont {P.}~\bibnamefont
  {Tzedakis}}, \bibinfo {author} {\bibfnamefont {M.}~\bibnamefont {Crucifix}},
  \bibinfo {author} {\bibfnamefont {T.}~\bibnamefont {Mitsui}},\ and\ \bibinfo
  {author} {\bibfnamefont {E.~W.}\ \bibnamefont {Wolff}},\ }\bibfield  {title}
  {\bibinfo {title} {A simple rule to determine which insolation cycles lead to
  interglacials},\ }\href@noop {} {\bibfield  {journal} {\bibinfo  {journal}
  {Nature}\ }\textbf {\bibinfo {volume} {542}},\ \bibinfo {pages} {427}
  (\bibinfo {year} {2017})}\BibitemShut {NoStop}%
\bibitem [{\citenamefont {Denton}\ \emph {et~al.}(2010)\citenamefont {Denton},
  \citenamefont {Anderson}, \citenamefont {Toggweiler}, \citenamefont
  {Edwards}, \citenamefont {Schaefer},\ and\ \citenamefont
  {Putnam}}]{denton2010last}%
  \BibitemOpen
  \bibfield  {author} {\bibinfo {author} {\bibfnamefont {G.~H.}\ \bibnamefont
  {Denton}}, \bibinfo {author} {\bibfnamefont {R.~F.}\ \bibnamefont
  {Anderson}}, \bibinfo {author} {\bibfnamefont {J.}~\bibnamefont
  {Toggweiler}}, \bibinfo {author} {\bibfnamefont {R.}~\bibnamefont {Edwards}},
  \bibinfo {author} {\bibfnamefont {J.}~\bibnamefont {Schaefer}},\ and\
  \bibinfo {author} {\bibfnamefont {A.}~\bibnamefont {Putnam}},\ }\bibfield
  {title} {\bibinfo {title} {The last glacial termination},\ }\href@noop {}
  {\bibfield  {journal} {\bibinfo  {journal} {science}\ }\textbf {\bibinfo
  {volume} {328}},\ \bibinfo {pages} {1652} (\bibinfo {year}
  {2010})}\BibitemShut {NoStop}%
\bibitem [{\citenamefont {Wolff}\ \emph {et~al.}(2009)\citenamefont {Wolff},
  \citenamefont {Fischer},\ and\ \citenamefont
  {R{\"o}thlisberger}}]{Wolff2009}%
  \BibitemOpen
  \bibfield  {author} {\bibinfo {author} {\bibfnamefont {E.}~\bibnamefont
  {Wolff}}, \bibinfo {author} {\bibfnamefont {H.}~\bibnamefont {Fischer}},\
  and\ \bibinfo {author} {\bibfnamefont {R.}~\bibnamefont
  {R{\"o}thlisberger}},\ }\bibfield  {title} {\bibinfo {title} {{Glacial
  terminations as southern warmings without northern control}},\ }\href@noop {}
  {\bibfield  {journal} {\bibinfo  {journal} {Nature Geoscience}\ }\textbf
  {\bibinfo {volume} {2}},\ \bibinfo {pages} {206} (\bibinfo {year}
  {2009})}\BibitemShut {NoStop}%
\bibitem [{\citenamefont {Petit}\ \emph {et~al.}(1999)\citenamefont {Petit},
  \citenamefont {Jouzel}, \citenamefont {Raynaud}, \citenamefont {Barkov},
  \citenamefont {Barnola}, \citenamefont {Basile}, \citenamefont {Bender},
  \citenamefont {Chappellaz}, \citenamefont {Davis}, \citenamefont {Delaygue}
  \emph {et~al.}}]{petit1999climate}%
  \BibitemOpen
  \bibfield  {author} {\bibinfo {author} {\bibfnamefont {J.-R.}\ \bibnamefont
  {Petit}}, \bibinfo {author} {\bibfnamefont {J.}~\bibnamefont {Jouzel}},
  \bibinfo {author} {\bibfnamefont {D.}~\bibnamefont {Raynaud}}, \bibinfo
  {author} {\bibfnamefont {N.~I.}\ \bibnamefont {Barkov}}, \bibinfo {author}
  {\bibfnamefont {J.-M.}\ \bibnamefont {Barnola}}, \bibinfo {author}
  {\bibfnamefont {I.}~\bibnamefont {Basile}}, \bibinfo {author} {\bibfnamefont
  {M.}~\bibnamefont {Bender}}, \bibinfo {author} {\bibfnamefont
  {J.}~\bibnamefont {Chappellaz}}, \bibinfo {author} {\bibfnamefont
  {M.}~\bibnamefont {Davis}}, \bibinfo {author} {\bibfnamefont
  {G.}~\bibnamefont {Delaygue}}, \emph {et~al.},\ }\bibfield  {title} {\bibinfo
  {title} {Climate and atmospheric history of the past 420,000 years from the
  vostok ice core, antarctica},\ }\href@noop {} {\bibfield  {journal} {\bibinfo
   {journal} {Nature}\ }\textbf {\bibinfo {volume} {399}},\ \bibinfo {pages}
  {429} (\bibinfo {year} {1999})}\BibitemShut {NoStop}%
\bibitem [{\citenamefont {Shackleton}(2000)}]{shackleton2000100}%
  \BibitemOpen
  \bibfield  {author} {\bibinfo {author} {\bibfnamefont {N.~J.}\ \bibnamefont
  {Shackleton}},\ }\bibfield  {title} {\bibinfo {title} {The 100,000-year
  ice-age cycle identified and found to lag temperature, carbon dioxide, and
  orbital eccentricity},\ }\href@noop {} {\bibfield  {journal} {\bibinfo
  {journal} {Science}\ }\textbf {\bibinfo {volume} {289}},\ \bibinfo {pages}
  {1897} (\bibinfo {year} {2000})}\BibitemShut {NoStop}%
\bibitem [{\citenamefont {Shakun}\ \emph {et~al.}(2012)\citenamefont {Shakun},
  \citenamefont {Clark}, \citenamefont {He}, \citenamefont {Marcott},
  \citenamefont {Mix}, \citenamefont {Liu}, \citenamefont {Otto-Bliesner},
  \citenamefont {Schmittner},\ and\ \citenamefont {Bard}}]{shakun2012global}%
  \BibitemOpen
  \bibfield  {author} {\bibinfo {author} {\bibfnamefont {J.~D.}\ \bibnamefont
  {Shakun}}, \bibinfo {author} {\bibfnamefont {P.~U.}\ \bibnamefont {Clark}},
  \bibinfo {author} {\bibfnamefont {F.}~\bibnamefont {He}}, \bibinfo {author}
  {\bibfnamefont {S.~A.}\ \bibnamefont {Marcott}}, \bibinfo {author}
  {\bibfnamefont {A.~C.}\ \bibnamefont {Mix}}, \bibinfo {author} {\bibfnamefont
  {Z.}~\bibnamefont {Liu}}, \bibinfo {author} {\bibfnamefont {B.}~\bibnamefont
  {Otto-Bliesner}}, \bibinfo {author} {\bibfnamefont {A.}~\bibnamefont
  {Schmittner}},\ and\ \bibinfo {author} {\bibfnamefont {E.}~\bibnamefont
  {Bard}},\ }\bibfield  {title} {\bibinfo {title} {Global warming preceded by
  increasing carbon dioxide concentrations during the last deglaciation},\
  }\href@noop {} {\bibfield  {journal} {\bibinfo  {journal} {Nature}\ }\textbf
  {\bibinfo {volume} {484}},\ \bibinfo {pages} {49} (\bibinfo {year}
  {2012})}\BibitemShut {NoStop}%
\bibitem [{\citenamefont {Abe-Ouchi}\ \emph {et~al.}(2013)\citenamefont
  {Abe-Ouchi}, \citenamefont {Saito}, \citenamefont {Kawamura}, \citenamefont
  {Raymo}, \citenamefont {Okuno}, \citenamefont {Takahashi},\ and\
  \citenamefont {Blatter}}]{Abe-Ouchi2013}%
  \BibitemOpen
  \bibfield  {author} {\bibinfo {author} {\bibfnamefont {A.}~\bibnamefont
  {Abe-Ouchi}}, \bibinfo {author} {\bibfnamefont {F.}~\bibnamefont {Saito}},
  \bibinfo {author} {\bibfnamefont {K.}~\bibnamefont {Kawamura}}, \bibinfo
  {author} {\bibfnamefont {M.~E.}\ \bibnamefont {Raymo}}, \bibinfo {author}
  {\bibfnamefont {J.}~\bibnamefont {Okuno}}, \bibinfo {author} {\bibfnamefont
  {K.}~\bibnamefont {Takahashi}},\ and\ \bibinfo {author} {\bibfnamefont
  {H.}~\bibnamefont {Blatter}},\ }\bibfield  {title} {\bibinfo {title}
  {{Insolation-driven 100,000-year glacial cycles and hysteresis of ice-sheet
  volume}},\ }\href@noop {} {\bibfield  {journal} {\bibinfo  {journal}
  {{Nature}}\ }\textbf {\bibinfo {volume} {500}},\ \bibinfo {pages} {190}
  (\bibinfo {year} {2013})}\BibitemShut {NoStop}%
\bibitem [{\citenamefont {Laskar}\ \emph {et~al.}(2004)\citenamefont {Laskar},
  \citenamefont {Robutel}, \citenamefont {Joutel}, \citenamefont {Gastineau},
  \citenamefont {Correia},\ and\ \citenamefont {Levrard}}]{Laskar2004long}%
  \BibitemOpen
  \bibfield  {author} {\bibinfo {author} {\bibfnamefont {J.}~\bibnamefont
  {Laskar}}, \bibinfo {author} {\bibfnamefont {P.}~\bibnamefont {Robutel}},
  \bibinfo {author} {\bibfnamefont {F.}~\bibnamefont {Joutel}}, \bibinfo
  {author} {\bibfnamefont {M.}~\bibnamefont {Gastineau}}, \bibinfo {author}
  {\bibfnamefont {A.}~\bibnamefont {Correia}},\ and\ \bibinfo {author}
  {\bibfnamefont {B.}~\bibnamefont {Levrard}},\ }\bibfield  {title} {\bibinfo
  {title} {A long-term numerical solution for the insolation quantities of the
  earth},\ }\href@noop {} {\bibfield  {journal} {\bibinfo  {journal} {Astronomy
  \& Astrophysics}\ }\textbf {\bibinfo {volume} {428}},\ \bibinfo {pages} {261}
  (\bibinfo {year} {2004})}\BibitemShut {NoStop}%
\bibitem [{\citenamefont {Huybers}\ and\ \citenamefont
  {Wunsch}(2005)}]{Huybers2005}%
  \BibitemOpen
  \bibfield  {author} {\bibinfo {author} {\bibfnamefont {P.}~\bibnamefont
  {Huybers}}\ and\ \bibinfo {author} {\bibfnamefont {C.}~\bibnamefont
  {Wunsch}},\ }\bibfield  {title} {\bibinfo {title} {{Obliquity pacing of the
  late Pleistocene glacial terminations}},\ }\href@noop {} {\bibfield
  {journal} {\bibinfo  {journal} {{Nature}}\ }\textbf {\bibinfo {volume}
  {434}},\ \bibinfo {pages} {491} (\bibinfo {year} {2005})}\BibitemShut
  {NoStop}%
\bibitem [{\citenamefont {Haaga}\ \emph {et~al.}(2018)\citenamefont {Haaga},
  \citenamefont {Brendryen}, \citenamefont {Diego},\ and\ \citenamefont
  {Hannisdal}}]{Haaga2018}%
  \BibitemOpen
  \bibfield  {author} {\bibinfo {author} {\bibfnamefont {K.~A.}\ \bibnamefont
  {Haaga}}, \bibinfo {author} {\bibfnamefont {J.}~\bibnamefont {Brendryen}},
  \bibinfo {author} {\bibfnamefont {D.}~\bibnamefont {Diego}},\ and\ \bibinfo
  {author} {\bibfnamefont {B.}~\bibnamefont {Hannisdal}},\ }\bibfield  {title}
  {\bibinfo {title} {{Forcing of late Pleistocene ice volume by spatially
  variable summer energy}},\ }\bibfield  {journal} {\bibinfo  {journal}
  {Scientific Reports}\ }\href {https://doi.org/10.1038/s41598-018-29916-3}
  {10.1038/s41598-018-29916-3} (\bibinfo {year} {2018})\BibitemShut {NoStop}%
\bibitem [{\citenamefont {Bereiter}\ \emph {et~al.}(2015)\citenamefont
  {Bereiter}, \citenamefont {Eggleston}, \citenamefont {Schmitt}, \citenamefont
  {Nehrbass-Ahles}, \citenamefont {Stocker}, \citenamefont {Fischer},
  \citenamefont {Kipfstuhl},\ and\ \citenamefont
  {Chappellaz}}]{bereiter2015revision}%
  \BibitemOpen
  \bibfield  {author} {\bibinfo {author} {\bibfnamefont {B.}~\bibnamefont
  {Bereiter}}, \bibinfo {author} {\bibfnamefont {S.}~\bibnamefont {Eggleston}},
  \bibinfo {author} {\bibfnamefont {J.}~\bibnamefont {Schmitt}}, \bibinfo
  {author} {\bibfnamefont {C.}~\bibnamefont {Nehrbass-Ahles}}, \bibinfo
  {author} {\bibfnamefont {T.~F.}\ \bibnamefont {Stocker}}, \bibinfo {author}
  {\bibfnamefont {H.}~\bibnamefont {Fischer}}, \bibinfo {author} {\bibfnamefont
  {S.}~\bibnamefont {Kipfstuhl}},\ and\ \bibinfo {author} {\bibfnamefont
  {J.}~\bibnamefont {Chappellaz}},\ }\bibfield  {title} {\bibinfo {title}
  {Revision of the epica dome c co2 record from 800 to 600 kyr before
  present},\ }\href@noop {} {\bibfield  {journal} {\bibinfo  {journal}
  {Geophysical Research Letters}\ }\textbf {\bibinfo {volume} {42}},\ \bibinfo
  {pages} {542} (\bibinfo {year} {2015})}\BibitemShut {NoStop}%
\bibitem [{\citenamefont {Bazin}\ \emph {et~al.}(2013)\citenamefont {Bazin},
  \citenamefont {Landais}, \citenamefont {Lemieux-Dudon}, \citenamefont
  {Toy{\'e} Mahamadou~Kele}, \citenamefont {Veres}, \citenamefont {Parrenin},
  \citenamefont {Martinerie}, \citenamefont {Ritz}, \citenamefont {Capron},
  \citenamefont {Lipenkov} \emph {et~al.}}]{bazin2013optimized}%
  \BibitemOpen
  \bibfield  {author} {\bibinfo {author} {\bibfnamefont {L.}~\bibnamefont
  {Bazin}}, \bibinfo {author} {\bibfnamefont {A.}~\bibnamefont {Landais}},
  \bibinfo {author} {\bibfnamefont {B.}~\bibnamefont {Lemieux-Dudon}}, \bibinfo
  {author} {\bibfnamefont {H.}~\bibnamefont {Toy{\'e} Mahamadou~Kele}},
  \bibinfo {author} {\bibfnamefont {D.}~\bibnamefont {Veres}}, \bibinfo
  {author} {\bibfnamefont {F.}~\bibnamefont {Parrenin}}, \bibinfo {author}
  {\bibfnamefont {P.}~\bibnamefont {Martinerie}}, \bibinfo {author}
  {\bibfnamefont {C.}~\bibnamefont {Ritz}}, \bibinfo {author} {\bibfnamefont
  {E.}~\bibnamefont {Capron}}, \bibinfo {author} {\bibfnamefont
  {V.}~\bibnamefont {Lipenkov}}, \emph {et~al.},\ }\bibfield  {title} {\bibinfo
  {title} {An optimized multi-proxy, multi-site antarctic ice and gas orbital
  chronology (aicc2012): 120-800 ka},\ }\href@noop {} {\bibfield  {journal}
  {\bibinfo  {journal} {Climate of the Past}\ }\textbf {\bibinfo {volume}
  {9}},\ \bibinfo {pages} {1715–} (\bibinfo {year} {2013})}\BibitemShut
  {NoStop}%
\bibitem [{\citenamefont {Veres}\ \emph {et~al.}(2013)\citenamefont {Veres},
  \citenamefont {Bazin}, \citenamefont {Landais}, \citenamefont {Toy{\'e}
  Mahamadou~Kele}, \citenamefont {Lemieux-Dudon}, \citenamefont {Parrenin},
  \citenamefont {Martinerie}, \citenamefont {Blayo}, \citenamefont {Blunier},
  \citenamefont {Capron} \emph {et~al.}}]{veres2013antarctic}%
  \BibitemOpen
  \bibfield  {author} {\bibinfo {author} {\bibfnamefont {D.}~\bibnamefont
  {Veres}}, \bibinfo {author} {\bibfnamefont {L.}~\bibnamefont {Bazin}},
  \bibinfo {author} {\bibfnamefont {A.}~\bibnamefont {Landais}}, \bibinfo
  {author} {\bibfnamefont {H.}~\bibnamefont {Toy{\'e} Mahamadou~Kele}},
  \bibinfo {author} {\bibfnamefont {B.}~\bibnamefont {Lemieux-Dudon}}, \bibinfo
  {author} {\bibfnamefont {F.}~\bibnamefont {Parrenin}}, \bibinfo {author}
  {\bibfnamefont {P.}~\bibnamefont {Martinerie}}, \bibinfo {author}
  {\bibfnamefont {E.}~\bibnamefont {Blayo}}, \bibinfo {author} {\bibfnamefont
  {T.}~\bibnamefont {Blunier}}, \bibinfo {author} {\bibfnamefont
  {E.}~\bibnamefont {Capron}}, \emph {et~al.},\ }\bibfield  {title} {\bibinfo
  {title} {The antarctic ice core chronology (aicc2012): an optimized
  multi-parameter and multi-site dating approach for the last 120 thousand
  years},\ }\href@noop {} {\bibfield  {journal} {\bibinfo  {journal} {Climate
  of the Past}\ }\textbf {\bibinfo {volume} {9}},\ \bibinfo {pages} {1733}
  (\bibinfo {year} {2013})}\BibitemShut {NoStop}%
\bibitem [{\citenamefont {Spratt}\ and\ \citenamefont
  {Lisiecki}(2016)}]{Spratt2016}%
  \BibitemOpen
  \bibfield  {author} {\bibinfo {author} {\bibfnamefont {R.~M.}\ \bibnamefont
  {Spratt}}\ and\ \bibinfo {author} {\bibfnamefont {L.~E.}\ \bibnamefont
  {Lisiecki}},\ }\bibfield  {title} {\bibinfo {title} {{A Late Pleistocene sea
  level stack}},\ }\href@noop {} {\bibfield  {journal} {\bibinfo  {journal}
  {Climate of the Past}\ }\textbf {\bibinfo {volume} {12}},\ \bibinfo {pages}
  {1079} (\bibinfo {year} {2016})}\BibitemShut {NoStop}%
\bibitem [{\citenamefont {Grant}\ \emph {et~al.}(2014)\citenamefont {Grant},
  \citenamefont {Rohling}, \citenamefont {Ramsey}, \citenamefont {Cheng},
  \citenamefont {Edwards}, \citenamefont {Florindo}, \citenamefont {Heslop},
  \citenamefont {Marra}, \citenamefont {Roberts}, \citenamefont {Tamisiea}
  \emph {et~al.}}]{grant2014sea}%
  \BibitemOpen
  \bibfield  {author} {\bibinfo {author} {\bibfnamefont {K.}~\bibnamefont
  {Grant}}, \bibinfo {author} {\bibfnamefont {E.}~\bibnamefont {Rohling}},
  \bibinfo {author} {\bibfnamefont {C.~B.}\ \bibnamefont {Ramsey}}, \bibinfo
  {author} {\bibfnamefont {H.}~\bibnamefont {Cheng}}, \bibinfo {author}
  {\bibfnamefont {R.}~\bibnamefont {Edwards}}, \bibinfo {author} {\bibfnamefont
  {F.}~\bibnamefont {Florindo}}, \bibinfo {author} {\bibfnamefont
  {D.}~\bibnamefont {Heslop}}, \bibinfo {author} {\bibfnamefont
  {F.}~\bibnamefont {Marra}}, \bibinfo {author} {\bibfnamefont
  {A.}~\bibnamefont {Roberts}}, \bibinfo {author} {\bibfnamefont {M.~E.}\
  \bibnamefont {Tamisiea}}, \emph {et~al.},\ }\bibfield  {title} {\bibinfo
  {title} {Sea-level variability over five glacial cycles},\ }\href@noop {}
  {\bibfield  {journal} {\bibinfo  {journal} {Nature communications}\ }\textbf
  {\bibinfo {volume} {5}},\ \bibinfo {pages} {1} (\bibinfo {year}
  {2014})}\BibitemShut {NoStop}%
\bibitem [{\citenamefont {Bossomaier}\ \emph {et~al.}(2016)\citenamefont
  {Bossomaier}, \citenamefont {Barnett}, \citenamefont {Harr{\'e}},\ and\
  \citenamefont {Lizier}}]{Bossomaier2016}%
  \BibitemOpen
  \bibfield  {author} {\bibinfo {author} {\bibfnamefont {T.}~\bibnamefont
  {Bossomaier}}, \bibinfo {author} {\bibfnamefont {L.}~\bibnamefont {Barnett}},
  \bibinfo {author} {\bibfnamefont {M.}~\bibnamefont {Harr{\'e}}},\ and\
  \bibinfo {author} {\bibfnamefont {J.~T.}\ \bibnamefont {Lizier}},\ }\bibfield
   {title} {\bibinfo {title} {An introduction to transfer entropy},\
  }\href@noop {} {\bibfield  {journal} {\bibinfo  {journal} {Cham, Germany:
  Springer International Publishing. Crossref}\ } (\bibinfo {year}
  {2016})}\BibitemShut {NoStop}%
\bibitem [{Note1()}]{Note1}%
  \BibitemOpen
  \bibinfo {note} {$TE_{x \to y}$ measures a TE-like quantity, except the
  causal direction flips and the conditioning in the argument of the logarithm
  occurs only on one variable, not on a mixture of the two variables, as for
  regular TE (eqs. \ref {eq:te_forwardlags_xtoy_pos} and \ref
  {eq:te_forwardlags_ytox_pos}).}\BibitemShut {Stop}%
\bibitem [{\citenamefont {Sauer}\ \emph {et~al.}(1991)\citenamefont {Sauer},
  \citenamefont {Yorke},\ and\ \citenamefont {Casdagli}}]{Sauer1991embedology}%
  \BibitemOpen
  \bibfield  {author} {\bibinfo {author} {\bibfnamefont {T.}~\bibnamefont
  {Sauer}}, \bibinfo {author} {\bibfnamefont {J.~A.}\ \bibnamefont {Yorke}},\
  and\ \bibinfo {author} {\bibfnamefont {M.}~\bibnamefont {Casdagli}},\
  }\bibfield  {title} {\bibinfo {title} {Embedology},\ }\href@noop {}
  {\bibfield  {journal} {\bibinfo  {journal} {Journal of Statistical Physics}\
  }\textbf {\bibinfo {volume} {65}},\ \bibinfo {pages} {579} (\bibinfo {year}
  {1991})}\BibitemShut {NoStop}%
\bibitem [{\citenamefont {Deyle}\ and\ \citenamefont
  {Sugihara}(2011)}]{Deyle2011generalized}%
  \BibitemOpen
  \bibfield  {author} {\bibinfo {author} {\bibfnamefont {E.~R.}\ \bibnamefont
  {Deyle}}\ and\ \bibinfo {author} {\bibfnamefont {G.}~\bibnamefont
  {Sugihara}},\ }\bibfield  {title} {\bibinfo {title} {Generalized theorems for
  nonlinear state space reconstruction},\ }\href@noop {} {\bibfield  {journal}
  {\bibinfo  {journal} {PLoS One}\ }\textbf {\bibinfo {volume} {6}},\ \bibinfo
  {pages} {e18295} (\bibinfo {year} {2011})}\BibitemShut {NoStop}%
\bibitem [{\citenamefont {Kantz}\ and\ \citenamefont
  {Schreiber}(2004)}]{Kantz2004nonlinear}%
  \BibitemOpen
  \bibfield  {author} {\bibinfo {author} {\bibfnamefont {H.}~\bibnamefont
  {Kantz}}\ and\ \bibinfo {author} {\bibfnamefont {T.}~\bibnamefont
  {Schreiber}},\ }\href@noop {} {\emph {\bibinfo {title} {Nonlinear time series
  analysis}}},\ Vol.~\bibinfo {volume} {7}\ (\bibinfo  {publisher} {Cambridge
  university press},\ \bibinfo {year} {2004})\BibitemShut {NoStop}%
\bibitem [{\citenamefont {Cover}\ and\ \citenamefont
  {Thomas}(2012)}]{Cover2012}%
  \BibitemOpen
  \bibfield  {author} {\bibinfo {author} {\bibfnamefont {T.~M.}\ \bibnamefont
  {Cover}}\ and\ \bibinfo {author} {\bibfnamefont {J.~A.}\ \bibnamefont
  {Thomas}},\ }\href@noop {} {\emph {\bibinfo {title} {Elements of information
  theory}}}\ (\bibinfo  {publisher} {John Wiley \& Sons},\ \bibinfo {year}
  {2012})\BibitemShut {NoStop}%
\bibitem [{\citenamefont {Kraskov}\ \emph {et~al.}(2004)\citenamefont
  {Kraskov}, \citenamefont {St{\"{o}}gbauer},\ and\ \citenamefont
  {Grassberger}}]{Kraskov2004}%
  \BibitemOpen
  \bibfield  {author} {\bibinfo {author} {\bibfnamefont {A.}~\bibnamefont
  {Kraskov}}, \bibinfo {author} {\bibfnamefont {H.}~\bibnamefont
  {St{\"{o}}gbauer}},\ and\ \bibinfo {author} {\bibfnamefont {P.}~\bibnamefont
  {Grassberger}},\ }\bibfield  {title} {\bibinfo {title} {{Estimating mutual
  information}},\ }\bibfield  {journal} {\bibinfo  {journal} {Physical Review E
  - Statistical Physics, Plasmas, Fluids, and Related Interdisciplinary
  Topics}\ }\href {https://doi.org/10.1103/PhysRevE.69.066138}
  {10.1103/PhysRevE.69.066138} (\bibinfo {year} {2004}),\ \Eprint
  {https://arxiv.org/abs/0305641} {arXiv:0305641 [cond-mat]} \BibitemShut
  {NoStop}%
\bibitem [{\citenamefont {P{\'{e}}guin-Feissolle}\ and\ \citenamefont
  {Ter{\"{a}}svirta}(1999)}]{Peguin-Feissolle1999}%
  \BibitemOpen
  \bibfield  {author} {\bibinfo {author} {\bibfnamefont {A.}~\bibnamefont
  {P{\'{e}}guin-Feissolle}}\ and\ \bibinfo {author} {\bibfnamefont
  {T.}~\bibnamefont {Ter{\"{a}}svirta}},\ }\href@noop {} {\emph {\bibinfo
  {title} {{A General Framework for Testing the Granger Noncausality
  Hypothesis}}}}\ (\bibinfo  {publisher} {Universites d'Aix-Marseille II et
  III},\ \bibinfo {year} {1999})\BibitemShut {NoStop}%
\bibitem [{\citenamefont {Ch{\'{a}}vez}\ \emph {et~al.}(2003)\citenamefont
  {Ch{\'{a}}vez}, \citenamefont {Martinerie},\ and\ \citenamefont {{Le Van
  Quyen}}}]{Chavez2003}%
  \BibitemOpen
  \bibfield  {author} {\bibinfo {author} {\bibfnamefont {M.}~\bibnamefont
  {Ch{\'{a}}vez}}, \bibinfo {author} {\bibfnamefont {J.}~\bibnamefont
  {Martinerie}},\ and\ \bibinfo {author} {\bibfnamefont {M.}~\bibnamefont {{Le
  Van Quyen}}},\ }\bibfield  {title} {\bibinfo {title} {{Statistical assessment
  of nonlinear causality: application to epileptic EEG signals}},\ }\href@noop
  {} {\bibfield  {journal} {\bibinfo  {journal} {Journal of Neuroscience
  Methods}\ }\textbf {\bibinfo {volume} {124}},\ \bibinfo {pages} {113}
  (\bibinfo {year} {2003})}\BibitemShut {NoStop}%
\bibitem [{\citenamefont {Krakovsk{\'a}}\ \emph {et~al.}(2015)\citenamefont
  {Krakovsk{\'a}}, \citenamefont {Mezeiov{\'a}},\ and\ \citenamefont
  {Bud{\'a}{\v{c}}ov{\'a}}}]{Krakovska2015f1nn}%
  \BibitemOpen
  \bibfield  {author} {\bibinfo {author} {\bibfnamefont {A.}~\bibnamefont
  {Krakovsk{\'a}}}, \bibinfo {author} {\bibfnamefont {K.}~\bibnamefont
  {Mezeiov{\'a}}},\ and\ \bibinfo {author} {\bibfnamefont {H.}~\bibnamefont
  {Bud{\'a}{\v{c}}ov{\'a}}},\ }\bibfield  {title} {\bibinfo {title} {Use of
  false nearest neighbours for selecting variables and embedding parameters for
  state space reconstruction},\ }\href@noop {} {\bibfield  {journal} {\bibinfo
  {journal} {Journal of Complex Systems}\ }\textbf {\bibinfo {volume} {2015}}
  (\bibinfo {year} {2015})}\BibitemShut {NoStop}%
\bibitem [{\citenamefont {Mooij}\ \emph {et~al.}(2016)\citenamefont {Mooij},
  \citenamefont {Peters}, \citenamefont {Janzing}, \citenamefont
  {Zscheischler},\ and\ \citenamefont {Sch{\"o}lkopf}}]{Mooij2016}%
  \BibitemOpen
  \bibfield  {author} {\bibinfo {author} {\bibfnamefont {J.~M.}\ \bibnamefont
  {Mooij}}, \bibinfo {author} {\bibfnamefont {J.}~\bibnamefont {Peters}},
  \bibinfo {author} {\bibfnamefont {D.}~\bibnamefont {Janzing}}, \bibinfo
  {author} {\bibfnamefont {J.}~\bibnamefont {Zscheischler}},\ and\ \bibinfo
  {author} {\bibfnamefont {B.}~\bibnamefont {Sch{\"o}lkopf}},\ }\bibfield
  {title} {\bibinfo {title} {Distinguishing cause from effect using
  observational data: methods and benchmarks},\ }\href@noop {} {\bibfield
  {journal} {\bibinfo  {journal} {The Journal of Machine Learning Research}\
  }\textbf {\bibinfo {volume} {17}},\ \bibinfo {pages} {1103} (\bibinfo {year}
  {2016})}\BibitemShut {NoStop}%
\end{thebibliography}%

\addtocontents{toc}{\protect\partbegin}

\end{document}